\documentclass[12pt,letterpaper]{report}

\pdfoutput=1

\providecommand{\version}{local-dev}

\usepackage{style/buthesis}
\usepackage[T2A]{fontenc}
\usepackage[utf8]{inputenc}
\usepackage[english]{babel}

\usepackage{savesym}
\usepackage[usenames,dvipsnames]{xcolor}
\usepackage[
	lambda,
	advantage,
	operators,
	sets,
	landau,
	probability,
	notions,
	logic,
	ff,
	mm,
	primitives,
	events,
	complexity,
	asymptotics,
	keys
]{cryptocode}
\usepackage{appendix}
\usepackage{adjustbox}
\usepackage[ruled]{algorithm}
\usepackage{algorithmicx}
\usepackage[noend]{algpseudocode}
\usepackage{amsmath}
\usepackage{amsfonts}
\usepackage{subcaption}
\usepackage{graphicx}
\usepackage{balance}
\usepackage{diagbox}
\usepackage{mathtools}
\usepackage{nicefrac}
\usepackage{booktabs}
\usepackage{multirow}
\usepackage{bm}
\usepackage{interval}
\usepackage{rotating}
\usepackage{fancybox}
\usepackage{frcursive}
\usepackage{inslrmin}
\usepackage{calligra}
\usepackage{wedn}
\usepackage{pbsi}
\usepackage{aurical}
\usepackage{inslrmin}
\usepackage{calligra}
\savesymbol{st} 
\usepackage{soul}
\restoresymbol{TXF}{st}
\PassOptionsToPackage{hyphens}{url}
\definecolor{BrightRed}{RGB}{204, 0, 0}
\definecolor{DarkRed}{RGB}{102, 0, 0}
\definecolor{DarkerRed}{RGB}{78, 0, 0}
\definecolor{MidRed}{RGB}{143, 0, 0}

\definecolor{MatplotlibOne}{HTML}{1F77B4}
\definecolor{MatplotlibTwo}{HTML}{FF7F0E}
\definecolor{MatplotlibThree}{HTML}{2CA02C}
\definecolor{MatplotlibFour}{HTML}{D62728}
\definecolor{MatplotlibFive}{HTML}{9467BD}
\definecolor{MatplotlibSix}{HTML}{8C564B}
\definecolor{MatplotlibSeven}{HTML}{E377C2}
\definecolor{MatplotlibEight}{HTML}{7F7F7F}
\definecolor{MatplotlibNine}{HTML}{BCBD22}
\definecolor{MatplotlibTen}{HTML}{17BECF}

\usepackage[
	colorlinks = true,
	linkcolor = DarkRed,
	urlcolor = MidRed,
	citecolor = BrightRed,
	hyperindex = true,
]{hyperref}
\usepackage[noabbrev,capitalise]{cleveref} 

\AtBeginEnvironment{appendices}{\crefalias{chapter}{appendix}}

\usepackage{color}
\usepackage{colortbl}
\usepackage{amsthm}
\usepackage{caption}
\usepackage[all]{nowidow}
\usepackage{hyphenat}
\usepackage[inline]{enumitem}
\usepackage[binary-units=true]{siunitx}
\usepackage[mathscr]{eucal}
\usepackage{longtable}
\usepackage{makecell}
\usepackage{threeparttable}
\usepackage{textcomp}
\usepackage{array}
\usepackage{calc}
\usepackage[final]{pdfpages}
\usepackage{pax}

\usepackage[
	backend=biber,
	style=alphabetic,
	giveninits=false,
	sorting=nyt,
	maxbibnames=1000,
	maxalphanames=4
]{biblatex}

\bibliography{bibfile}

\graphicspath{{./graphics/}}

\ifdefined\draft%
	\usepackage{epstopdf}
	\DeclareGraphicsExtensions{.png,.pdf}
\fi

\usepackage[
	acronym,
	nonumberlist,
	nopostdot,
	nogroupskip
]{glossaries}

\newglossarystyle{gloTable}
{%
	\newlength\aglslong%
	\settowidth\aglslong{\widthof{\textbf{EMP-toolkit}}}

	\newlength\aglsleft%
	\setlength{\aglsleft}{\dimexpr\aglslong+0.1em\relax}

	\newlength\aglsright%
	\setlength{\aglsright}{\widthof{Bidirectional Encoder Representation from Transformer}}

	\newlength\aglscenter%
	\setlength{\aglscenter}{\dimexpr\linewidth-\aglsright-\aglsleft\relax}

	{
		\begin{center}
			\begin{longtable}{
				>{\raggedright}p{\aglsleft}
				p{\aglscenter}
				>{\raggedright}p{\aglsright}
			}
	}%
	{
			\end{longtable}
		\end{center}
	}%
	\renewcommand*{\glsgroupheading}[1]{}%
	\ifglsnogroupskip%
	\else
	\fi
}
\setglossarystyle{gloTable}

\makeglossaries%
\glsaddall%

\renewcommand*{\glstextformat}[1]{\textcolor{DarkerRed}{#1}}

\newcommand{\crypte}{Crypt\ensuremath{\epsilon}}
\newcommand{\epsolute}{\ensuremath{\mathcal{E}}p\-so\-lu\-te}
\newcommand{\kanon}{\ensuremath{k\text{-a}n\text{o}n}}

\newcommand{\record}{\ensuremath{r}}
\newcommand{\recordID}{\ensuremath{\record^\mathsf{ID}}}

\newcommand{\querySet}{\ensuremath{\mathcal{Q}}}
\newcommand{\queryKey}{\ensuremath{K}}

\newcommand{\domainSize}{\ensuremath{N}}
\newcommand{\dataSize}{\ensuremath{n}}
\newcommand{\oramsNumber}{\ensuremath{m}}

\newcommand{\user}{\ensuremath{\mathscr{U}}}
\newcommand{\client}{\ensuremath{\mathscr{C}}}
\newcommand{\server}{\ensuremath{\mathscr{S}}}

\newcommand{\protocol}{\ensuremath{\Pi}}
\newcommand{\protocolSetup}{\ensuremath{\protocol_{\mathsf{setup}}}}
\newcommand{\protocolQuery}{\ensuremath{\protocol_{\mathsf{query}}}}
\newcommand{\protocolNoGamma}{\ensuremath{\protocol_{\mathsf{no-}\gamma}}}
\newcommand{\protocolGamma}{\ensuremath{\protocol_\gamma}}

\newcommand{\searchKey}{\textsf{SK}}
\newcommand{\searchKeyDomain}{\ensuremath{\mathcal{X}}}

\newcommand{\serverDS}{\ensuremath{\mathcal{DS}}}
\newcommand{\indexI}{\ensuremath{\mathcal{I}}}
\newcommand{\database}{\ensuremath{\mathcal{D}}}
\newcommand{\databaseDef}{\ensuremath{\database = \allowbreak \{(\record_1, \allowbreak \recordID_1, \allowbreak \searchKey_1), \allowbreak \ldots, \allowbreak (\record_\dataSize, \allowbreak \recordID_\dataSize, \allowbreak \searchKey_\dataSize)\}}}
\newcommand{\fanout}{\ensuremath{k}}

\newcommand{\oram}{\ensuremath{\textsc{ORAM}}}
\newcommand{\oramProgram}{\ensuremath{\mathbf{y}}}
\newcommand{\oramRead}{\ensuremath{\mathbf{r}}}
\newcommand{\oramWrite}{\ensuremath{\mathbf{w}}}

\newcommand{\efficiencyCoefficient}{\ensuremath{a_1}}
\newcommand{\efficiencyOffset}{\ensuremath{a_2}}

\DeclareDocumentCommand{\algo}{ m g }{%
	{%
		\textsc{#1}%
		\IfNoValueF{#2}{\ensuremath{\left( #2 \right)}}%
	}%
}

\DeclareDocumentCommand{\query}{ g g }{%
	{%
		\IfValueTF{#2}%
			{\ensuremath{q_{\interval{#1}{#2}}}}%
			{
				\IfValueTF{#1}%
					{\ensuremath{q_{#1}}}%
					{\ensuremath{q}}
			}%
	}%
}

\newcommand{\adversary}{\ensuremath{\mathcal{A}}}

\renewcommand{\simulator}{\textsc{Sim}}
\DeclareDocumentCommand{\view}{ g g }{%
	{%
		\IfValueTF{#2}%
			{\ensuremath{\algo{View}_{#1} \left( #2 \right)}}%
			{
				\IfValueTF{#1}%
					{\ensuremath{\algo{View}_{#1}}}%
					{\ensuremath{\algo{View}}}
			}%
	}%
}

\providecommand{\bigTheta}[1]{\ensuremath{\Theta \left( #1 \right)}}

\newcommand{\fromNtoM}[3]{\ensuremath{#1_#2, \allowbreak \ldots, \allowbreak #1_#3}}
\newcommand{\probability}[1]{\ensuremath{\textnormal{Pr}\left[ #1 \right]}} 
\newcommand{\efficiency}[2]{\ensuremath{\left( \bigO{ #1 }, \ifthenelse{\equal{#2}{0}}{#2}{\bigO{ #2 }} \right)}}

\newcommand{\BPlus}{B\raisebox{.35\height}{\scalebox{.8}{+}}}

\newtheorem{definition}{Definition}[section]
\newtheorem{remark}{Remark}[section]

\newtheorem{theorem}{Theorem}[section]

\newtheorem{corollary}[theorem]{Corollary}

\makeatletter
	\newcommandx*{\sendmessageboth}[2][1=<->]{%
		\sendmessage{#1}{#2}%
	}
	\WithSuffix\newcommand\sendmessageboth*[2][\pcdefaultmessagelength]{%
		\begingroup%
			\renewcommand{\@pcsendmessagetop}{\let\halign\@pc@halign$\begin{aligned}#2\end{aligned}$}
			\sendmessage{<->}{length=#1}%
		\endgroup%
	}
\makeatother

\newcommand{\pcinput}[1]{\textbf{Input:}\ #1}
\newcommand{\pcouput}[1]{\textbf{Output:}\ #1}

\definecolor{lightGrey}{RGB}{220,220,220}

\makeatletter
	\let\orgdescriptionlabel\descriptionlabel%
	\renewcommand*{\descriptionlabel}[1]{%
		\let\orglabel\label%
		\let\label\@gobble
		\phantomsection%
		\protected@edef\@currentlabel{#1\unskip}
		\let\label\orglabel
		\orgdescriptionlabel{#1}%
	}
\makeatother

\AtBeginDocument{
	\pcfixhyperref%
	\makeatletter
		\crefformat{@pclinenumber}{line~#2#1#3}
		\Crefformat{@pclinenumber}{Line~#2#1#3}
		\crefrangeformat{@pclinenumber}{lines~#3#1#4 to~#5#2#6}
		\Crefrangeformat{@pclinenumber}{Lines~#3#1#4 to~#5#2#6}
	\makeatother
}

\newcommand{\domain}{\mathcal{D}}
\newcommand{\leak}{\mathcal{L}}
\newcommand{\setup}{\algo{KGen}}
\newcommand{\encrypt}{\algo{Enc}}
\newcommand{\compare}{\algo{Cmp}}

\newcommand{\Csharp}{%
	{\settoheight{\dimen0}{C}C\kern-.05em \resizebox{!}{\dimen0}{\raisebox{\depth}{\#}}} 
}

\newcommand{\FOne}{\ensuremath{\text{F}_1}}

\newcommand{\printpublication}[1]{\AtNextCite{\defcounter{maxnames}{99}}\fullcite{#1}}

\newcommand{\readerC}[5]{
	\ifdefined\emptyApproval%
		\reader{#1}{#2}{#3}{#4}{\rule{0.75\textwidth}{0.1mm}}
	\else
		\readerSigned{#1}{#2}{#3}{#4}{#5}
	\fi
}

\newacronym%
	{vm}
	{VM}
	{Virtual Machine}

\newacronym%
	{ore}
	{ORE}
	{Order-Revealing Encryption}

\newacronym%
	{prf}
	{PRF}
	{Pseudo-Random Function}

\newacronym%
	{prp}
	{PRP}
	{Pseudo-Random Permutation}

\newacronym%
	{sse}
	{SSE}
	{Searchable Symmetric Encryption}

\newacronym%
	{prg}
	{PRG}
	{Pseudo-Random Generator}

\newacronym%
	{pph}
	{PPH}
	{Property-Preserving Hash}

\newacronym%
	{ope}
	{OPE}
	{Order-Preserving Encryption}

\newacronym%
	{dp}
	{DP}
	{Differential Privacy}

\newacronym%
	{oram}
	{ORAM}
	{Oblivious Random Access Machine}

\newacronym%
	{hg}
	{HG}
	{Hypergeometric Probability Distribution}

\newacronym%
	{lpa}
	{LPA}
	{Laplacian Perturbation Algorithm}

\newacronym%
	{sql}
	{SQL}
	{Structured Query Language}

\newacronym%
	{io}
	{I/O}
	{Input / Output}

\newacronym%
	{rdbms}
	{RDBMS}
	{Relational Database Management System}

\newacronym%
	{knn}
	{\ensuremath{k}NN}
	{$k$-nearest-neighbor}

\newacronym%
	{dcpe}
	{DCPE}
	{Distance Comparison Preserving Encryption}

\newacronym%
	{cdpodb}
	{CDP-ODB}
	{Computationally \acrshort{dp} Outsourced Database System}

\newacronym%
	{trec}
	{TREC}
	{Text Retrieval Conference}

\newacronym%
	{mrr}
	{MRR}
	{Mean Reciprocal Rank}

\newacronym%
	{ndcg}
	{nDCG}
	{Normalized Discounted Cumulative Gain}

\newacronym%
	{faiss}
	{FAISS}
	{Facebook AI Similarity Search}

\newacronym%
	{gpu}
	{GPU}
	{Graphics Processing Unit}

\newacronym%
	{hipaa}
	{HIPAA}
	{Health Insurance Portability and Accountability Act}

\newacronym%
	{tee}
	{TEE}
	{Trusted Execution Environment}

\newacronym%
	{sgx}
	{SGX}
	{Software Guard Extensions}

\newacronym%
	{ap}
	{AP}
	{Access Pattern}

\newacronym%
	{cv}
	{CV}
	{Communication Volume}

\newacronym%
	{fpga}
	{FPGA}
	{Field-Programmable Gate Array}

\newacronym%
	{cdf}
	{CDF}
	{Cumulative Distribution Function}

\newacronym%
	{aspe}
	{ASPE}
	{Asymmetric Scalar-Product-Preserving Encryption}

\newacronym%
	{cpu}
	{CPU}
	{Central Processing Unit}

\newacronym%
	{kvs}
	{KVS}
	{Key-Value Store}

\newacronym%
	{pbc}
	{PBC}
	{Pairing-Based Crypto}

\newacronym%
	{emp}
	{EMP-toolkit}
	{Efficient MultiParty Computation Toolkit}

\newacronym%
	{brc}
	{BRC}
	{Best Range Cover}

\newacronym%
	{aws}
	{AWS}
	{Amazon Web Services}

\newacronym%
	{gcp}
	{GCP}
	{Google Cloud Platform}

\newacronym%
	{id}
	{ID}
	{Identifier}

\newacronym%
	{xml}
	{XML}
	{Extensible Markup Language}

\newacronym%
	{ssd}
	{SSD}
	{Solid State Drive}

\newacronym%
	{aes}
	{AES}
	{Advanced Encryption Standard}

\newacronym%
	{des}
	{DES}
	{Data Encryption Standard}

\newacronym%
	{sha}
	{SHA}
	{Secure Hash Algorithm}

\newacronym%
	{ecb}
	{ECB}
	{Electronic Codebook}

\newacronym%
	{cbc}
	{CBC}
	{Cipher Block Chaining}

\newacronym%
	{gcm}
	{GCM}
	{Galois / Counter}

\newacronym%
	{xts}
	{XTS}
	{\acrshort{xex}-based tweaked-codebook with ciphertext stealing}

\newacronym%
	{xex}
	{XEX}
	{XOR-Encrypt-XOR}

\newacronym%
	{ctr}
	{CTR}
	{Counter}

\newacronym%
	{iv}
	{IV}
	{Initialization Vector}

\newacronym%
	{nist}
	{NIST}
	{National Institute of Standards and Technology}

\newacronym%
	{ind-cpa}
	{IND-CPA}
	{Indistinguishability under \acrlong{cpa}}

\newacronym%
	{cpa}
	{CPA}
	{Chosen Plaintext Attack}

\newacronym%
	{cca}
	{CCA}
	{Chosen Ciphertext Attack}

\newacronym%
	{ind-cca}
	{IND-CCA}
	{Indistinguishability under \acrlong{cca}}

\newacronym%
	{lstm}
	{LSTM}
	{Long Short-Term Memory}

\newacronym%
	{ram}
	{RAM}
	{Random Access Memory}

\newacronym%
	{os}
	{OS}
	{Operating System}

\newacronym%
	{ml}
	{ML}
	{Machine Learning}

\newacronym%
	{bert}
	{BERT}
	{Bidirectional Encoder Representation from Transformer}

\newacronym%
	{nlp}
	{NLP}
	{Natural Language Processing}

\newacronym%
	{api}
	{API}
	{Application Programming Interface}

\newacronym%
	{mlc}
	{MLC}
	{Multi-Label Classification}

\newacronym%
	{msp}
	{MSP}
	{Multi-Set Prediction}

\begin{document}


\title{Secure and Efficient Query Processing in Outsourced Databases}
\author{Dmytro Bogatov}

\degree=2

\prevdegrees{
	B.S., Worcester Polytechnic Institute, 2017 \\
	M.S., Boston University, 2019
}

\department{Graduate School of Arts and Sciences}

\defenseyear{2022}
\degreeyear{2022}

\readerC{First}{George Kollios, Ph.D.}{Professor of Computer Science}{}{{\Fontskrivan\slshape\setul{0.1ex}{}\ul{~~Approved~~}}} 
\readerC{Second}{Leonid Reyzin, Ph.D.}{Professor of Computer Science}{}{{\calligra\setul{0.1ex}{}\ul{~~Approved~~}}}
\readerC{Third}{Manos Athanassoulis, Ph.D.}{Assistant Professor of Computer Science}{}{{\cursive\setul{0.1ex}{}\ul{~~Approved~~}}}
\readerC{Fourth}{Adam O'Neil, Ph.D.}{Assistant Professor of Computer Science}{University of Massachusetts --- Amherst}{{\wedn\setul{0.1ex}{}\ul{~~Approved~~}}}

\numadvisors=1
\majorprof{George Kollios, Ph.D.}{Professor of Computer Science}


\hypersetup{
	pdfauthor   = {Dmytro Bogatov <dmytro@bu.edu>},
	pdftitle    = {Secure and Efficient Query Processing in Outsourced Databases},
	pdfsubject  = {Doctoral Dissertation},
	pdfkeywords = {OPE, ORE, Range Query Protocols, Epsolute, kNN},
	pdfcreator  = {LaTeX with hyperref package},
	pdfproducer = {dvips + ps2pdf}
}

	\ifthenelse{1=1}{
		\maketitle
	}{
		Not particularly proud, but it seems necessary.
		I use pandoc to convert Tex to text and then run some plain text analysis.
		For some reason pandoc fails for the \\maketitle statement.
		However, it is not smart enough to deduce that this condition is always true, while Tex is.
	}
	\cleardoublepage%

	\copyrightpage%
	\cleardoublepage%

	\approvalpage%

	\cleardoublepage%

	\IfFileExists%
		{frontmatter/quote.tex}
		{
			\ifx\noAcks\undefined%
				\newpage
				{
	\null\vfill

	\newlength\longest%
	\settowidth\longest{\LARGE\itshape{} Ask not what your country can do for you ---} 
	\centering
	\parbox{\longest}{%
		\raggedright{\LARGE\itshape%
			Ask not what your country can do for you --- \\
			ask what you can do for your country.\/\par\bigskip
		}
		\raggedleft\Large\MakeUppercase{John Fitzgerald Kennedy}\par%
	}

	\vfill\vfill
}

				\cleardoublepage%
			\fi
		}
		{}

	\IfFileExists%
		{frontmatter/dedication.tex}
		{
			\ifx\noAcks\undefined%
				\newpage
				\section*{\centerline{Dedication}}

	I dedicate my dissertation work to my wife, my parents and the brave defenders of Ukraine.
	Dasha, love of my life, thank you for supporting me since the very first day of the program, through the long distance, once-in-a-century world-wide health crisis and the war.
	It was with your help that I was able to come so far.
	A special feeling of gratitude to my loving parents, Kostiantyn and Nataliia, whose help early in my academic career was critical and enabled my doctorate program in the first place, and whose support never waned to this day.
	Lastly, my heart and my thoughts go to the brave people of Ukraine, who are heroically defending the Motherland from the Russian invasion with their lives.
	We are with you, physically and virtually, in Lviv and Mariupol, in Kharkiv and in Kyiv.
	My sincere hope, my wish, is that this work will contribute, however little, to the cause of my beloved nation.
	Слава Україні!

				\cleardoublepage%
			\fi
		}
		{}

	\IfFileExists%
		{frontmatter/acknowledgments.tex}
		{
			\ifx\noAcks\undefined%
				\newpage
				\section*{\centerline{Acknowledgments}}

	First and foremost, I want to thank my advisor, George Kollios, for his guidance and continued support throughout my doctorate journey, from the day he accepted me to his lab to this moment.
	I am honored to call him and his family my friends.
	I also thank Leo Reyzin for the time and effort he spent helping me with my work, and for the many cups of coffee and long life talks we have had over these years.
	I want to especially note the many hours-long philosophical discussions with Manos Athanassoulis --- the discussions I truly enjoyed.
	In the heat of our debates we have lived up to the Ph.\ part of our degrees.
	My special thanks go to Adam O'Neil for his countless contributions to my works.
	He has been an invaluable co-author, contributor, reviewer and friend.

	I would like to express my gratitude to my closest ally in this life, my beloved wife, Daria Bogatova.
	There is not a single piece of writing I did without her review, not a single figure I crafted without her keen eye refining the tiniest details, from fonts to colors.
	Even the source code of this very thesis has been branched off from her Bachelor work \cite{dasha-thesis}, and the names of my systems, \epsolute{} and \kanon{} are her inventions.
	She has always been the first to listen to my ideas, and the first to support them unconditionally.

	I would particularly like to thank my bosom friend, roommate and fellow Doctor, Oleksandr Narykov.
	Living in Boston will not be the same without him, and I hope his career will bring him here in no time.
	I also want to thank my good friend and fellow BU alumnus, Vasili Ramanishka.
	I miss our arguments and discussions in the middle of a night.

	My deepest appreciation goes to my colleagues and co-authors, Georgios Kellaris, Björn Tackmann, Kaoutar Elkhiyaoui, Angelo De Caro, Kobbi Nissim and Hamed Zamani.
	It has been a great pleasure and a rewarding experience working with you.
	Your input in my work has been indispensable.
	I am deeply grateful to my fellow Amazonians, Kiran Chinta, Sriram Krishnamurthy, Naresh Chainani, Ramchandra Kulkarni and Abhishek Rai Sharma with whom I shared a number of internships.
	I am very excited to join the team full-time!

	I owe my deepest gratitude to my family, my parents, grandma and a little sister.
	I thank you for always being there for me.
	For attending my defense and admiring my presentation even though the content may have been very technical and in a foreign language.
	I am also infinitely glad that my family has expanded during my doctorate program.
	I thank Dasha's side of the family, in the similar composition of parents, grandma and a little sister, for accepting me and supporting me through my journey.

	Last but not least, I want to note my little fluffy friend, Pixel.
	My years of the program would not have been the same without him sleeping on my keyboard.

				\cleardoublepage%
			\fi
		}
		{}

	\begin{abstractpage}
		As organizations struggle with processing vast amounts of information, outsourcing sensitive data to third parties becomes a necessity.
Various cryptographic techniques are used in outsourced database systems to ensure data privacy while allowing for efficient querying.
This thesis proposes a definition and components of a new secure and efficient outsourced database system, which answers various types of queries, with different privacy guarantees in different security models.

This work starts with the survey of five order-preserving and order-revealing encryption schemes that can be used directly in many database indices, such as the B+ tree, and five range query protocols with various tradeoffs in terms of security and efficiency.
The survey systematizes the state-of-the-art range query solutions in a snapshot adversary setting and offers some non-obvious observations regarding the efficiency of the constructions.

The thesis then proceeds with \epsolute{} --- an efficient range query engine in a persistent adversary model.
In \epsolute{}, security is achieved in a setting with a much stronger adversary where she can continuously observe everything on the server, and leaking even the result size can enable a reconstruction attack.
\epsolute{} proposes a definition, construction, analysis, and experimental evaluation of a system that provably hides both access pattern and communication volume while remaining efficient.

The dissertation concludes with \kanon{} --- a secure similarity search engine in a snapshot adversary model.
The work presents a construction in which the security of \acrshort{knn} queries is achieved similarly to \acrshort{ope} / \acrshort{ore} solutions --- encrypting the input with an approximate \acrlong{dcpe} scheme so that the inputs, the points in a hyperspace, are perturbed, but the query algorithm still produces accurate results.
Analyzing the solution, we run a series of experiments to observe the tradeoff between search accuracy and attack effectiveness.
We use \acrshort{trec} datasets and queries for the search, and track the rank quality metrics such as \acrshort{mrr} and \acrshort{ndcg}.
For the attacks, we build an \acrshort{lstm} model that trains on the correlation between a sentence and its embedding and then predicts words from the embedding.
We conclude on viability and practicality of the solution.

	\end{abstractpage}
	\cleardoublepage%

	\setcounter{tocdepth}{4}
	\phantomsection%
	\addcontentsline{toc}{chapter}{Contents}
	\tableofcontents
	\cleardoublepage%

	\newpage
	\listofalgorithms%
	\addcontentsline{toc}{chapter}{\listalgorithmname}
	\cleardoublepage%

	\newpage
	\listoftables
	\addcontentsline{toc}{chapter}{\listtablename}
	\cleardoublepage%

	\newpage
	\listoffigures
	\addcontentsline{toc}{chapter}{\listfigurename}

	\addtocontents{toc}{\protect\contentsline{chapter}{\protect{}List of Acronyms}{xx}{link}}
	\printglossaries%
	\cleardoublepage%

	\newpage
	\endofprelim%
	\cleardoublepage%

	\chapter{Introduction}
\thispagestyle{myheadings}

	As the organizations struggle with demands for storage and processing of their data, they increasingly turn to third parties for outsourcing capabilities.
	A number of companies including Amazon (\acrshort{aws}), Microsoft (Azure) and Google (\acrshort{gcp}) offer outsourced database solutions to individuals and other businesses.
	This model is lucrative because not only do the clients pay exactly for what they use in terms of pure computational resources, but also the cloud takes care of the entire deployment process, including availability, scalability, replication, and, most importantly, security.
	The cloud business model provides resources on-demand --- from bare-bones \acrshortpl{vm} to database-as-a-service products.

	While the cloud providers typically have strict customer data privacy policies and even offer server-side encryption-at-rest services, the clients still have to trust the provider with their plaintext data.
	Server-side encryption-at-rest by definition requires the provider to know the encryption key to manipulate the data, even if the key is ephemeral and is not stored in the cloud permanently.
	Moreover, cloud providers in general, and customers' \acrshortpl{vm} in particular, may be vulnerable to external attacks --- from snapshot-level attacks, in which the adversary obtains a copy of the \acrshort{vm} memory, to more devastating persistent attacks, in which the adversary continuously monitors the \acrshort{vm} processes.

	Protecting the private information beyond cloud provider guarantees typically requires encrypting it in a way that preserves the ability to process it.
	A line of research targets securing outsourced database systems, but often achieves protection at the cost of efficiency too high for a solution to be viable for practical applications.
	In this thesis, we will cover the constructions that are used to answer different types of database queries in the outsourced model while providing both provable security and practical efficiency guarantees.

	\section{Model}

		In this work, we consider \emph{an outsourced database system} model, a notion first introduction in \cite{outsourced-db-intro}, adapted from \cite{generic-attacks-kellaris} and \cite{epsolute}.

		\subsection{Outsourced database model}\label{section:introduction:model:odb}

			Similar to \cite{epsolute}, a database is abstracted as a collection of \dataSize{} records \record{}, each with a unique identifier \recordID{}, associated with search keys \searchKey{}: \databaseDef{}.
			All records are assumed to have an identical fixed bit-length, and the search keys are elements of some domain \searchKeyDomain{}.
			A query is modeled as a predicate $\query \in \querySet: \searchKeyDomain \to \bin$.
			Evaluating a query \query{} on a database \database{} results in $\query( \database ) = \{ \record_i : \query( \searchKey_i ) = 1 \}$, all records whose search keys satisfy \query{}.

			Formally, an outsourced database system consists of two protocols between a stateful user \user{}, who owns the data, and an untrusted stateful server \server{}, to whom these data are outsourced.
			In setup protocol \protocolSetup{}, \user{} receives as input a database \databaseDef{} and \server{} may optionally output a data structure \serverDS{}.
			In query protocol \protocolQuery{}, \user{} receives a query $\query \in \querySet$, \server{} receives \serverDS{} produced in the setup protocol, and \user{} outputs the result of the query $\query( \database )$.
			Both parties may update their internal states.
			We call a system \emph{correct} if it holds with overwhelming probability over the randomness of the above runs that running \protocolSetup{} and  \protocolQuery{} on the corresponding inputs yields the correct result $\{ \record_i : \query( \searchKey_i ) = 1 \}$.

		\subsection{Security model}

			In an outsourced database system we are not aiming for perfect security, because the stronger the data protection guarantees are the harder it is to manipulate these data, the less efficient and functional the system becomes.
			There will always be some leakage, and our goal is to quantify, analyze and reduce it, while retaining the system's performance and usability.
			In terms of security models, we define two types of adversaries --- a \emph{snapshot} and a \emph{persistent} adversary.

			As the name suggests, a snapshot adversary can see a ``snapshot'' of the server's data at multiple points in time.
			One can think of such an attack as if someone steals a hard drive or accesses a backup.
			Formally, \adversary{} knows \server{} state at all stages of the protocol.

			A persistent adversary is stronger in that she monitors the server continuously.
			Therefore, she can see the same information as the snapshot adversary plus the network traffic and the access pattern.
			Such adversary can be thought of as a malicious software (virus) that runs as a background process with broad permissions.
			Formally, on top of the \server{} state, \adversary{} knows the size and content of \server{} communication and the sequence of accesses \server{} makes to its internal state at all protocol stages.

			Intuitively, one can think that encrypting records should protect the data.
			Depending on how the records are encrypted (i.e., whether the symmetric or property-preserving encryption is used), this approach can mitigate the snapshot adversary.
			Persistent adversary, however, can observe the communication size even if the traffic itself is encrypted, and can see the access pattern even if the records are protected.
			It has been shown that the knowledge of access pattern \cite{multidimensional-range-queries, inference-attack-islam-14, leakage-abuse-attacks-cash-15, inference-attacks-naveed-15, generic-attacks-kellaris, attacks-tao-of-inference, grubbs-attacks, access-pattern-disclosure, attacks-improved-reconstruction} or communication volume \cite{generic-attacks-kellaris, state-of-uniform, attacks-improved-reconstruction, pump-volume-attacks, volume-range-attacks} alone can enable a series of reconstruction attacks.

			Note that both adversaries are \emph{honest-but-curios} --- they only observe and never interfere.
			Denial-of-service attacks and integrity protection are out of scope of this work.

		\subsection{Query types}\label{section:intro:model:query-types}

			The type of query \query{} is deliberately left abstract.
			The outsourced database system assumes that a query contains a way (a predicate) to select only the records whose search keys satisfy it.
			In this work, we consider the following types of queries.

			\paragraph*{A point query.}
				This query selects records whose key is equal to a given value.
				The domain of the point value does not have to be ordered; it can be categorical, like color names.
				The relevant \acrshort{sql} query can be

				\begin{center}
					\texttt{SELECT * FROM t1 WHERE color = 'blue'}. 
				\end{center}

			\paragraph*{A range query.}
				This query selects records whose keys lie between two values from an ordered domain.
				The relevant \acrshort{sql} query can be

				\begin{center}
					\texttt{SELECT * FROM t1 WHERE age BETWEEN 18 and 65}.
				\end{center}

			\paragraph*{A \acrshort{knn} query.}
				\acrlong{knn} query selects $k$ records whose keys are ``closest'' to a given value.
				This query type requires a definition of distance metric over the domain of search keys, for example, simple Euclidean distance.
				The relevant \acrshort{sql} query can be

				\begin{center}
					\texttt{SELECT * FROM t1 ORDER BY location <-> '(29.9691,-95.6972)' LIMIT 5}. 
				\end{center}

	\section{Thesis structure}

		In \cref{section:background}, we will cover the building blocks that are used in the outsourced database systems.
		These blocks include symmetric encryption, \acrshort{oram} and \acrlong{dp}.
		\cref{section:related-work} includes the overview of approaches that provide the privacy and/or security in the outsourced setting.
		The chapter also includes the overview of the attacks against the mechanisms.
		\cref{section:range-snapshot} analyses in detail the range query mechanisms in the snapshot adversary model.
		The chapter offers a comparative evaluation of five \acrlong{ore} schemes and five secure range query protocols \cite{ore-benchmark-17}.
		\cref{section:range-persistent} proposes a novel solution for the range queries in the persistent adversary model, \epsolute{} \cite{ore-benchmark-17}.
		In \cref{section:knn-snapshot}, we describe \kanon{} \cite{k-anon} --- a mechanism to answer similarity search (i.e., \acrlong{knn}) queries in a snapshot adversary setting using a type of property-preserving encryption similar to \acrshort{ope}.
		In the chapter we describe the encryption method, the set of experiments to empirically measure the search accuracy and the level of protection against the attacks.
		The source code of this thesis is publicly available.\footnote{
			The thesis has been compiled
			\ifthenelse%
				{\equal{\version}{local-dev}}
				{locally}
				{from \href{https://github.com/dbogatov/doctoral-thesis/commit/\version}{\texttt{dbogatov/doctoral-thesis@\version}}}
			on \today{}.
		}

		\subsection{Works completed during the Ph.D. program}

			\newlength{\citationLength}
			\settowidth{\citationLength}{\cite{k-anon}}
			\setlength{\citationLength}{\dimexpr\citationLength+0.9em\relax}
			\begin{description}[
				leftmargin=\dimexpr\citationLength+0.7em\relax,
				labelindent=0pt,
				labelwidth=\citationLength%
			]
				\item[\cite{ore-benchmark-17}] \printpublication{ore-benchmark-17}
				\item[\cite{epsolute}] \printpublication{epsolute}
				\item[\cite{k-anon}] \printpublication{k-anon}
				\item[\cite{bogatov-idemix-2020}] \printpublication{bogatov-idemix-2020}
				\item[\cite{dispot}] \printpublication{dispot}
			\end{description}

			The works \cite{ore-benchmark-17,epsolute,k-anon} are discussed in \cref{section:range-snapshot,section:range-persistent,section:knn-snapshot} respectively.
			The other two works \cite{bogatov-idemix-2020,dispot} fall outside of scope of this thesis.
			\cite{bogatov-idemix-2020} proposes an imporved anonymous delegatable credential scheme and its novel extensions, auditability and revocation, along with the instantiation and comprehensible set of experiments, see the abstract in \cref{section:appendix:idemix-abstract}.
			\cite{dispot} is from the domain of Bioinformatics, it offers a knowledge-based statistical potential that estimates the propensity of an interaction between a pair of protein domains, see the abstract in \cref{section:appendix:dispot-abstract}.
			I also want to note the works I have completed on the route to my doctorate program \cite{bogatov-ipe-journal-2017,bogatov-wpi-library-2016,nurbekov-wpi-library-2015}.

	\cleardoublepage%
	\chapter{Background}\label{section:background}
\thispagestyle{myheadings}

	In this section, we will go over the building blocks required to construct the outsourced database systems and their components that we will discuss in the next chapters.
	These prerequisites include the symmetric encryption, \acrshortpl{oram} and PathORAM \cite{path-oram} in particular, \acrlong{dp} and \acrshort{dp} sanitizers, and finally, \acrlongpl{tee}.
	\emph{Some of the following sections were paraphrased or taken verbatim from my published work \cite{ore-benchmark-17,epsolute}.}

	\section{Symmetric encryption}\label{section:background:encryption}

		Symmetric encryption scheme is a tuple of algorithms $\algo{E} = \{ \algo{KeyGen}, \algo{Enc}, \algo{Dec} \}$ with the following properties.
		$\algo{E.KeyGen} \left( \secparam \right) \to \key$ is a \emph{randomized} algorithm that on a security parameter $\secparam$ returns a key that will be used for both encryption and decryption.
		$\algo{E.Enc} \left( m \right) \to c$ is a \emph{randomized} algorithm that on a plaintext message $m \in \bin^*$ produces its ciphertext $c \in \bin^*$.
		$\algo{E.Dec} \left( c \right) \to m$ is a \emph{deterministic} algorithm that on a ciphertext $c \in \bin^*$ produces its original plaintext message $m \in \bin^*$.

		\subsection{Security}

			The security of the symmetric encryption is typically defined as the indistinguishability under a certain attack.
			The definition is structured around the game between the challenger and the adversary \adversary{}.
			The challenger fixes one of the two ``worlds'', left or right, and the adversary wins the game if she can reliably tell which world it was.

			A weaker security definition, \acrfull{ind-cpa}, intuitively, requires that the ciphertext leaks nothing about the plaintext.
			To formalize the requirement, the adversary can give the challenger a set of plaintext pairs to encrypt, and the challenger responds with a set of ciphertexts where the left or the right part was encrypted.
			The adversary can then use any (polynomial-time) algorithm over the ciphertexts and make a guess of whether the left or the right part was encrypted.
			The claim is, if there is anything that the ciphertext leaks about the plaintext, there exists an adversary who will win.
			The security claim is therefore contrapositive --- the scheme is \acrshort{ind-cpa} secure iff there is no such adversary that wins the game.

			\acrshort{ind-cpa} security is not by itself sufficient since it does not account for the decryption part of the scheme.
			There are known attacks that decrypt the plaintext (i.e., defeat the encryption) if the adversary can trigger the decryption and observe the process or the result (see padding attack \cite{padding-attack} and \acrshort{xml} encryption attack \cite{xml-break-encryption}).

			The stronger definition, \acrfull{ind-cca}, captures the decryption component.
			It extends the \acrshort{ind-cpa} game in that the adversary can now request the challenger to decrypt \emph{any} ciphertext of her choice \emph{except} the ones that the challenger himself encrypted for the adversary.
			The adversary still outputs a guess of the two worlds and wins if reliably guesses correctly.
			Note that in this game if the decryption can fail for any reason, or even if \adversary{} can observe any difference in execution for different inputs, the scheme is insecure.
			Therefore, \acrshort{ind-cca} immediately rules out the aforementioned attacks \cite{padding-attack,xml-break-encryption}.

		\subsection{Components}

			Note that for the practical purposes we define the encryption algorithm as randomized --- producing different ciphertext for the same plaintext on every invocation.
			While this is how the symmetric encryption scheme is used in applications, formally, producing the deterministic ciphertext and randomizing it are different operations.

			The randomness is produced independently, typically using a \acrfull{prg}, and is used for both secrecy and integrity (which is necessary for \acrshort{cca} security).
			After obtaining the random bits, the algorithm repeatedly uses the block cipher (formally, a \acrfull{prp}), with the number of invocations linear in the message length.
			How the randomness and the message are combined is defined by the \emph{mode of operation} and differs from one encryption scheme to another.
			Typically, the randomness comes in a form of a block, an \acrfull{iv}, filled with random bits.
			The mode of operation then defines how the blocks and the \acrshort{iv} are combined together.

			In practical systems, the ciphertext is then broken up into components, like the ciphertext material itself, the \acrshort{iv} (varies by the scheme), the version of the key, etc.
			Also note that the encryption scheme key has a maximum number of times it can be used for encryption (its \emph{operational lifetime}).

			When it comes to the real-world encryption systems, we use standardized primitives --- a block cipher (\acrshort{prp}), a \acrlong{prg} and a mode of operation.
			\acrfull{aes} \cite{aes-nist} is a \acrshort{nist}-standardized block cipher, which operates on 128-bit blocks.
			\acrshort{nist} also offers recommendations for random number generator mechanisms \cite{nist-prg-mechanism}, constructions \cite{nist-prg-constructions} and sources of entropy \cite{nist-prg-entropy}.
			Lastly, some of the commonly used modes of operation are \acrshort{cbc} and \acrshort{ctr} \cite{nist-modes} modes for general-purpose encryption, \acrshort{gcm} \cite{nist-gcm} for an authenticated encryption and \acrshort{xts} \cite{ieee-xts} for encryption of data on block-oriented devices (e.g., disks).

	\section{\texorpdfstring{\acrlong{oram}}{Oblivious Random Access Machine}}\label{section:background:oram}

		Informally, \acrfull{oram} is a mechanism that lets the users hide their access pattern to remote storage.
		An adversarial server can monitor the actual accessed locations, but she cannot tell a read from a write, the content of the block or even whether the same logical location is being referenced.
		The notion was first defined by \textcite{oram-theory} and \textcite{oram-original}.

		More formally, a $(\eta_1, \eta_2)$-\acrshort{oram} protocol is a two-party protocol between a client \client{} and a server \server{} who maintains the storage in a form of array of blocks.
		In each round, the client \client{} has input $(o, a, d)$, where $o$ is an access type (\oramRead{} or \oramWrite{}), $a$ is a storage block address and $d$ is a new data value, or $\bot$ for read operation.
		The input of \server{} is the current storage array.
		Via the protocol, the server updates the storage or returns to \user{} the data stored at the requested block, respectively.
		We speak of a sequence of such operations as a program \oramProgram{} being \emph{executed under the \acrshort{oram}}.

		An \acrshort{oram} protocol must satisfy correctness and security.
		Correctness requires that \client{} obtains the correct output of the computation except with at most probability $\eta_1$.
		For security, we require that for every client \client{} there exists a simulator $\simulator_\oram$ which provides a simulation of the server's view in the above experiment given only the number of operations.
		That is, the output distribution of $\simulator_\oram (c)$ is indistinguishable from $\algo{View}_\server$ with probability at most $\eta_2$ after $c$ protocol rounds.
		Note that the \acrshort{oram} protocols typically differ in the way the storage is organized and manipulated, but are similar in that the records or blocks are symmetrically encrypted (see \cref{section:background:encryption}).

		\acrshort{oram} protocols are generally stateful, after each execution the client and server states are updated.
		\emph{For brevity, throughout the thesis we will assume the \acrshort{oram} state updates are implicit, including the encryption key generated and maintained by the client.}

		Some existing efficient \acrshort{oram} protocols are Square Root \acrshort{oram} \cite{oram-theory}, Hierarchical \acrshort{oram} \cite{oram-original}, Binary-Tree \acrshort{oram} \cite{binary-tree-oram}, Interleave Buffer Shuffle Square Root \acrshort{oram} \cite{shortest-path-oram}, TP-ORAM \cite{tp-oram}, PathORAM \cite{path-oram} and TaORAM \cite{taostore}.
		For detailed descriptions of each protocol, we recommend the work of \textcite{oram-survey-feifei}.
		The latter three \acrshortpl{oram} achieve the lowest communication and storage overheads, $\bigO{\log \dataSize}$ and \bigO{\dataSize}, respectively.

		\subsection{PathORAM}

			PathORAM \cite{path-oram} is one of the most commonly used \acrshort{oram} protocols due to its efficiency and simplicity.
			In this section we briefly describe this construction as it is used as an \acrshort{oram} instantiation in the rest of the thesis.

			In the PathORAM, both the client \client{} and the server \server{} are stateful.
			The server stores the encrypted records (blocks) grouped in buckets, and the buckets form a binary tree.
			The client's storage, although asymptotically linear in the data size, is relatively small in practice.
			The client stores the \emph{position map} that maps the record \acrshort{id} to a leaf in the server tree storage and a small amount of stash, which can store some plaintext blocks on the client side.

			The main invariant of the protocol is that at all times the ciphertext of the record $a$ is stored somewhere on the \emph{path} from the root to the leaf that is mapped to $a$, or in the stash (hence the name of the construction).

			The \acrshort{oram} is initialized with a binary tree of buckets with all buckets containing valid encryptions of dummy records.
			The position map is sampled at random (filled with permuted distinct numbers).

			Main routine of the \acrshort{oram} is an access sub-protocol, which is similar for read \oramRead{} and write \oramWrite{} types of access (remember, \acrshort{oram} hides the type of access from the curious server).
			In PathORAM, the access $(o, a, d)$ consists of four steps.
			First, remap the current leaf $x$ for $a$ to a new random leaf $x^\prime$.
			Second, read the entire path to leaf $x$ (all buckets from root to leaf) into the client stash.
			Third, the client updates the block value to $d$ if the access is a write \oramWrite{}.
			Finally, write back the path to leaf $x$ filling the buckets with all blocks from stash in a way that maintains the invariant.

			The newly updated block $a$ with the new value $d$ may be included in the new path, or it may stay in the stash.
			It is important that the stash size be provably bounded.
			\cite[Theorem 1]{path-oram} does exactly that --- at least for the bucket size of 5, the probability of stash overflow and its size are related as in \cref{equation:path-oram-stash}.
			\begin{equation}\label{equation:path-oram-stash}
				\probability{ \mathsf{stash\ size} > x } \leq 14 \cdot (0.6002) ^ x
			\end{equation}
			If the probability of a protocol failure is thought of as the adversary's advantage (the probability of breaking the security), then the stash size equivalent to 128-bit security is about 100 blocks for the bucket of size 5 \cite[Figure 5]{path-oram}.

	\section{\texorpdfstring{\acrlong{dp}}{Differential Privacy}}

		\acrfull{dp} is a guarantee on a mechanism that takes a dataset and returns some result.
		The guarantee states that for two neighboring databases (that differ in exactly one record), the probability that the adversary will understand by looking at the output, which of the two databases was used as an input, is bounded.
		More formally, \acrlong{dp} is defined in \cref{definition:dp}.

		\begin{definition}[\acrlong{dp}, adapted from \cite{our-data-ourselves, differential-privacy-original}]\label{definition:dp}
			A randomized algorithm \algo{A} is $(\epsilon, \delta)$-differentially private if for all $\database_1 \sim \database_2 \in \searchKeyDomain^\dataSize$, and for all subsets $\mathcal{O}$ of the output space of \algo{A},
			\[
				\probability{ \algo{A}{ \database_1 } \in \mathcal{O} } \leq \exp(\epsilon) \cdot \probability{ \algo{A}{ \database_2 } \in \mathcal{O} } + \delta \; .
			\]
		\end{definition}

		One way to interpret this definition is the following.
		Probabilities are taken over the coins of algorithm \algo{A}, which answers a query based on a dataset.
		A natural instantiation of \algo{A} is a view of a distinguishing adversary \adversary{}, who tries to guess which of the two datasets was used.
		The expression in \cref{definition:dp} then bounds the advantage of \adversary{} with $\epsilon$ and $\delta$ parameters.
		Note that $\exp( x ) \approx 1 + x + \frac{x^2}{2!}$, and for sufficiently small $x$ the last term is negligible.
		If we put $\epsilon + 1$ in place of $\exp( \epsilon )$, it becomes clear that $\epsilon$ is the exact value by which two probabilities are allowed to differ.
		For $\epsilon = 0$, they have to be equal, for $\epsilon = 0.01$, probabilities may differ by \SI{1}{\percent}.
		Therefore, $\epsilon$ is called \emph{a privacy budget} of a \acrshort{dp} system.
		$\delta$ term is additive and therefore must be small by itself.
		This term is essentially a probability that the entire system fails.
		For example, if \algo{A} is randomized and fails with a certain chance, this probability will be $\delta$.
		For instance, a PathORAM \cite{path-oram} algorithm can have a stash overflow with a bounded probability \cite[Theorem 1]{path-oram} and it will cause the entire system to fail.
		If PathORAM is used in a \acrshort{dp} system then this probability, however small, bounded and negligible, will have to be accounted for in $\delta$.

		Note that \cref{definition:dp} describes a property of \algo{A} and not a construction method.
		To construct \algo{A}, the seminal work of \textcite{differential-privacy-original} offers an algorithm called \acrfull{lpa}.
		The idea is to tune the noise sampled from the Laplacian distribution to the \emph{sensitivity} of a query, defined as the change of output with respect to change in input.
		For example, if a change in one record of the dataset causes a change in the output value of at most one (e.g., a count query), then the sensitivity is 1.
		\cite{differential-privacy-original} proves that if one adds $\algo{Laplace}{0, \frac{\mathsf{sensitivity}}{\epsilon}}$ to the real result of a query, the resulting mechanism is $\epsilon$-\acrshort{dp}.

		\subsection{\texorpdfstring{\acrshort{dp}}{DP} sanitizers}\label{section:background:dp-sanitizers}

			While the \acrlong{lpa} is an effective and simple way of answering a single count query, we will need to answer a sequence of count queries, ideally, without imposing a bound on the length of this sequence.
			We will hence make use of \emph{sanitization} algorithms.

			\begin{definition}\label{definition:dp-danitizer}
				Let \querySet{} be a collection of queries.
				An $(\epsilon, \delta, \alpha, \beta)$-differentially private sanitizer for \querySet{} is a pair of algorithms $(\algo{A}, \algo{B})$ such that:
				\begin{itemize}
					\item $A$ is $(\epsilon, \delta)$-differentially private, and
					\item on input a dataset $\database = \fromNtoM{d}{1}{\dataSize} \in \searchKeyDomain^\dataSize$, \algo{A} outputs a data structure \serverDS{} such that with probability $1 - \beta$ for all $\query \in \querySet$, $\abs{ \algo{B}{ \serverDS, \query } - \sum_i \query(d_i) } \leq \alpha$.
				\end{itemize}
			\end{definition}

			In \cref{definition:dp-danitizer}, the query is a predicate, which is defined as in \cref{section:introduction:model:odb}, and it returns a binary 0 or 1 when executed over a search key (an attribute) $d \in \searchKeyDomain$.
			$\algo{B}{ \serverDS, \query }$ then returns a count, a scalar value which bounds the number of search keys that satisfy the query.
			For example, a query \query{} can be a range query, then $ \sum_i \query(d_i) $ is the number of records in the range, and $ \algo{B}{ \serverDS, \query } $ returns a scalar value of noise (the fake records), such that it is within $\alpha$ from the true count.

			\begin{remark}\label{remark:dp-sanitizer-guarantees}
				Given an $(\epsilon, \delta, \alpha, \beta)$-\acrshort{dp} sanitizer as in \cref{definition:dp-danitizer} one can replace the answer $\algo{B}{ \serverDS, \query }$ with $\textsc{B}^\prime ( \serverDS, \allowbreak \query ) = \algo{B}{ \serverDS, \query } + \alpha$.
				Hence, with probability $1 - \beta$, for all $\query \in \querySet$, $0 \leq \algo{\ensuremath{\textsc{B}^\prime}}{ \serverDS, \query } - \sum_i \query(d_i) \leq 2 \alpha$.
				We will hence assume from now on that sanitizers have this latter guarantee on their error.
			\end{remark}

			The main idea of \emph{sanitization} (a.k.a.\ private data release) is to release specific noisy statistics on a private dataset once, which can then be combined in order to answer an arbitrary number of queries without violating privacy.
			Depending on the query type (see \cref{section:intro:model:query-types}) and the notion of \acrlong{dp} (i.e., pure or approximate), different upper bounds on the error have been proven.
			Omitting the dependency on $\epsilon$ and $\delta$, in case of point queries over domain size \domainSize{}, pure \acrlong{dp} results in $\alpha = \bigTheta{\log \domainSize}$ \cite{bounds-on-sample-complexity}, while for approximate \acrlong{dp} $\alpha = \bigO{1}$ \cite{private-learning-and-sanitization}.
			Note that the sanitizers add noise to a count of records, and for the point queries the count is the number of records with a given categorical value (i.e., number of female students in a class).
			For range queries over domain size \domainSize{}, these bounds are $\alpha = \bigTheta{\log \domainSize}$ for pure \acrlong{dp} \cite{non-interactive-database-privacy,dp-under-observation}, and $\alpha = \bigO{(\log^{*} \domainSize)^{1.5}}$ for approximate \acrlong{dp} (with an almost matching lower bound of $\alpha = \bigOmega{\log^{*} \domainSize}$) \cite{private-learning-and-sanitization, dp-release, privately-learning-thresholds}.
			More generally, \textcite{non-interactive-database-privacy} showed that any finite query set \querySet{} can be sanitized, albeit non-efficiently.

			\subsubsection{Answering point and range queries with differential privacy}

				Utilizing the \acrshort{lpa} for answering point queries results in error $\alpha = \bigO{\log \domainSize}$.
				A practical solution for answering range queries with error bounds very close to the optimal ones is the hierarchical method \cite{dp-under-observation, accuracy-dp-histograms, dp-wavelet}.
				The main idea is to build an aggregate tree on the domain, and add noise to each node proportional to the tree height (i.e., noise scale logarithmic in the domain size \domainSize{}).
				Then, every range query is answered using the minimum number of tree nodes.
				\textcite{hierarchical-methods-for-dp} showed that the hierarchical algorithm of \textcite{accuracy-dp-histograms}, when combined with their proposed optimizations, offers the lowest error.

			\subsubsection{Composition}

				Finally, in this thesis we will make use of a \emph{composition} theorem (adapted from \cite{privacy-integrated-queries}) based on \cite{differential-privacy-original,our-data-ourselves}. 
				It concerns executions of multiple \acrshort{dp} mechanisms on non-disjoint and disjoint inputs.

				\begin{theorem}\label{theorem:composition}
					Let \fromNtoM{\algo{A}}{1}{r} be mechanisms, such that each $\algo{A}_i$ provides $\epsilon_i$-\acrshort{dp}.
					Let \fromNtoM{\database}{1}{r} be pairwise non-disjoint (resp., disjoint) datasets.
					Let $\algo{A}$ be another mechanism that executes $\algo{A}_1(\database_1), \ldots, \algo{A}_r(\database_r)$ using independent randomness for each $\algo{A}_i$, and returns their outputs.
					Then, mechanism $\algo{A}$ is $\left( \sum_{i=1}^r \epsilon_i \right)$-\acrshort{dp} (resp., $\left( \max_{i=1}^r \epsilon_i \right)$-\acrshort{dp}).
				\end{theorem}

	\section{\texorpdfstring{\acrlongpl{tee}}{Trusted Execution Environments}}

\begin{figure}[!ht]
	\centering
	\includegraphics[width=\textwidth]{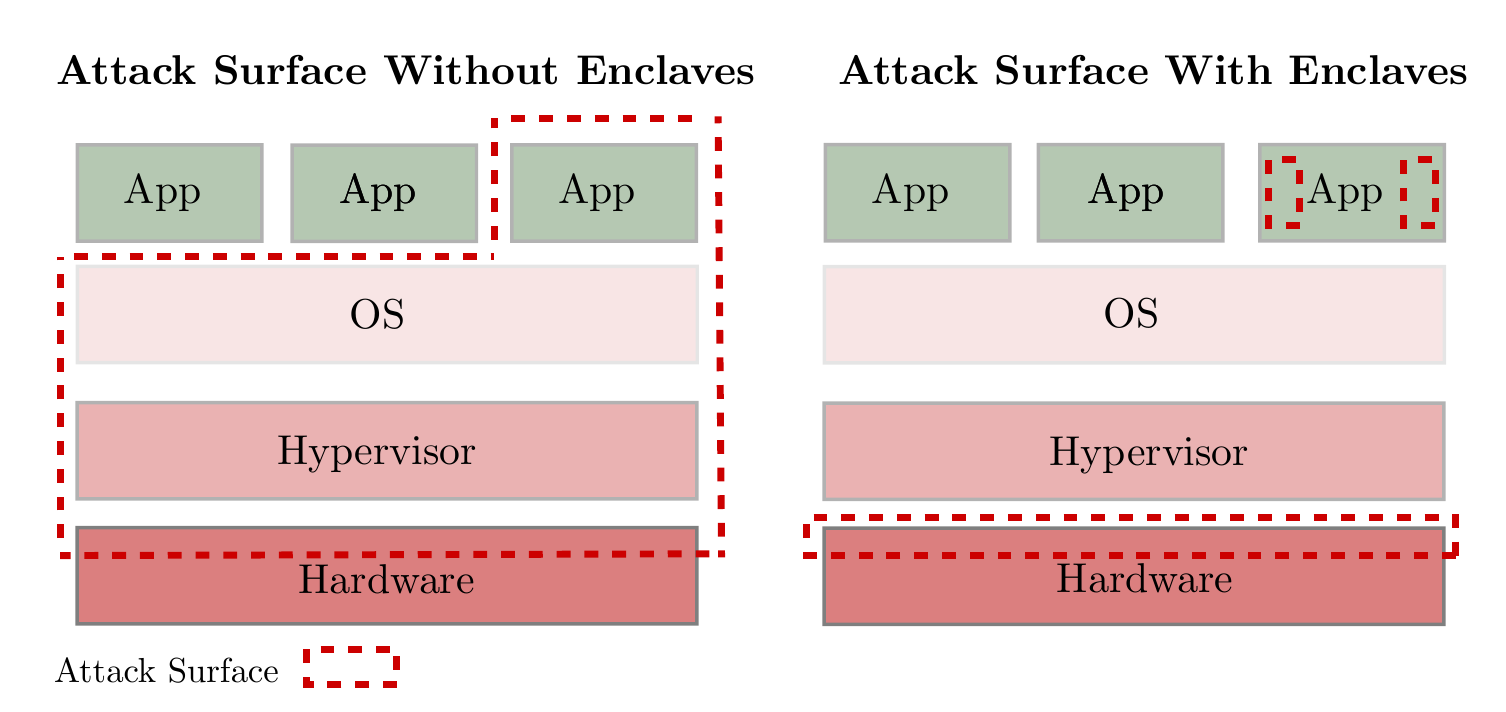}
	\caption[Attack surface with \acrshort{tee}]{
		Attack surface with \acrshort{tee}.
		Adapted from \url{sgx101.gitbook.io}.
	}%
	\label{figure:enclave}
\end{figure}

		\acrlongpl{tee} is a generalized term for a ``protected'' part of a processing engine.
		Security in this setting means a combination of confidentiality, integrity, secrecy, isolation and protection against side-channel attacks.
		\acrshort{tee} cannot be entered through system calls, jumps or register manipulations.
		Environment's memory content and integrity are protected, and neither \acrshort{os}, nor a hypervisor can access it.
		Main purpose of the \acrshort{tee} is to vastly reduce the attack surface, see \cref{figure:enclave}.

		While \acrshort{tee} is a concept defining an execution environment, specific solutions include a hardware security module (a plug-in device with a protected memory and a crypto-optimized processing unit), an \acrshort{fpga}, and a set of extensions to an existing processor architecture, such as an \acrshort{sgx} for Intel x86, Secure Encrypted Virtualization \cite{amd-memory-encryption} for AMD and Apple's Secure Enclave\footnote{\url{https://support.apple.com/guide/security/secure-enclave-sec59b0b31ff/web}}.
		Of all these, Intel \acrshort{sgx} is the most widely used technology.

		\subsection{\texorpdfstring{\acrlong{sgx}}{Software Guard Extensions}}

			\acrfull{sgx} is a set of instructions for the Intel x86 architecture that allow a user or an \acrlong{os} to define a region of protected memory, called the \emph{enclave}, and interact with it.
			The enclave can only be accessed using the \acrshort{sgx} instructions (i.e, regular \texttt{mov} instruction would not work), and all pages of the enclave are symmetrically encrypted and physically protected.
			\acrshort{sgx} guarantees the integrity and security of the memory pages within the enclave.
			Although the size of enclave memory is very limited, \acrshort{sgx} can use regular \acrshort{ram} by transparently swapping the pages between the trusted and untrusted memory.
			The pages are then encrypted with integrity protection when placed in the \acrshort{ram} (\emph{sealing} in \acrshort{sgx} terms).

			An \acrshort{sgx}-enabled application declares its trusted and untrusted components upfront.
			Trusted part will live entirely in the enclave, while the untrusted part is a normal process that runs within the \acrshort{os}.
			The application has to be digitally signed for the enclave to accept it, and the enclave itself can authenticate to the user via an \emph{attestation} process.
			Conceptually, the simplest \acrshort{sgx} application in an outsourced database system can be seen as a trusted component that operates over sensitive material (e.g., keys, tokens, plaintext user data), a remote trusted client application that communicates with the enclave, and a layer of code that passes requests through (untrusted part of an \acrshort{sgx} application).
			Cipherbase \cite{cipherbase-daas} and StealthDB \cite{stealth-db} are good examples of such approach.

		\subsection{Issues with \acrshort{sgx}}

			\acrshort{sgx}, as the closest instantiation of \acrshort{tee} available, has been extensively targeted.
			The attacks include Foreshadow \cite{foreshadow}, Prime+Probe cache attack \cite{prime-probe-sgx-attack}, an attack from within the enclave \cite{enclave-sgx-attack}, Spectre line of attacks that can bypass \acrshort{sgx} \cite{spectre-sgx-attack}, replay attack \cite{replay-sgx-attack}, Plundervolt attack \cite{plundervolt-sgx-attack}, Load Value Injection attack \cite{lvi-sgx-attack} and SGAxe attack \cite{sgaxe-sgx-attack}.
			Due to its many security issues, \acrshort{sgx} has been officially discontinued.\footnote{As of $11^\text{th}$ Generation Intel Core processor.}

			Besides the attacks against the \acrshort{sgx}, the design itself has a number of restrictions.
			First of all, the enclave memory is capped at \SI{128}{\mega\byte}, part of which is occupied by the \acrshort{sgx} control structures, leaving the application about \SI{96}{\mega\byte}.
			\acrshort{sgx} allows the use of external memory pages through the sealing mechanism, but it imposes high overhead of re-encryption and crossing the enclave physical boundary.
			Second, the code execution inside the enclave is significantly slower.
			Third, by design, only the untrusted application component can interact with the \acrshort{os}, for example, make network or storage \acrshort{io} requests.
			Finally, from the security standpoint the enclave is vulnerable against the side-channel attacks, most of all, access pattern leakage.
			Such leakage implies that the normal database application cannot be placed directly in the enclave and be deemed secure, because access pattern has been effectively exploited up to a reconstruction attack \cite{generic-attacks-kellaris}.
			One then has to design the application specifically to conceal the access pattern.
			For example, ZeroTrace \cite{zerotrace} is a variant of PathORAM \cite{path-oram} that is internally oblivious and thus can work in \acrshort{sgx}.
			Oblix \cite{oblix} is another example of a structure that is, in \textcite{oblix} own terms, doubly-oblivious --- internally in how it manages memory and registers, and externally in how it interacts with a storage.

	\cleardoublepage%
	\chapter{Related work}\label{section:related-work}
\thispagestyle{myheadings}

	In this section we will go over the relevant work in the area.
 	The discussion is grouped by the query and adversary type, consistent with the overall thesis structure.
	\emph{Some of the following sections were paraphrased or taken verbatim from my published work \cite{ore-benchmark-17,epsolute}.}

	\section{Range query security in a snapshot model}

		In \cref{section:range-snapshot} we explore the methods of answering range queries in a snapshot adversary setting.
		We group these methods into those relying on \acrshort{ope} / \acrshort{ore} schemes coupled with regular database range indices, such as \BPlus{} tree, and other mostly interactive constructions that use custom data structures to support secure indexing.

		The list of recently proposed \acrfull{ope} schemes includes \cite{ope-original, anti-tamper-dbs, bclo-ope, ope-leakage, ope-beyond-one-wayness, ope-early-fh-ope, ope-beyond-ideal-object, ope-mv-opes, fh-ope, ope-mv-popes, ope-multi-user, ope-note, ope-for-encrypted-dbs, ope-structure, ope-non-linear, ope-mdope, ope-outsourced-database, fh-ope-imporved-update}. 
		\textcite{ope-ideal-security-protocol} present a nice analysis of these schemes and they are the first to show that using a stateful scheme one can achieve the ideal security guarantees for \acrshort{ope}.
		In addition, there are a number of \acrfull{ore} schemes \cite{ore-original, clww-ore, lewi-wu-ore, parameter-hiding-ore, parameter-hiding-ore, ore-learning, ore-partial, ore-multi-client,  delegatable-ore, multi-client-ore} and related protocols \cite{ore-sorel} that have been proposed. 
		After the introduction of \acrshort{ope} / \acrshort{ore} as the means of protecting range indices, a line of attacks \cite{leakage-abuse-attacks-cash-15,attacks-what-else-revealed,inference-attacks-naveed-15, ore-file-injection-attack} have emerged.

		The second group of approaches assumes an outsourced setting where the client may have to communicate with the server during query processing \cite{pope, florian-protocol, secure-queries-overview, practical-range-search}.
		We would like to point out that there are some other methods that can be used to run range queries on encrypted data that use different types of schemes and techniques.
		CryptDB \cite{crypt-db} is a seminal work in this direction offering computations over encrypted data.
		Arx \cite{arx} provides strictly stronger security guarantees by using only semantically secure primitives.
		Seabed \cite{seabed} uses an additively symmetric homomorphic encryption scheme for aggregates and certain filter queries.
		\textcite{ppqed} offer a method to verify and apply a predicate (a junction of conditions) using garbled circuits or homomorphic encryption without revealing the predicate itself.
		SisoSPIR \cite{sisospir} presents a mechanism to build an oblivious index tree such that neither party learns the pass taken.
		See \cite{secure-queries-overview} and \cite{protocols-survey} for an overview of other methods. 
		We also note a work on theoretical analysis of \acrshort{ore} security \cite{ore-theory-security}.

	\section{Range query security in a persistent model}

		For the persistent adversary setting, we group the related secure databases, engines, and indices into two categories
		\begin{enumerate*}[label={(\roman*)}]
			\item systems that are oblivious or volume-hiding and do not require \acrfull{tee}, and
			\item constructions that rely on \acrshort{tee} (usually, Intel \acrshort{sgx}).
		\end{enumerate*}
		In this section, where relevant, we aim to compare the related work to our own point and range query solution, \epsolute{} \cite{epsolute}, see \cref{section:range-persistent}.
		\emph{We claim that \epsolute{} is the most secure and practical range- and point-query engine in the outsourced database model, that protects both \acrfull{ap} and \acrfull{cv} using \acrlong{dp}, while not relying on \acrshort{tee}, linear scan or padding result size to the maximum.}

		\subsection{Obliviousness and volume-hiding without enclave}

			This category is the most relevant to \epsolute{}, wherein the systems provide either or both \acrshort{ap} and \acrshort{cv} protection without relying on \acrshort{tee}.
			\crypte{} \cite{crypte} is a recent end-to-end system executing ``\acrshort{dp} programs''.
			\crypte{} has a different model than \epsolute{} in that it assumes two non-colluding servers, an adversarial querying user (the analyst), and it uses \acrshort{dp} to protect the privacy of an individual in the database, which includes volume-hiding for aggregate queries.
			\crypte{} also does not consider oblivious execution and attacks against the \acrshort{ap}.
			Shrinkwrap \cite{shrinkwrap} (and its predecessor SMCQL \cite{smcql}) is an excellent system designed for complex queries over federated and distributed data sources.
			In Shrinkwrap, \acrshort{ap} protection is achieved by using oblivious operators (linear scan and sort) and \acrshort{cv} is concealed by adding fake records to intermediate results with \acrshort{dp}.
			Padding the result to the maximum size first and doing a linear scan over it afterwards to ``shrink'' it using \acrshort{dp}, is much more expensive than in \epsolute{}, however.
			In addition, in processing a query, the worker nodes are performing an $O(n \log{n})$ cost oblivious sorting, where $n$ is the maximum result size (whole table for range query), since they are designed to answer more general complex queries.
			SEAL \cite{seal} offers adjustable \acrshort{ap} and \acrshort{cv} leakages, up to specific bits of leakage.
			SEAL builds on top of Logarithmic-SRC \cite{practical-range-search}, splits storage into multiple \acrshortpl{oram} to adjust \acrshort{ap}, and pads results size to a power of 2 to adjust \acrshort{cv}.
			\epsolute{}, on the other hand, fully hides the \acrshort{ap} and uses \acrshort{dp} with its guarantees to pad the result size.
			PINED-RQ \cite{pined-rq} samples Laplacian noise right in the B+ tree index tree, adding fake and removing real pointers according to the sample.
			Unlike \epsolute{}, PINED-RQ allows false negatives (i.e., result records not included in the answer), and does not protect against \acrshort{ap} leakage.
			On the theoretical side, \textcite{differential-obliviousness} (followed by \textcite{differential-obliviousness-followup}) treat the \acrshort{ap} itself as something to protect with \acrshort{dp}.
			\cite{differential-obliviousness} introduces a notion of differential obliviousness that is admittedly weaker than the full obliviousness used in \epsolute{}. 
			Most importantly, \cite{differential-obliviousness} ensures differential privacy with respect to the \acrshort{oram} only, while \epsolute{} ensures \acrshort{dp} with respect to the entire view of the adversary. 

		\subsection{Enclave-based solutions}

			Works in this category use trusted execution environment (usually, \acrshort{sgx} enclave).
			These works are primarily concerned with the \acrshort{ap} protection for both trusted and untrusted memory, unlike \epsolute{} which also protects \acrshort{cv}.
			Cipherbase \cite{cipherbase-daas} was a pioneer introducing the idea of using \acrshort{tee} (\acrshort{fpga} at that time) to assist with \acrshort{rdbms} security.
			HardIDX \cite{hardidx} simply puts the B+ tree in the enclave, while StealthDB \cite{stealth-db} symmetrically encrypts all records and brings them in the enclave one at a time for processing.
			EnclaveDB \cite{enclave-db} assumes somewhat unrealistic \SI{192}{\giga\byte} enclave and puts the entire database in it.
			ObliDB \cite{oblidb} and Opaque \cite{opaque} assume fully oblivious enclave memory (not available as of today) and devise algorithms that use this fully trusted portion to obliviously execute common \acrshort{rdbms} operators, like filters and joins.
			Oblix \cite{oblix} provides a multimap that is oblivious both in and out of the enclave.
			HybrIDX claims protection against both \acrshort{ap} and \acrshort{cv} leakages, but unlike \epsolute{} it only obfuscates them.
			\epsolute{} offers an indistinguishability guarantee for \acrshort{ap} and a \acrshort{dp} guarantee for \acrshort{cv}, while HybrIDX hides the exact result size and only obfuscates the \acrshort{ap}.
			Lastly, Hermetic \cite{hermetic} takes on the \acrshort{sgx} side-channel attacks, including \acrshort{ap}.
			It provides oblivious primitives, however, it only offers protection against software and not physical attacks (e.g., it trusts a hypervisor to disable interrupts).

	\section{\texorpdfstring{\acrshort{knn}}{kNN} query security in a snapshot model}\label{section:related-work:knn}

		In this section, where relevant, we aim to compare the related work to our own \acrshort{knn} query solution, \kanon{} \cite{k-anon}, see \cref{section:knn-snapshot}.

		A work immediately related to ours is QuickN by \textcite{quick-n}.
		QuickN offers an adaptation of nearest-neighbor search algorithm in conventional tree data structures (i.e., R-trees) to well-established \acrfull{ope} schemes.
		Unlike our solution that involves a novel property-preserving encryption scheme specifically designed for high-dimensional vectors, QuickN encrypts each vector dimension separately with \acrshort{ope}.
		\cite{quick-n} includes the experiments with attacks against their solution (attack by \textcite{leakage-abuse-grubs-2017}) and report a high degree of protection against them (at most \SI{3.7}{\percent} matching rate for the inference attacks against coordinates represented as pairs of longitude and latitude).

		QuickN approach of applying \acrshort{ope} to an R-tree, however, has some disadvantages.
		First, an ideal stateless \acrshort{ope} has been shown inferior (\cite{ope-leakage}) to its counterpart, an \acrfull{ore} in which the comparison over ciphertexts is defined explicitly.\footnote{
			\cite{quick-n} uses mOPE \cite{ope-ideal-security-protocol} which is an interactive protocol and not a traditional lightweight stateless \acrshort{ope} like \cite{bclo-ope}.
			Since mOPE is an ideal (though stateful) \acrshort{ope}, \textcite{quick-n} do not include \acrshort{ope} leakage in their security definition.
		}
		An \acrshort{ore}, in turn, can have a varying level of security, with the higher security level translating into lower comparison performance \cite{ore-benchmark-17}.
		In QuickN, an R-tree-based nearest-neighbor algorithm involves a very high number of comparisons, linear in data dimensionality.
		With the cost of comparison no longer negligible, the overhead of a query over 2D or 3D is already high, saying nothing of 768-dimensional vectors that our work targets.
		Second, QuickN protocol is not single-round (i.e., it takes two roundtrips per query) and it returns a large number of false positives even for a minimal $k$ (\num{4000} false positives for $10^6$ dataset and $k = 1$).

		\textcite{knn-aspe} propose a novel scheme, \acrshort{aspe}, that preserves a special type of scalar product.
		Namely, \acrfull{aspe} scheme preserves the scalar product of a database point and a query point, but not the product of a database point with itself or another database point.
		\acrshort{aspe} is then naturally integrated in existing \acrshort{knn} algorithms that use such asymmetric scalar product in their indices.
		The work of \textcite{knn-aspe} is similar to \kanon{} in that we also apply a property-preserving encryption scheme to existing \acrshort{knn} algorithms.
		\acrshort{aspe} is different in that it preserves a scalar product while we preserve an $\text{L}_2$ distance comparison, and \acrshort{aspe} has been broken in \cite{secure-nn-revisited-break-aspe} (although the attack is a chosen plaintext attack, i.e., one cannot decrypt a random ciphertext).

		Other works either target different aspects of query security, like integrity and soundness of results \cite{knn-integrity-soundness,svknn}, or involve mechanisms other than property-preserving encryption \cite{seceqp,practical-approx-knn,knn-sharing-keys,knn-mult-data-owners,knn-over-encrypted,knn-paillier,knn-blind,knn-homomorphism,knn-strong-location-privacy,knn-no-anonymizers,knn-efficient,knn-new-casper}.

	\cleardoublepage%
	\chapter{Range queries in the snapshot model}\label{section:range-snapshot}
\thispagestyle{myheadings}

	\acrfull{ope} and \acrfull{ore} are two important encryption schemes that have been proposed in the area of query evaluation over encrypted data.
	These schemes can provide very efficient query execution but at the same time may leak some information to adversaries.
	In this chapter, we present the first comprehensive comparison among a number of important \acrshort{ope} and \acrshort{ore} schemes using a framework that we developed.
	We evaluate protocols that are based on these schemes as well.
	We analyze and compare them both theoretically and experimentally and measure their performance over database indexing and query evaluation techniques using not only execution time but also \acrshort{io} performance and usage of cryptographic primitive operations.
	Our comparison reveals some interesting insights concerning the relative security and performance of these approaches in database settings.
	Furthermore, we propose a number of improvements for some of these scheme and protocols.
	Finally, we provide a number of suggestions and recommendations that can be valuable to database researchers and users.

	\emph{Some of the following sections were paraphrased or taken verbatim from the following published work.}

	\cite{ore-benchmark-17} \printpublication{ore-benchmark-17}

	\section{Introduction}

	\acrfull{ope} was proposed by~\textcite{ope-original} in their seminal paper.
	The main idea is to ``encrypt'' numerical values into ciphertexts that have the same order as the original plaintexts.
	This is a very useful primitive since it allows a database system to make comparisons between chiphertexts and get the same results as if it had operated on plaintexts.
	A scheme was proposed in \cite{ope-original} but no security analysis was given.

	\textcite{bclo-ope} were the first to treat \acrshort{ope} schemes from a cryptographic point of view, providing security models and rigorous analysis.
	The ideal functionality of such a scheme is to leak only the order of the plaintexts and nothing more.
	However, it was shown by \textcite{bclo-ope} that the ideal functionality is not achievable if the scheme is \emph{stateless} and \emph{immutable}.
	In order to achieve the ideal functionality, \textcite{ope-ideal-security-protocol} proposed a mutable scheme that constructs a binary tree on plaintexts and uses paths as ciphertexts.
	This tree is the encrypted full state of the dataset, and once an insertion or a deletion rebalances the tree, multiple ciphertexts get mutated.
	\textcite{fh-ope} proposed an improvement on this scheme that also hides the frequency of each plaintext (how many times a given value appears).

	Furthermore, in order to improve the security of these schemes, \textcite{ore-original} proposed to generalize \acrshort{ope} to \acrfull{ore}.
	In \acrshort{ore}, ciphertexts have no particular order and look more like typical semantically secure encryptions.
	The database system has a special comparison function that can be used to compare two ciphertexts.
	These schemes are more secure than \acrshort{ope} schemes, although they still leak some information, and in general are more expensive to compute.
	Since these schemes leak some information, a number of recent works considered attacks on systems that may use these schemes \cite{access-pattern-disclosure, inference-attack-islam-14, inference-attacks-naveed-15, grubbs-attacks, generic-attacks-kellaris, leakage-abuse-attacks-cash-15, attacks-what-else-revealed, attacks-improved-reconstruction, attacks-tao-of-inference, attacks-ore-injection}.
	Most of these attacks assume the attacker possesses \emph{auxiliary information} and no other protections are available.

	\acrshort{ope} / \acrshort{ore} schemes can be used with almost no changes to the underlying database engine.
	To provide greater security, a number of more complex protocols for protecting data in outsourced databases have been proposed.
	These constructions are often interactive, rely on custom data structures and are optimized for certain tasks, such as range queries.
	Naturally, the more secure the protocol is, the larger performance overhead it incurs.
	The most secure of these --- \acrfull{oram} based protocol --- provides strong, well-understood, cryptographic privacy guarantees with no information leakage.

	Applications that can benefit from such schemes and protocols include cloud access security brokers and financial and banking applications.
	Indeed, a number of commercial brokers including Skyhigh Networks\footnote{\url{https://www.skyhighnetworks.com/}} and CipherCloud\footnote{\url{https://www.ciphercloud.com/}} have been using some form of \acrshort{ope} or \acrshort{ore} schemes in their systems.
	In addition, financial institutions may be able to encrypt their data using the aforementioned schemes in order to provide another layer of security, assuming that the performance overhead is acceptable.
	For many of these applications the auxiliary information that is needed for the attacks mentioned above is either unavailable or difficult to get.

	Currently, it is a very challenging task for users to choose an appropriate data privacy approach for their application, because the security and performance tradeoff is not well understood.
	Both security and performance of every approach need to be thoroughly evaluated.
	Characterizing security benefits of different approaches remains an open problem, unlikely to be solved in the immediate future.
	However, it is possible to evaluate the performance of each approach, so as to enable better-informed decisions about whether the improved performance of some schemes is worth the uncertainty about the security they achieve.

	We emphasize that it is not trivial to evaluate the performance of these schemes.
	Many of the papers presenting the above approaches provide only a theoretical treatment and concentrate more on the security definitions and analysis and less on the performance.
	Some of these constructions have not been even implemented properly.
	Furthermore, even though the main target of these schemes and protocols are database applications, most of them have not been evaluated in database settings.

	To address this problem, in this paper we design a new framework that allows for systematic and extensive comparison of \acrshort{ope} and \acrshort{ore} schemes and secure range query protocols in the context of database applications.
	We employ these schemes in database indexing techniques (i.e. {\BPlus} trees) and query protocols and we report various costs including \acrshort{io} complexity.

	The main contribution of this work is to present an experimental evaluation using both real and synthetic datasets using our new framework that tracks not only time but also primitive usage, \acrshort{io} complexity, and communication cost.
	In the process, we present improvements for some of the schemes that make them more efficient and/or more secure.
	To make understanding of these schemes easier for the reader, we present the main ideas behind these constructions, discuss their security definitions and leakage profiles, and provide an analysis of implementation challenges for each one.

	\section{Security Perspective}\label{section:range-snapshot:security}

	Each scheme and protocol we analyze has its own security definition, which captures different leakage levels.
	We attempt to unify these definitions and analyze them under a common framework.
	We also attempt to assess relative security of these definitions and analyze their leakages.

	In this work we mostly consider the snapshot model, where the attacker can observe all the database contents at one time instant.
	Note that this excludes timing attacks such as measuring encryption time.
	All security definitions of the schemes and protocols that we discuss here are based on this model.
	Also, the snapshot attacker is the most common attacker that we face today \cite{secure-queries-overview}.
	The idea is that a hacker or an insider can steal the entire encrypted database and all its contents at some point in time.

	Beyond the snapshot model, it is also possible to consider a stronger adversary who can track communication volume and data access patterns in real time.
	Approaches that help protect against such an attacker include \acrshort{oram} for protection against access pattern leakage and differential privacy for protection against communication volume leakage.
	Although this model is not a primary target of this paper, our benchmark includes a protocol (\cref{section:range-snapshot:oram}) that is secure in this setting to show the cost of adding such protection.

	We wanted to specifically comment on a work of \textcite{leakage-abuse-grubs-2017}, which demonstrates a series of attacks against \acrshort{ope} and \acrshort{ore} schemes.
	The attacks can be very successful, but they depend on certain prerequisites.
	First, all attacks assume the existence of a well-correlated auxiliary dataset.
	Second, the binomial attack, which works against a ``perfectly secure frequency-hiding scheme'', reliably recovers only high-frequency elements.
	Finally, the attacks are specifically devastating against encrypted strings (e.g.\ first and last names) as opposed to numerical data, and we also do not recommend using \acrshort{ope} / \acrshort{ore} for strings (see \cref{section:range-snapshot:variable-inputs}).
	One of the conclusions of our work is that security is negatively correlated with performance and it is up to a practitioner to trade off security and performance constraints.

	\subsection{A note on variable-length inputs}\label{section:range-snapshot:variable-inputs}

		A generic \acrshort{ope} / \acrshort{ore} scheme accepts bit-strings of any length as inputs, and treats them as numbers or processes them bit-by-bit.
		We warn against supplying raw bytes of variable length (e.g.\ encoded strings) to \acrshort{ope} and \acrshort{ore} schemes, as such an approach will introduce both performance and security challenges.

		From the performance standpoint, the complexity of \acrshort{ope} / \acrshort{ore} schemes usually depends on the input length at least linearly (see \cref{table:primitive-usage-theory}).
		32-bit numbers already introduce a noticeable overhead for some (usually more secure) schemes, and supplying arbitrary-length inputs may worsen performance by at least an order of magnitude.

		Security of such a construction will be minimal as most schemes leak some information about the magnitude of the difference, and longer inputs will naturally be treated as larger numbers.
		Thus, the difference between long and short inputs will be apparent.
		We refer to the work of~\textcite{leakage-abuse-grubs-2017} as they have a practically supported discussion of security consequences of using \acrshort{ope} / \acrshort{ore} with arbitrary strings.

		On the other hand, other protocols in our benchmark can usually handle variable-length inputs as long as they fit into a single block for the underlying block cipher.

	\section{\texorpdfstring{\acrshort{ope}}{OPE} and \texorpdfstring{\acrshort{ore}}{ORE} Schemes}

	An \acrlong{ore} scheme is a triple of polynomial\hyp{}time algorithms $\setup$, $\encrypt$ and $\compare$.
	$\setup$ generates a key of parameterized length (the $\lambda$ parameter).
	$\encrypt$ takes a numerical input (as a bit string) and produces a ciphertext.
	$\compare$ takes two ciphertexts generated by the scheme and outputs whether the first plaintext was strictly less than the second.
	Note that being able to check this condition is enough to apply all other comparison operators ($<$, $\le$, $=$, $\ge$, $>$).
	Also note that an \acrshort{ore} scheme does not include a decryption algorithm, because one can simply append a symmetric encryption of the plaintext to the produced ciphertext and use it for decryption.\footnote{
		\emph{Given the secret key}, it is possible to decrypt a ciphertext by doing binary search on the plaintext domain: encrypting known values and comparing them against the target ciphertext, until the target plaintext is found.
		However, this would require $\bigO{\log{|\domain|}}$ encryption and comparison operations.
	}
	An \acrfull{ope} scheme is a particular case of an \acrshort{ore} scheme where ciphertexts are numerical and thus $\compare$ routine is trivial (the numerical order of ciphertexts is the same as underlying plaintexts).
	\acrshort{ope} may optionally include a decryption algorithm, since appending a symmetric ciphertext is no longer possible.

	Both \acrshort{ope} and \acrshort{ore} schemes by definition allow to totally order the ciphertexts.
	This is their inherent leakage (by design) and all the \acrshort{ope} / \acrshort{ore} security definitions account for this and possibly additional leakage.

	We proceed by describing and analyzing the \acrshort{ope} / \acrshort{ore} schemes we have benchmarked.
	All plaintexts are assumed to be 32-bit signed integers, or $n$-bit inputs in complexity analysis.
	\acrshort{ope} ciphertexts are assumed to be 64-bit signed integers.

	From here, we will use the term \acrshort{ore} to refer to both \acrshort{ope} and \acrshort{ore}, unless explicitly stated otherwise.
	Each scheme has its own subsection where the first part is the construction overview followed by security discussion, and the second part is our theoretical and experimental analysis.

	\subsection{BCLO \texorpdfstring{\acrshort{ope}}{OPE} \texorpdfstring{\cite{bclo-ope}}{}}

	The \acrshort{ope} scheme by \textcite{bclo-ope} was the first \acrshort{ope} scheme that provided formal security guarantees and was used in one of the first database systems that executes queries over encrypted data (CryptDB \cite{crypt-db}).
 	The core principle of their construction is the natural connection between a random order-preserving function and the hypergeometric probability distribution.

	The encryption algorithm works by splitting the domain into two parts according to a value sampled from the \acrfull{hg}, and splitting the range in half recursively.
	When the domain size contains a single element, the corresponding ciphertext is sampled uniformly from the current range.

	All pseudo-random decisions are made by an internal \acrshort{prg} (refereed to as \algo{TapeGen} in \cite{bclo-ope}).
	This way not only the algorithm is deterministic, but also decryption is possible.
	The decryption procedure takes the same ``path'' of splitting domain and range, and when the domain size reaches one, the only value left is the original plaintext.

	\subsubsection{Security}
		This scheme is POPF-CCA secure \cite{bclo-ope}, meaning that it is as secure as the underlying ideal object --- randomly sampled order-preserving function from a certain domain to a certain range.
		For practical values of the parameters, \textcite{ope-leakage} showed that the distance between the plaintexts can be approximated to an accuracy of about the square root of the domain size.
		In other words, approximately, half of the bits (the most significant) are leaked.
		\textcite{leakage-abuse-grubs-2017} showed that this leakage allows to almost entirely decrypt the ciphertexts (given auxiliary data with a similar distribution) and encrypting strings (rather than numbers) with this scheme is especially dangerous (see \cref{section:range-snapshot:variable-inputs}).

	\subsubsection{Analysis and implementation challenges}

		Efficient sampling from the hypergeometric distribution is a challenge by itself.
		Authors suggest using a randomized yet exact (not approximate) Fortran algorithm by \textcite{hg-sampler}.
		It should be noted that the algorithm relies on infinite precision floating-point numbers, which most regular frameworks do not have.
		The security consequences of finite precision computations is actually an open question.
		The complexity of this randomized algorithm is hard to analyze; however, we empirically verified that its running time is no worse than linear in the input bit length.
		The authors also suggest a different algorithm for smaller inputs \cite{hg-sampler-small}.

		On average, encryption and decryption algorithms make $n$ calls to \acrshort{hg}, which in turn consumes entropy generated by the internal \acrshort{prg}.
		The entropy, and thus the number of calls to \acrshort{prg}, needed for one \acrshort{hg} run is hard to analyze theoretically.
		However, we derived this number experimentally (see \cref{section:range-snapshot:evaluation}).

		BCLO has been implemented in numerous languages and has been deployed in a number of secure systems.
		We add {\Csharp} implementation to the list, and recommend using a library that supports infinite precision floating-point numbers when building the hypergeometric sampler.

	\subsection{CLWW \texorpdfstring{\acrshort{ore}}{ORE} \texorpdfstring{\cite{clww-ore}}{}}\label{section:range-snapshot:clww}

	The \acrshort{ore} scheme by \textcite{clww-ore}, which authors call ``Practical ORE'', is the first efficient \acrshort{ore} implementation based on \acrshortpl{prf}.

	On encryption, the plaintext is split into $n$ values in the following way.
	For each bit, a value is this bit concatenated with all more significant bits.
	This value is given to a keyed \acrshort{prf} and the result is numerically added to the next less significant bit.
	This resulting list of $n$ elements is the ciphertext.

	The comparison routine traverses two lists in-order looking for the case when one value is greater than the other by exactly one, revealing location and value of the first differing bit.
	If no such index exists, the plaintexts are equal.

	\subsubsection{Security}

		A generic \acrshort{ore} security definition was introduced along with the scheme \cite{clww-ore}.
		\acrshort{ore} leakage is more clearly quantified than in \acrshort{ope}.
		The definition says that the scheme is secure with a leakage $\leak(\cdot)$ if there exists an algorithm (simulator) that has access to the leakage function and can generate output indistinguishable from the one generated by the real scheme.
		This scheme satisfies \acrshort{ore} security definition with the leakage $\leak(\cdot)$ of the location and value of the first differing bit of every pair of plaintexts.
		Note that the most significant differing bit also leaks the approximate distance between two values.

	\subsubsection{Analysis and implementation challenges}

		On encryption the algorithm makes $n$ calls to \acrshort{prf} and the comparison procedure does not use any cryptographic primitives.
		Ciphertext is a list of length $n$, where each element is an output of a \acrshort{prf} modulo 3.
		The authors claim that the ciphertext's size is $n \log_2 3$, just $1.6$ times larger than the plaintext's size.
		While this may be true for large enough $n$ if ternary encoding is used, we found that in practice the ciphertext size is still $2n$.
		$1.6 n$ for 32-bit words is $51.2$ bits, which will have to occupy one 64-bit word, or two 32-bit words, therefore resulting in $2n$ anyway.

	\subsection{Lewi-Wu \texorpdfstring{\acrshort{ore}}{ORE} \texorpdfstring{\cite{lewi-wu-ore}}{}}

	\textcite{lewi-wu-ore} presented an improved version of the CLWW scheme \cite{clww-ore} which leaks strictly less.

	The novel idea was to use the ``left / right framework'' in which two ciphertexts get generated --- left and right.
	The right ciphertexts are semantically secure, so an adversary will learn nothing from them.
	Comparison is only defined between the left ciphertext of one plaintext and the right ciphertext of another plaintext.

	The approach is to split the plaintext into blocks of $d$ bits.
	The ciphertext is computed block-wise.
	For the right side, the algorithm compares the given block value to all $2^d$ possible block values; each comparison result is added (modulo 3) to a \acrshort{prf} of the previous blocks.
	All $2^d$ comparison results go into the right ciphertext.
	The left side, which is shorter, is produced in such a way as to reveal the correct comparison result.
	This way the location of the differing bit inside the block is hidden, but the location of the first differing block is revealed.

	\subsubsection{Security}

		This scheme satisfies the \acrshort{ore} security definition introduced by~\textcite{clww-ore} with the leakage $\leak(\cdot)$ of the location of the first differing \emph{block}.
		This property allows a practitioner to set performance-security tradeoff by tuning the block size.
		Left / right framework is particularly useful in a {\BPlus} tree since it is possible to store only one (semantically secure) side of a ciphertext in the structure (see \cref{section:range-snapshot:ore-to-protocol}).

	\subsubsection{Analysis and implementation challenges}

		Let $n$ be the size of input in bits (for example, 32) and $d$ be the number of bits per block (for example, 2).

		Left encryption loops $\frac{n}{d}$ times making one \acrshort{prp} call and two \acrshort{prf} calls each iteration.
		Right encryption loops $\frac{n}{d} 2^d$ times making one \acrshort{prp} call, one hash call and two \acrshort{prf} calls each iteration.
		Comparison makes $\frac{n}{d}$ calls to hash at worst and half of that number on average.
		Please note that the complexity of right encryption is exponential in the block size.
		In the \cref{table:primitive-usage-theory} the \acrshort{prp} usage is linear due to our improvement.
		The ciphertext size is no longer negligible --- $\frac{n}{d} \left(\lambda + n + 2^{d + 1} \right) + \lambda$, for $\lambda$ being \acrshort{prf} output size.

		The implementation details of this approach raise an interesting security question.
		Although the authors suggest using 3-rounds Feistel networks \cite{unbalanced-feistel} for \acrshort{prp} and use it in their implementation, it may not be secure for small input sizes.
		Feistel networks security depends on the input size \cite{feistel-security} --- exponential in the input size.
		However, the typical input for \acrshort{prp} in their scheme is 2--8 bits, thus even exponential number is small.

		We have considered multiple \acrshort{prp} implementations to use instead of the Feistel networks.
		Because the domain size is small (from $2^2$ to $2^8$ elements), we have decided to build a \acrshort{prp} by simply using the key as an index into the space of all possible permutations on the domain, where a permutation is obtained from the key via Knuth shuffle (this approach was mentioned in \cite{knuth-shuffle-security}).
		Another important aspect of the implementation is that for each block we need to compute the permutation of all the values inside the block.
		This operation applied many times can be expensive.
		To address this, we propose to generate a \acrshort{prp} table once for the whole block and then use this table when one needs to compute the location of an element of permutation.
		This can reduce the \acrshort{prp} usage (indeed, we observe a reduction from 80 to 32 calls in our case).
		We evaluate this improved approach in our experimental section.

	\subsection{CLOZ \texorpdfstring{\acrshort{ore}}{ORE} \texorpdfstring{\cite{parameter-hiding-ore}}{}}

	\textcite{parameter-hiding-ore} introduced a new \acrshort{ore} scheme that provably leaks less than any previous scheme.
	The idea is to use \textcite{clww-ore} construction (see \cref{section:range-snapshot:clww}), but permute the list of \acrshort{prf} outputs.
	The original order of those outputs is not necessary, as one can simply find a pair that differs by one.
	This is not enough to reduce leakage, however, since an adversary can count how many elements two ciphertexts have in common.

	To address this problem, the authors define a new primitive they call a \acrfull{pph}.
	A \acrshort{pph} as defined and used in \cite{parameter-hiding-ore}, allows one to expose a property (specifically $y \overset{?}{=} x + 1$) of two (numerical) elements such that nothing else is leaked.
	In particular, the outputs are randomized, so same element hashed twice will have different hashes.
	Please refer to the original paper \cite{parameter-hiding-ore} for formal correctness and security definitions.

	Equipped with the \acrshort{pph} primitive, the algorithm ``hashes'' the elements of the ciphertexts before outputting them.
	Due to security of \acrshort{pph}, the adversary would not be able to count how many elements two ciphertexts have in common, thus, would not be able to tell the location of differing bit.

	\subsubsection{Security}

		The strong side of the scheme is its security.
		The scheme leaks $\leak(\cdot)$ an \emph{equality pattern} of the most-significant differing bits (satisfying \textcite{clww-ore} definition).
		As defined in \cite{parameter-hiding-ore}, the intuition behind equality pattern is that for any triple of plaintexts $m_1$, $m_2$, $m_3$, it leaks whether $m_2$ differs from $m_1$ before $m_3$ does. 
		We do not know of any attacks against this construction (partially because no implementation exists yet, see next subsection), but it is inherently vulnerable to frequency attacks that apply to all frequency-revealing \acrshort{ore} schemes (see \cref{section:range-snapshot:security}).

	\subsubsection{Analysis and implementation challenges}

		On encryption, the scheme makes $n$ calls to \acrshort{prf}, $n$ calls to \acrshort{pph} \algo{Hash} and one call to \acrshort{prp}.
		Comparison is more expensive, as the scheme makes $n^2$ calls to \acrshort{pph} \algo{Test}.

		The scheme has two limitations that make it impractical.
		The first one is the square number of calls to \acrshort{pph}, which is around $1024$ for a single comparison.

		The second problem is the \acrshort{pph} itself.
		Authors suggest a construction based on bilinear maps.
		The hash of an argument is an element of a group, and the test algorithm is computing a pairing.
		This operation is very expensive --- order of magnitude more expensive than any other primitive we have implemented for other schemes.

		We have implemented this scheme in C++ using the \acrshort{pbc} library \cite{pbc} to empirically assess schemes's performance, and on our machine (see \cref{section:range-snapshot:evaluation}), a single comparison takes 1.9 seconds on average.
		Although we have produced the first (correct and secure) real implementation of this scheme in C++, it is infeasible to use it in the benchmark (it will take years to complete a single run).
		Therefore, for the purposes of our benchmark, we implemented a ``fake'' version of \acrshort{pph} --- correct, but insecure, which does not use pairings.
		Consequently, in our analysis we did not benchmark the speed of the scheme, but measured all other data.

	\subsection{FH-OPE \texorpdfstring{\cite{fh-ope}}{}}

	Frequency-hiding \acrshort{ope} by \textcite{fh-ope} is a stateful scheme that hides the frequency of the plaintexts, so the adversary is unable to construct a frequency histogram.

	This scheme is stateful, which means that the client needs to keep a data structure and update it with every encryption and decryption.
	The data structure is a binary search tree where the node's value is the plaintext and node's position in a tree is the ciphertext.
	For example, consider the range $[1, 128]$.
	Any plaintext that happens to arrive first (for example, $6$), will be the root, and thus the ciphertext is $64$.
	Then any plaintext smaller than the root, say $3$, will become the left child of the root, and will produce the ciphertext $32$.
	To encrypt a value, the algorithm traverses the tree until it finds a spot for the new plaintext, or finds the same plaintext.
	If the same plaintext is found, the traversal pseudo-randomly passes to the left or right child, up to the leaf.
	This way, the invariant of the tree --- intervals of the same plaintexts do not overlap --- is maintained.
	The ciphertext generated from the new node's position is returned.

	Due to randomized ciphertexts, the comparison algorithm is more complicated than in the regular deterministic \acrshort{ope}.
	To properly compare ciphertexts, the algorithm needs to know the boundaries --- the minimum and maximum ciphertexts for a particular plaintext.
	The client is responsible for traversing the tree to find the plaintext for the ciphertext and then minimum and maximum ciphertext values.
	Having these values, the comparison is trivial --- equality is a check that the value is within the boundaries, and other comparison operators are similar.

	Authors have designed a number of heuristics to minimize the state size, however, these are mostly about compacting the tree and the result depends highly on the tree content.
	In our analysis, we consider the worst case performance without the use of heuristics.
	In our experimental evaluation, however, we did implement compaction.

	\subsubsection{Security}

		The security of the scheme relies on the large range size to domain size ratio.
		Authors recommend at least 6 times longer ciphertexts than the plaintexts in bit-length, which means ciphertexts should be 192-bit numbers that are not commonly supported.
		It is possible to operate over arbitrary-length numbers, but the performance overhead would be substantial.
		We did a quick micro-benchmark in {\Csharp} and the overhead of using \texttt{BigInteger} is 15--20 times for basic arithmetic operations.

		This scheme satisfies IND-FAOCPA definition (introduced along with the scheme \cite{fh-ope}), meaning that it does not leak the equality or relative distance between the plaintexts.
		This definition has been criticized in \cite{florian-def-critique}, who claim that the definition is imprecise and propose an enhanced definition along with a small change to construction to satisfy this new definition.
		Both schemes leak the insertion order, because it affects the tree structure.
		We do not know of any attacks against this leakage, but it does not mean they cannot exist.
		\textcite{leakage-abuse-grubs-2017} describe an attack against this scheme (binomial attack), but it applies to any perfectly secure (leaking only total order) frequency-hiding \acrshort{ope}.

	\subsubsection{Analysis and implementation challenges}

		If the binary tree grows in only one direction, at some point it will be impossible to generate another ciphertext.
		In this case, the tree has to be rebalanced.
		This procedure will invalidate all ciphertexts already generated.
		This property makes the scheme difficult to use in some protocols since they usually rely on the ciphertexts on the server being always valid.
		The authors explicitly mention that the scheme works under the assumption of uniform input.
		However, the rebalancing will be caused by insertion of just 65 consecutive input elements for 64-bit integer range.

		The scheme makes one tree traversal on encryption and decryption.
		Comparison is trickier as it requires one traversal to get the plaintext, and two traversals for minimum and maximum ciphertexts.
		We understand that it is possible to get these values in fewer than three traversals, but we did not optimize the scheme for the analysis and evaluation.

		For practitioners we note that the stateful nature of the scheme implies that the client storage is no longer negligible as the state grows proportionally to the number of encryptions.
		We also note that implementing compaction extensions will affect code complexity and performance.
		Finally, we stress again that some non-uniform inputs can break the scheme by causing all ciphertexts to be invalid.
		It is up to the users of the scheme to ensure uniformity of the input, which poses serious restrictions on the usage of the scheme.


\begin{sidewaystable}
	\renewcommand{\arraystretch}{1.5}
	\centering
	\captionsetup{width=\textwidth}
	\caption[Primitive usage by \acrshort{ope} / \acrshort{ore} schemes]{
		\cite[Tables 1 and 4]{ore-benchmark-17}.
		Primitive usage by \acrshort{ope} / \acrshort{ore} schemes.
		Ordered by security rank --- most secure below.
		$n$ is the input length in bits, $d$ is a block size for Lewi-Wu \cite{lewi-wu-ore} scheme, $\lambda$ is a \acrshort{prf} output size, $N$ is a total data size, \textbf{\acrshort{hg}} is a hyper-geometric distribution sampler, \textbf{\acrshort{pph}} is a property-preserving hash with $h$-bit outputs built with bilinear maps and \textbf{bolded} are weak points of the schemes.
		Values in parentheses are simulation-derived. $N = 10^3$, $n = 32$, $d = 2$, $\lambda = 128$ and $h = 128$ in this simulation.
	}\label{table:primitive-usage-theory}
	\begin{tabular*}{\linewidth}{ !{\extracolsep\fill} l c c c c c } 

		\toprule

		\multirow{2}{*}{Scheme}							& \multicolumn{2}{c}{Primitive usage (number of invocations)}																				& \multirow{2}{*}{\makecell{Ciphertext size, \\ or state size (bits)}}					&  \multirow{2}{*}{\makecell{Leakage \\ (on top of total order)}}							\\ \cline{2-3}
		\rule{0pt}{10pt}								& Encryption																& Comparison													& 																						& 																							\\

		\toprule

		\cite{bclo-ope}									& $\bm{\approx n}$ \textbf{(41) \acrshort{hg}}								& none															& $2n$ (64)																				& \textbf{$\approx$ top half of the bits}													\\

		\midrule

		\cite{clww-ore}									& $n$ (32) \acrshort{prf} 													& none															& $2n$ (64)																				& \textbf{\makecell{Most-significant \\ differing bit}}										\\

		\midrule

		\multirow{3}{*}{\cite{lewi-wu-ore}}				& \boldmath{} $\nicefrac{2n}{d}$ \unboldmath{} \textbf{(32) \acrshort{prp}}	& \multirow{3}{*}{$\frac{n}{2d}$ (9) Hash}						& 																						& \multirow{3}{*}{\makecell{Most-significant \\ differing block}}							\\
														& $2 \frac{n}{d} \left( 2^d + 1 \right)$ (160) \acrshort{prf}				&																& $\frac{n}{d} \left(\lambda + n + 2^{d + 1} \right) + \lambda$							&																							\\
														& $\frac{n}{d} 2^d$ (64) Hash												&																& (2816)																				&																							\\

		\midrule

		\multirow{3}{*}{\cite{parameter-hiding-ore}}	& $n$ (32) \acrshort{prf}													& \multirow{3}{*}{$\bm{n^2}$ \textbf{(1046) \acrshort{pph}}}	& \multirow{3}{*}{$n \cdot h$ (4096)}													& \multirow{3}{*}{\makecell{Equality pattern \\ of the most-significant \\ differing bit}}	\\
														& $n$ (32) \acrshort{pph}													&																&																						& 																							\\
														& 1 \acrshort{prp}															&																&																						& 																							\\

		\midrule

		\cite{fh-ope}									& 1 Traversal																& 3 Traversals													& $\bm{3 \cdot n \cdot N}$ \textbf{(86842)}												& Insertion order																			\\

		\bottomrule

	\end{tabular*}
\end{sidewaystable}

\section{Secure Range Query Protocols}

	We proceed by describing and analyzing the range query protocols we have chosen.
	For the purpose of this paper, a secure range-query protocol is defined as a client-server communication involving construction and search stages.
	Communication occurs between a client, who owns some sensitive data, and an honest server, who securely stores it.
	In construction stage, a client sends the server the encrypted datapoints (index-value tuples) and the server stores them in some internal data structure.
	In search stage, a client asks the server for a range (usually specifying it with encrypted endpoints) and the server returns a set of encrypted records matching the query.
	Note that the server may interact with the client during both stages (e.g.\ ask the client to sort a small list of ciphertexts).
	Also note that we do not allow batch insertions as it would limit the use cases (e.g.\ client may require interactive one-by-one insertions).

	The first protocol is a family of constructions where a data structure ({\BPlus} tree in this case) uses \acrshort{ore} schemes internally.
	Then, we present alternative solutions with varying performance and security profiles, not relying on \acrshort{ore}.
	Finally, we introduce two baseline solutions we will use in the benchmark --- one that achieves the best performance and the other that achieves the maximal security.

	\subsection{Range query protocol from \texorpdfstring{\acrshort{ore}}{ORE}}\label{section:range-snapshot:ore-to-protocol}

	So far we have analyzed \acrshort{ope} and \acrshort{ore} schemes without much context.
	One of the best uses of an \acrshort{ore} is within a secure protocol.
	In this section we provide a construction of a search protocol built with a {\BPlus} tree working on top of an \acrshort{ore} scheme and analyze its security and performance.

	The general idea is to consider some data structure that is optimized for range queries, and to modify it to change all comparison operators to \acrshort{ore} scheme's $\compare$ calls.
	This way the data structure can operate only on ciphertexts.
	Performance overhead would be that of using the \acrshort{ore} scheme's $\compare$ routine instead of a plain comparison.
	Space overhead would be that of storing ciphertexts instead of plaintexts.

	In this paper, we have implemented a typical {\BPlus} tree \cite{b-tree} (with a proper deletion algorithm \cite{b-plus-tree-deletion}) as a data structure.

	For protocols, we also analyze the \acrshort{io} performance and the communication cost.
	In particular, we are interested in the expected number of \acrshort{io} requests the server would have made to the secondary storage, and the number and size of messages parties would have exchanged.

	The relative performance of the {\BPlus} tree depends only on the page capacity (the longer the ciphertexts, the smaller the branching factor). 	Therefore, the query complexity is $\bigO{\log_B \left( \nicefrac{N}{B} \right) + \nicefrac{r}{B}}$, where $B$ is the number of records (ciphertexts) in a block, $N$ is the number of records (ciphertexts) in the tree and $r$ is the number of records (ciphertexts) in the result (none for insertions).

	Communication amount of the protocol is relatively small as its insertions and queries require at most one round trip.

	\subsubsection{Security}
		The leakage of this protocol consists of leakage of the underlying \acrshort{ore} scheme plus whatever information about insertion order is available in the {\BPlus} tree.
		Please note that Lewi-Wu \cite{lewi-wu-ore} \acrshort{ore} is particularly well-suited in this construction with its left / right framework, because only the semantically secure side of the ciphertext is stored in the structure.
		In this case, the \acrshort{ore} leakage becomes only the total order and the security of the protocol is comparable with other non-\acrshort{ore} constructions.

	\subsection{Kerschbaum-Tueno \texorpdfstring{\cite{florian-protocol}}{}}

	\textcite{florian-protocol} proposed a new data structure, which satisfies their own definitions of security (IND-CPA-DS) and efficiency (search operation has poly\hyp{}logarithmic running time and linear space complexity).

	In short, the idea is to maintain a (circular) array of symmetrically encrypted ciphertexts in order.
	On insertion, the array is rotated around a uniformly sampled offset to hide the location of the smallest element.
	Client interactively performs a binary search requesting an element, decrypting it and deciding which way to go.

	\subsubsection{Security}

		Authors prove that this construction is IND-CPA-DS secure (defined in the same paper \cite{florian-protocol}).
		The definition assumes an array data structure and therefore serves specifically this construction (as opposed to being generic).
		It provably hides the frequency due to semantic encryption and hides the location of the first element due to random rotations.
		Leakage-wise this construction is strictly better than {\BPlus} tree with \acrshort{ore} --- they both leak total order, but \cite{florian-protocol} hides distance information and smallest / largest elements.
		Specifically, for all pairs of consecutive elements $e_i$ and $e_{i+1}$ it is revealed that $e_{i+1} \ge e_i$ except for one pair of smallest and largest elements in the set.

	\subsubsection{Analysis and implementation challenges}

		Insertions are \acrshort{io}-heavy because they involve rotation of the whole data structure.
		All records will be read and written, thus the complexity is $\bigO{\nicefrac{N}{B}}$.
		Searches are faster since they involve logarithmic number of blocks.
		The first few blocks can be cached and the last substantial number of requests during the binary search will target a small number of blocks.
		The complexity is then $\bigO{\log_2 \nicefrac{N}{B}}$.

		Communication volume is small as well.
		Insertion requires $\log_2 N$ messages from each side.
		Searches require double that number because separate protocol is run for both endpoints.

		The data structure is linear in size, and the client storage is always small.
		Sizes of messages are also small as only a single ciphertext is usually transferred.

		For practitioners we have a few points.
		The construction in the original paper \cite{florian-protocol} contains a typo as $m$ and $m^\prime$ must be swapped in the insertion algorithm.
		Also, we have found some rare edge cases; when duplicate elements span over the modulo, the algorithm may not return the correct answer.
		Both inconsistencies can be fixed however.
		This protocol is not optimized for \acrshort{io} operations for insertions, and thus would be better suited for batch uploads.

	\subsection{POPE \texorpdfstring{\cite{pope}}{}}

	\textcite{pope} presented a protocol, direct improvement over mOPE \cite{ope-ideal-security-protocol}, which is especially suitable for large number of insertions and small number of queries.
	The construction is heavily based on buffer trees \cite{buffer-tree} to support fast insertion and lazy sorting.

	The idea is to maintain a POPE tree on the server and have the client manipulate that tree.
	POPE tree is similar to B-tree, in that the nodes have multiple children and nodes are sorted on each level.
	Each node has an ordered list of \emph{labels} of size $L$ and an unbounded unsorted set of encrypted data called buffer.
	Parameter $L$ controls the list size, the leaf's buffer size, and the size of client's working set.
	The insertion procedure simply adds an encrypted piece of data to the root's buffer, thus we do not concentrate on insertion analysis in this section.

	The query procedure is more complex.
	To answer a query, the server interacts with the client to split the tree according to the query endpoints.
	On a high level, for each endpoint the buffers are cleared (content pushed down to leaves), and nodes in the paths are split.
	After that, answering a query means replying with all ciphertexts in all buffers between the two endpoint leaves.

	The authors provide cost analysis of their construction.
	Search operations are expected to require $\bigO{\log_L n}$ rounds.
	It must be noted that the first queries will require many more rounds, since large buffers must be sorted.

	\subsubsection{Security}

		This construction satisfies the security definition of frequency\hyp{}hiding partial order-preserving (FH-POP) protocol (introduced in the paper \cite{pope}).
		According to \cite[Theorem~3]{pope}, after $n$ insertions and $m$ queries with local storage of size $L$, where $m L \in o(n)$, the POPE scheme is frequency-hiding partial order-preserving with $\bigOmega{ \frac{n^2}{mL \log_L n} - n }$ incomparable pairs of elements. 
		Simply put, the construction leaks pairwise order of a \emph{bounded} number of elements.
		Aside from this, the construction provably hides the frequency (i.e., equality) of the elements.

	\subsubsection{Analysis and implementation challenges}

		In our analysis we count each request-response communication as a round.
		This is different from \cite{pope} where they use \emph{streaming} a number of elements as a single round. 
		The rationale for our approach is that if we allow persistent channels additionally to messages, then any protocol can open a channel for each operation.
		Thus, we do not allow channels for all protocols in our analysis.

		Also, as noted by the authors, if $L = n^{\epsilon}$ for $0 < \epsilon < 1$, then the amortized costs become $\bigO{1}$.
		While this is true, in our analysis the choice of $L$ depends on the storage volume block size for \acrshort{io} optimizations, instead of the client's volatile storage capacity.
		Thus, the costs remain logarithmic.

		Search bandwidth depends heavily on the current state of the tree.
		When the tree is completely unsorted (the first query), all elements of the tree will be transferred to split the large root, then possibly internal node will have to be split requiring sending of $\frac{N}{L}$ elements, and so on, thus $\bigO{N + r}$.
		When the tree is completely sorted (after a large number of uniform queries), the bandwidth will be similar to that of a standard {\BPlus} tree --- $\bigO{L \log_L N + r}$.
		The average case is hard to compute; however, authors prove an upper bound on bandwidth after $n$ insertions and $m$ queries --- $\bigO{m L \log_L n + n \log_L m + n \log_L (\lg n) }$.

		POPE tree is not optimized for \acrshort{io} the way B-tree is.
		Search complexity is hard to analyze as is bandwidth complexity.
		In the worst-case (first query), all blocks need to be accessed $\bigO{\frac{N}{B} + \frac{r}{B}}$.
		In the best-case all nodes occupy exactly one block and \acrshort{io} complexity is the same as with {\BPlus} tree $\bigO{\log_L \frac{N}{B} + \frac{r}{B}}$.
		The average case is in between and matters get worse as the node is not guaranteed to occupy a single block due to the buffers of arbitrary size.

		Client's persistent storage is negligibly small --- it stores the encryption key.
		Volatile storage is bounded by $L$.

		For practitioners we present a number of things to consider.
		Buffer within one node is unsorted, so in the worst-case, $L$-sized chunks remain unordered.
		Due to this property, the query result may contain up to $2 (L - 1)$ extra entries, which the client will have to discard from the response.

		The first query after a large number of insertions will result in client sorting the whole $N$ elements, and thus, POPE has different performance for cold and warm start.
		Also, even to navigate an already structured tree, the server has to send to the client the whole $L$ elements and ask where to go on all levels.

		Furthermore, \cite{pope} does not stress the fact that after alternating insertions and queries, it may happen that some intermediate buffers are not empty, thus returning buffers between endpoints must include intermediate buffers as well. 
		The consequence is that the whole subtree is traversed between paths to endpoints, unlike the {\BPlus} tree case where only leaves are involved.

		Finally, POPE tree is not optimized for \acrshort{io} operations.
		Even if $L$ is chosen so that the node fits in the block, only leaves and only after some number of searches will optimally fit in blocks.
		Arbitrary sized buffers of intermediate nodes and the lack of underflow requirement do not allow for \acrshort{io} optimization.

	\subsection{Logarithmic-BRC \texorpdfstring{\cite{practical-range-search}}{}}

	\textcite{practical-range-search} introduced a novel protocol called ``Logarithmic-BRC'' whose \acrshort{io} complexity depends only on the result size, regardless of the database size.
	The core primitive for their construction is a \acrfull{sse} scheme.
	An \acrshort{sse} scheme is a server-client protocol in which the server stores a specially encrypted keywords-to-documents map, and a client can query documents with keywords while the server
	learns neither keywords nor the documents.
	Note that the map stores short document identifiers instead of the actual documents, and we will use the term ``documents'' to mean ``document identifiers'' or ``record \acrshortpl{id}'' in this section.

	The construction treats record values as documents and index ranges as keywords so that records can be retrieved by the ranges that include them.
	Specifically, a client builds a virtual binary tree over the domain of indices and assigns each record a set of keywords, which is the path from that record to the root.
	This way, the root keyword is associated with all documents and the leaf keyword is associated with only one record.

	Upon query, a client computes a cover --- a set of nodes whose sub-trees cover the requested range.
	A client sends these keywords to the \acrshort{sse} server, which returns encrypted documents --- result values.
	Of the several covering techniques suggested in the protocol \cite{practical-range-search} we have chosen the \acrfull{brc}, because it results in fewest nodes and does not return false-positives.
	\textcite{brc} have proven that the worst-case number of nodes for domain of size $N$ is $\bigO{\log N}$ and presented an efficient \acrshort{brc} algorithm.

	\subsubsection{Security}

		In a snapshot setting, this construction's security is that of the \acrshort{sse}.
		We have used \cite{cjjkrs-13} and \cite{cjjjkrs-14} \acrshort{sse} schemes; their leakage in a snapshot setting is the database size and at most some initialization parameters.
		Thus, the security of these schemes is high enough to call them \emph{fully hiding} in our setting.
		Additional access pattern leakage comes up during queries; exact implications of this leakage remain an open research problem but it is known that it can be harmful \cite{generic-attacks-kellaris}.

	\subsubsection{Analysis and implementation challenges}

		Communication involves a client sending at worst $\log_2 N$ keywords and server responding with the exact result.

		For each keyword in the query set, server will query the \acrshort{sse} scheme, which will return $r$ documents.
		Therefore, server's \acrshort{io} complexity is that of \acrshort{sse}.

		\textcite{practical-range-search} have used \cite{cjjkrs-13} \acrshort{sse} scheme in their implementation, but we have found it slow it terms of \acrshort{io}.
		Instead, we have implemented an improved scheme \cite{cjjjkrs-14}, which directly addresses \acrshort{io} optimization.

		Both \acrshort{sse} schemes' \acrshort{io} complexity is linear with the result size $r$.
		\cite{cjjjkrs-14} scheme makes at most one \acrshort{io} per result document in the worst-case and there are extensions to significantly improve \acrshort{io} complexity. 
		We have implemented the \texttt{pack} extension, which packs documents in blocks to fit the \acrshort{io} pages.
		We note that this extension can dramatically reduce the \acrshortpl{io} (see \cref{section:range-snapshot:results-protocols} and \cref{figure:protocols-query-sizes}).

		Logarithmic-BRC is very scalable as its performance does not depend on total data size and only degrades with the result size.
		Storage overhead, however, is significant.
		Each record is associated with the whole path in the binary tree --- $\log_2 N$ nodes (keywords).
		The storage complexity is therefore $\bigO{N \log N}$, and the overhead is then a factor of $\log N$.

		Updates, while addressed in the original protocol, are not very practical in this construction.
		Authors suggest using bulk-loading for updates, maintaining merge trees, and requiring the client to do a merge once in a while.
		The \acrshort{io} complexity of such approach is unclear.
		In our implementation we perform the construction stage only in batch mode, and thus do not include it in the analysis.
		We also emphasize that the update routine was not implemented for evaluation in the original paper.

	\subsection{The two extremes}

	To put the aforementioned protocols in a context we introduce the baselines --- an efficient and insecure construction we will refer to as \emph{no encryption} and maximal security protocol we refer to as \emph{\acrshort{oram}}.

	\subsubsection{No encryption}

		This protocol is a regular {\BPlus} tree \cite{b-tree} without any \acrshort{ore} in it.
		It is the construction one can expect to see in almost any general-purpose database.

		In terms of security it provides no guarantees --- all data is in the clear.
		In terms of efficiency it is optimal.
		{\BPlus} tree data structure is optimal in \acrshort{io} usage, indices inside nodes are smallest possible (integers) and there is no overhead in comparing elements inside the nodes as opposed to working with \acrshort{ore} ciphertexts.

	\subsubsection{\texorpdfstring{\acrshort{oram}}{ORAM}}\label{section:range-snapshot:oram}

		\acrfull{oram} is a construction that additionally to semantic security of a snapshot setting (see \cref{section:range-snapshot:security}) provably hides the access pattern --- a sequence of reads and writes to particular memory locations.
		With \acrshort{oram} an adversary would not be able to recognize a series of accesses to the same location and will not differentiate reads versus writes.
		\acrshort{oram} was introduced by~\textcite{oram-original} who also proved its lower bound (strengthened in \cite{oram-tighter-lower-bound}) --- logarithmic overhead per request.
		A number of efficient \acrshort{oram} constructions were designed (see \cite{oram-survey-feifei} for a good survey) and we use the state-of-the-art construction, PathORAM \cite{path-oram}.

		A generic \acrshort{oram} server responds to read and write requests for a particular address.
		In our baseline protocol we store {\BPlus} tree nodes in \acrshort{oram}.
		A client works with the tree as it normally would except each time it needs to access a node, it communicates with \acrshort{oram}.

		In terms of security this protocol is fully hiding in the snapshot model and provably hides the access pattern.
		We note that one can improve security even further by adding noise to the result obscuring communication volume.
		We also note that a practitioner can use a similar protocol with \acrshort{oram} replaced with a trivial data store and have the tree nodes encrypted.
		It would be fully hiding in a snapshot setting, but we prefer the baseline that covers more than only the snapshot model.

		In terms of performance this construction incurs some noticeable overhead.
		Regardless of specific \acrshort{oram} being used, each access incurs at least logarithmic overhead according to lower bounds \cite{oram-original}.
		Combined with logarithmic complexity of the {\BPlus} tree itself, the complexity, both \acrshort{io} and communication, is $\bigO{\log^2 N}$.
		We found that PathORAM has good \acrshort{io} performance, as its internal tree structure translates into good cache affinity.
		Unlike in other protocols in our benchmark, \acrshort{oram} client does most of the computational work.
		While the server only makes \acrshort{io} requests, the client handles encryption, shuffling, and request logic.

		We present this protocol as a baseline solution in terms of security over efficiency.
		We have not implemented stand-alone PathORAM, but rather a simulator which correctly reports \acrshort{io}, communication and primitive usage.
		Surprisingly, we found that \acrshort{oram} protocol's overhead, although higher than in \acrshort{ore}-based protocols, is in-line with the most secure protocols in our benchmark.


\begin{sidewaystable}
	\renewcommand{\arraystretch}{1.5}
	\centering
	\captionsetup{width=\textwidth}
	\caption[Performance of the range query protocols]{
		\cite[Tables 2 and 3]{ore-benchmark-17}.
		Performance of the range query protocols.
		Ordered by security rank --- most secure below.
		$N$ is a total data size, $B$ is an \acrshort{io} page size, $L$ is a POPE tree branching factor, $r$ is the result size in records and \textbf{bolded} are weak points of the protocols.
		The cell content is structured as follows: top value is the analytical result in $\mathcal{O}$ notation, bottom value is the number of requests for \acrshort{io} requests or number of messages and their total size for communication.
		In these experiments, $N = 247K$, $B = \SI{4}{\kilo\byte}$, $r \approx 247K \cdot \SI{0.5}{\percent} = 1235$, and $L = 60$.
	}\label{table:protocols}
	\begin{tabular*}{\linewidth}{ !{\extracolsep\fill} c c c c c c } 

		\toprule

		\multirow{2}{*}{Protocol}						& \multicolumn{2}{c}{\acrshort{io} requests}																																			& \multirow{2}{*}{Leakage}						& \multicolumn{2}{c}{Communication}																																										\\ \cline{2-3} \cline{5-6}
		\rule{0pt}{10pt}								& Construction														& Query																												&												& Construction																						& Query 																							\\

		\toprule

		\makecell{{\BPlus} tree \\ with \acrshort{ore}}	& \makecell{$\log_B \frac{N}{B}$ \\ 3 requests}						& \makecell{$\log_B \frac{N}{B} + \frac{r}{B}$ \\ 44 requests}														& \textbf{Same as \acrshort{ore}}				& \makecell{$1$ \\ 2 / \SI{177}{\byte}}																& \makecell{$1$ \\ 2 / \SI{342}{\byte}}																\\

		\midrule

		\cite{florian-protocol}							& \makecell{$\bm{\frac{N}{B}}$ \\ \textbf{494 requests}}			& \makecell{$\log_2 \frac{N}{B} + \frac{r}{B}$ \\ 17 requests}														& \textbf{Total order}							& \makecell{$\log_2 N$ \\ 40 / \SI{671}{\byte}}														& \makecell{$\log_2 N$ \\ 86 / \SI{1453}{\byte}}													\\

		\midrule

		\makecell{\cite{pope} \\ warm}					& \multirow{2}{*}{\makecell{$1$ \\ 1 request}}						& \makecell{$\log_L \frac{N}{B} + \frac{r}{B}$ \\ 300 requests}														& \textbf{Partial order}						& \multirow{2}{*}{\makecell{$1$ \\ 2 / \SI{32}{\byte}}}												& \makecell{$\log_L N$ \\ 914 / \SI{347}{\kilo\byte}}												\\ \cline{1-1} \cline{3-3} \cline{6-6}

		\makecell{\cite{pope} \\ cold}					& 																	& \makecell{$\bm{{\nicefrac{N}{B}}}$ \\ \textbf{2175 requests}}														& Fully hiding									& 																									& \makecell{$\bm{N}$ \\ \textbf{498K / \SI[detect-all=true]{9}{\mega\byte}}}						\\

		\midrule

		\cite{practical-range-search}					& \textbf{---}														& \makecell{$\bm{r}$ \\ \textbf{40 requests}}																		& Same as \acrshort{sse}						& \textbf{---}																						& \makecell{$\log_2 N$ \\ 2 / \SI{391}{\byte}}														\\

		\midrule

		\acrshort{oram}									& \makecell{$\bm{{ \log^2 \frac{N}{B} }}$ \\ \textbf{31 requests}}	& \makecell{$\bm{{ \log_2 \frac{N}{B} \left( \log_B \frac{N}{B} + \frac{r}{B} \right) }}$ \\ \textbf{185 requests}}	& \makecell{Fully hiding \\ (access pattern)}	& \makecell{$\bm{{ \log^2 \frac{N}{B} }}$ \\ \textbf{143 / \SI[detect-all=true]{18}{\kilo\byte}}}	& \makecell{$\bm{{ \log^2 \frac{N}{B} }}$ \\ \textbf{490 / \SI[detect-all=true]{63}{\kilo\byte} }}	\\

		\bottomrule

	\end{tabular*}
\end{sidewaystable}

	\section{Evaluation}\label{section:range-snapshot:evaluation}

	All experiments were conducted on a single machine.
	We use macOS 10.14.2 with 8-Core \SI{3.2}{\giga\hertz} Intel Xeon W processor, \SI{32}{\giga\byte} DDR4 ECC main memory and \SI{1}{\tera\byte} \acrshort{ssd} disk.
	The main code is written in {\Csharp} and runs on {.NET Core 2.1.3}.

	\subsubsection*{Interactive website}\label{section:range-snapshot:website}

		Additionally to making our source code, compiled binaries and Docker images available, we want to let researchers interactively run small-sized simulations.
		We host a website\footnote{\url{https://ore.dbogatov.org/}} where one can select a protocol (including baselines, CLOZ \cite{parameter-hiding-ore} and both \acrshort{sse} schemes), cache size and policy and \acrshort{io} page parameter; supply one's own data and query sets, and run the simulations.
		Simulations are run one at a time and usually complete within seconds.
		The user is then able to view the result --- tables, plots, values and raw logs, which we used to build plots for this paper.
		Input size on the website is limited for practical purposes and users are encouraged to run arbitrary-size simulations using our binaries or Docker images.

	\subsection{Implementation}

		We have implemented most of the primitives, data structures, and constructions ourselves.
		For some primitives and all schemes we provided the first open-sourced cross-platform {\Csharp} implementation.
		We note that neither primitives, nor schemes are production-ready; however, we believe they can be used in research projects and prototypes.
		We also emphasize that the {\BPlus} tree implementation we are using, although our own with instrumentation in it, is not custom in any way, but rather standard as defined in the original paper \cite{b-tree} with deletion algorithm by \cite{b-plus-tree-deletion}.

		This software project (22K lines of code, third of which are tests) is documented and tested (over \SI{97}{\percent} coverage).
		All code including primitives, data structures, schemes, protocols, simulation logic, benchmarks, build scripts and tests is published on GitHub\footnote{\url{https://github.com/dbogatov/ore-benchmark}} under CC BY-NC 4.0 license.
		Additionally, we have published parts of the project as stand-alone {.NET Core} (nuGet) packages, and we host a web-server where users can run simulations for small inputs (see previous subsection).

		\subsubsection{Primitives}

			All schemes and protocols use the same primitives, most of which we implemented ourselves.
			All primitives rely on the default {.NET Core} \acrshort{aes} implementation.
			{.NET Core} uses platform-specific implementation of \acrshort{aes}, thus leverages \texttt{AES-NI} \acrshort{cpu} instruction.
			In our project all key sizes are 128 bits, as is \acrshort{aes} block size.

			We implemented \acrshort{aes}-based \acrshort{prg}, which uses \acrshort{aes} \cite{aes-nist} in \acrshort{ctr} mode \cite{nist-modes} and caches unused entropy (as suggested in \cite{aes-ctr-rfc}).
			For \acrshort{prf}, since we need only 128-bit inputs and outputs, we used one application of \acrshort{aes} \cite[Proposition 3.27]{intro-to-modern-crypto}.
			For symmetric encryption we use \acrshort{aes} with a random initialization vector in \acrshort{cbc} mode \cite[Section 3.6.2]{intro-to-modern-crypto}.
			For hash we use default {.NET Core} \acrshort{sha}-2 implementation.
			For \acrshort{prp}, we implemented unbalanced Feistel networks \cite{unbalanced-feistel} for large inputs and Knuth shuffle \cite{knuth-shuffle} for small inputs.
			Please see the README of project's repository\footnote{\url{https://github.com/dbogatov/ore-benchmark}} for low-level details.

			\begin{figure}[!ht]
	\centering
	\begin{minipage}[t]{0.48\columnwidth}
		\captionsetup[figure]{justification=centering}
		\centering
		\includegraphics[width=\linewidth]{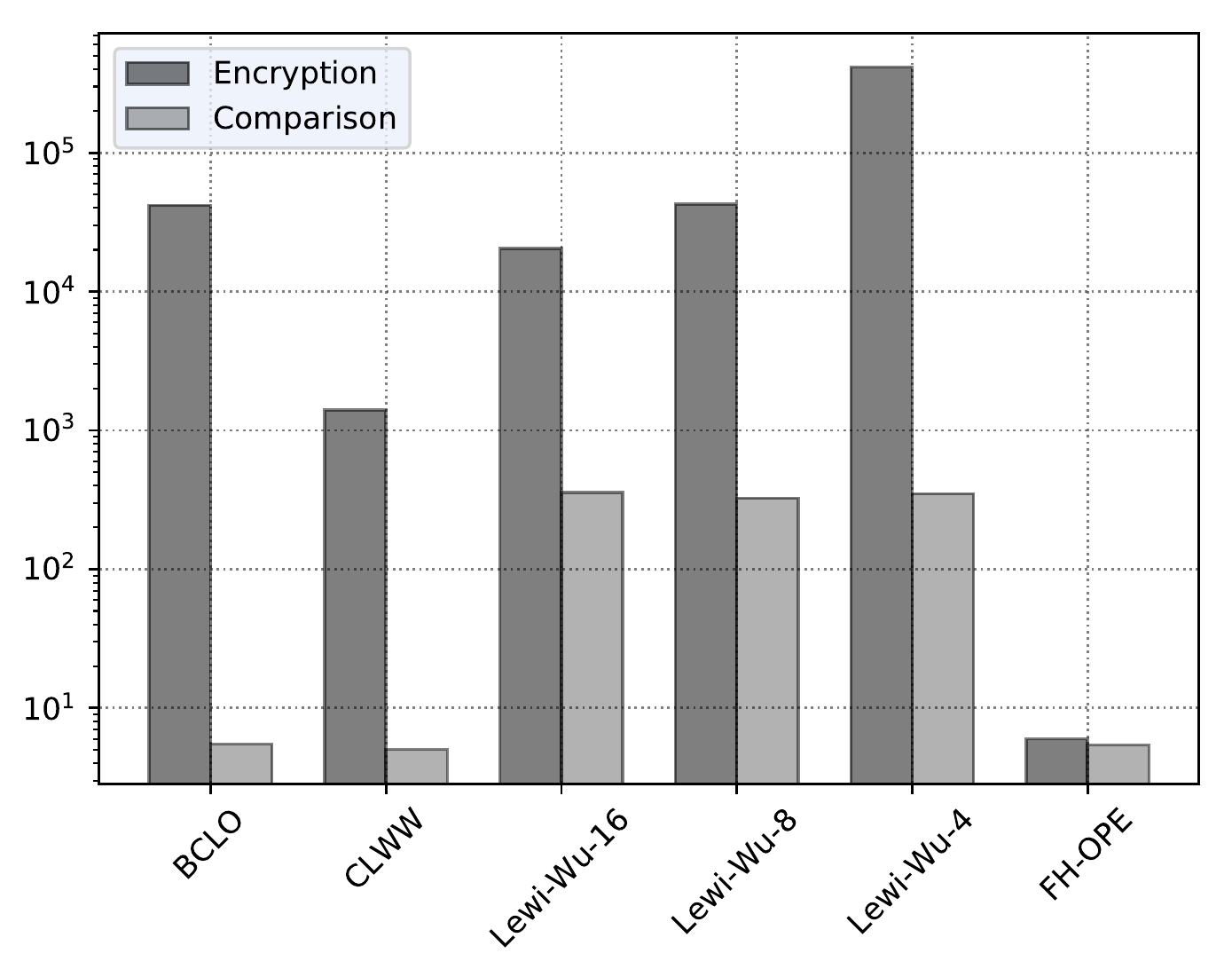}
		\captionof{figure}[\acrshort{ope} / \acrshort{ore} schemes benchmark]{
			Schemes benchmark (time in microseconds, log scale).
			Lewi-Wu parameter is the number of blocks.
		}%
		\label{figure:benchmarks:schemes}
	\end{minipage}
	\hfill
	\begin{minipage}[t]{0.48\columnwidth}
		\captionsetup[figure]{justification=centering}
		\centering
		\includegraphics[width=\linewidth]{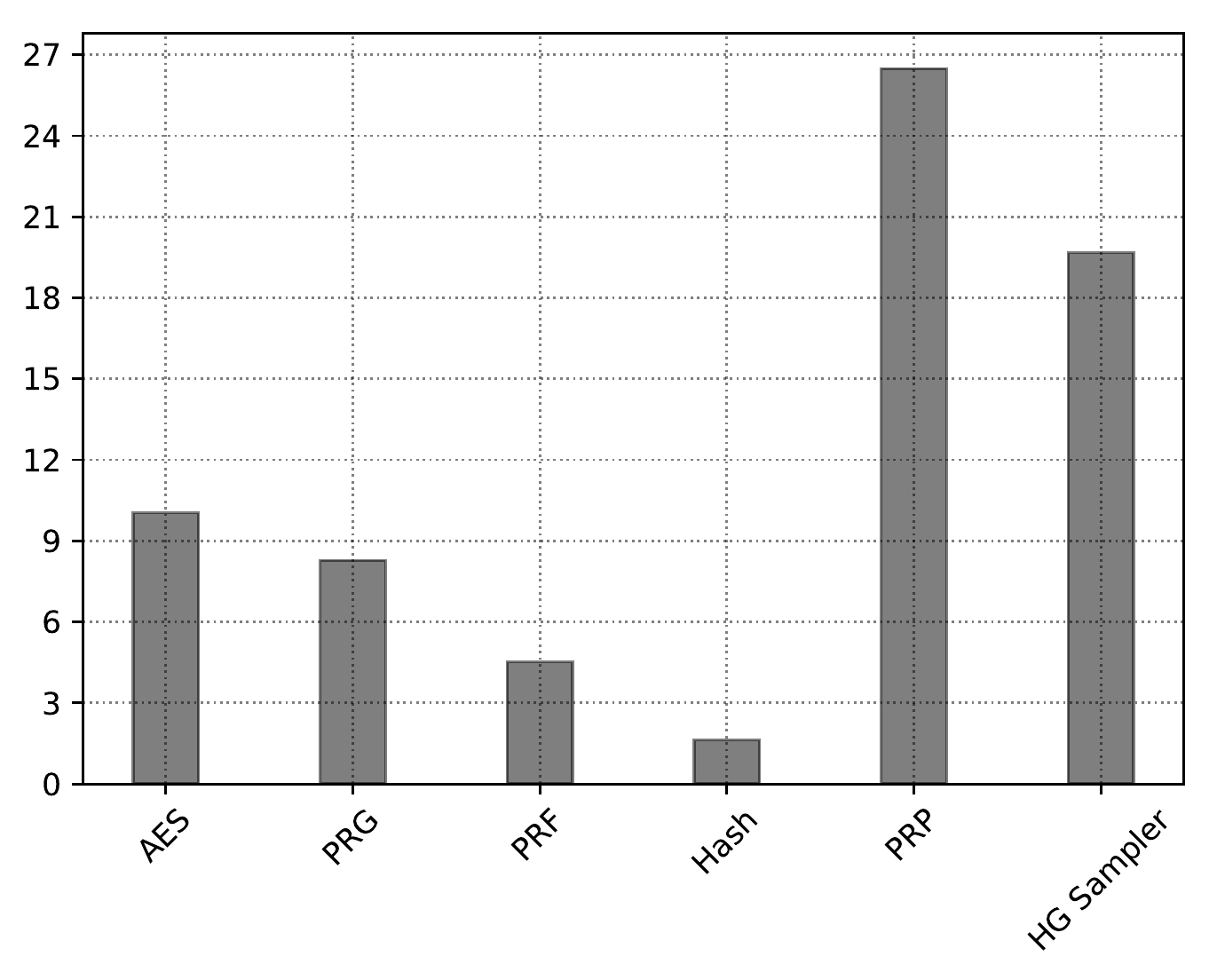}
		\captionof{figure}[Cryptographic primitives benchmark]{Primitives benchmark (time in microseconds)}%
		\label{figure:benchmarks:primitives}
	\end{minipage}
\end{figure}

		\subsubsection{Schemes and protocols}

			We implemented schemes and protocols precisely as in the original papers.
			When we found problems or improvements, we described them in implementation challenges notes, but did not alter the original designs in our code, unless explicitly stated.
			Each \acrshort{ore} scheme implements a {\Csharp} interface; thus our own implementation of {\BPlus} tree operates on a generic \acrshort{ore}.
			For the \emph{no encryption} baseline, we have a stub implementation of the interface, which has identity functions for encryption and decryption.
			It is important to note that all schemes and protocols use exclusively our implementations of primitives.
			Thus we rule out the possible bias of one primitive implementation being faster than the other.

		\subsubsection{Simulations}

			We have four types of simulations.

			Protocol simulation runs both protocol stages --- construction and search --- on supplied data for all protocols including all schemes coupled with {\BPlus} tree.
			In this simulation we measure the primitive usage, number of \acrshort{ore} scheme operations (when applies), communication volume and size, and the number of \acrshort{io} requests.
			We intentionally do not measure elapsed time, since it would be extremely inaccurate in this setting --- simulation and measurement routines take substantial fraction of time.

			Scheme simulation runs all five \acrshort{ore} schemes and tracks only the primitive usage.

			The scheme benchmark, however, is designed to track time.
			We use Benchmark.NET \cite{benchmark-net} to ensure that the reported time is accurate.
			This tool handles issues like cold / warm start, elevating process' priority, and performing enough runs to draw statistically sound conclusions.
			This benchmark reports elapsed time up to nanoseconds for all four schemes (excluding CLOZ \cite{parameter-hiding-ore}) and their variants.

			\begin{figure*}[ht!]
	\captionsetup[subfigure]{justification=centering}
	\centering
	\begin{subfigure}[t]{0.5\textwidth}
		\centering
		\includegraphics[width=\linewidth]{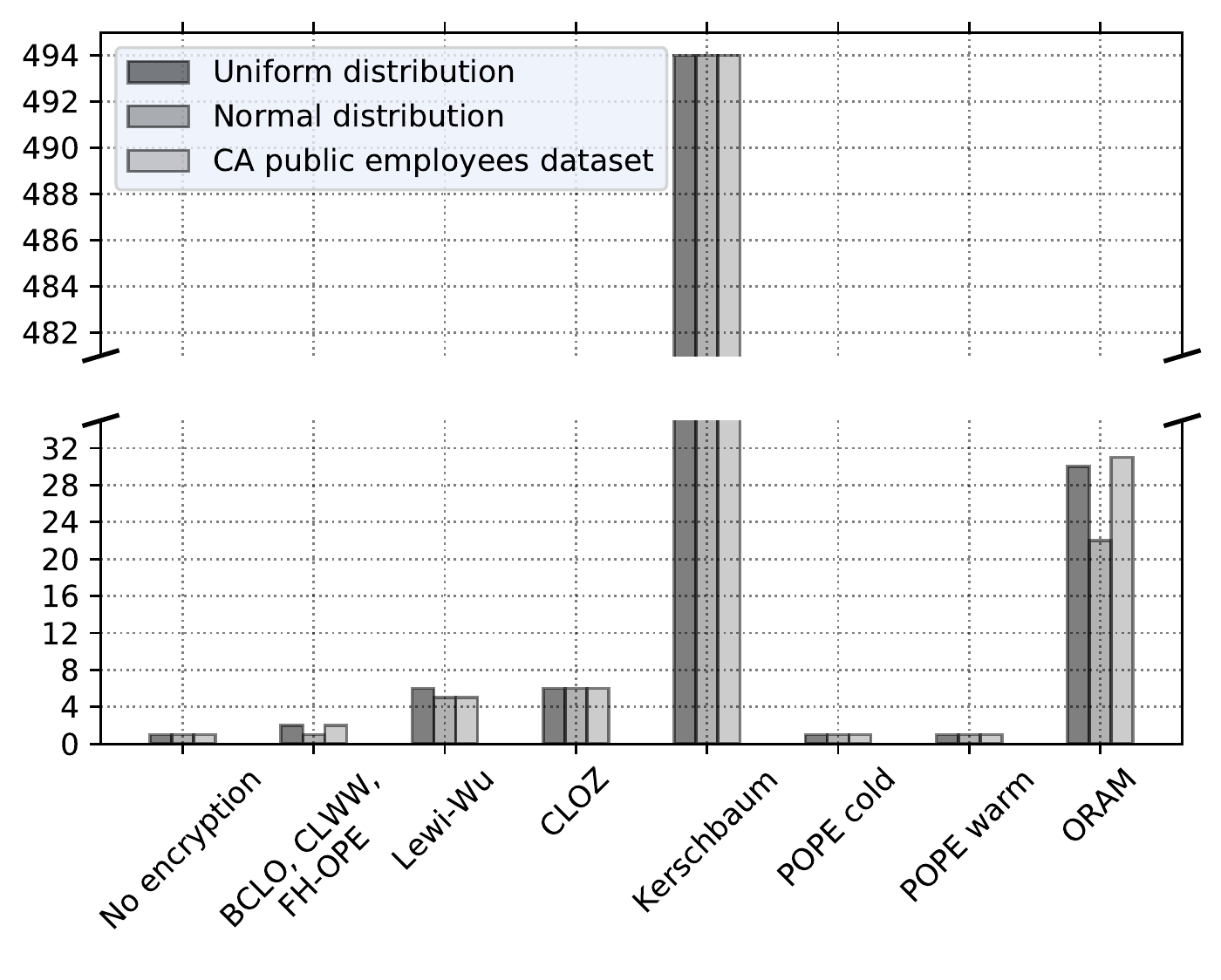}
		\caption{Construction stage number of \acrshort{io} requests}%
		\label{figure:protocols-ios:c}
	\end{subfigure}%
	\hfill
	\begin{subfigure}[t]{0.5\textwidth}
		\centering
		\includegraphics[width=\linewidth]{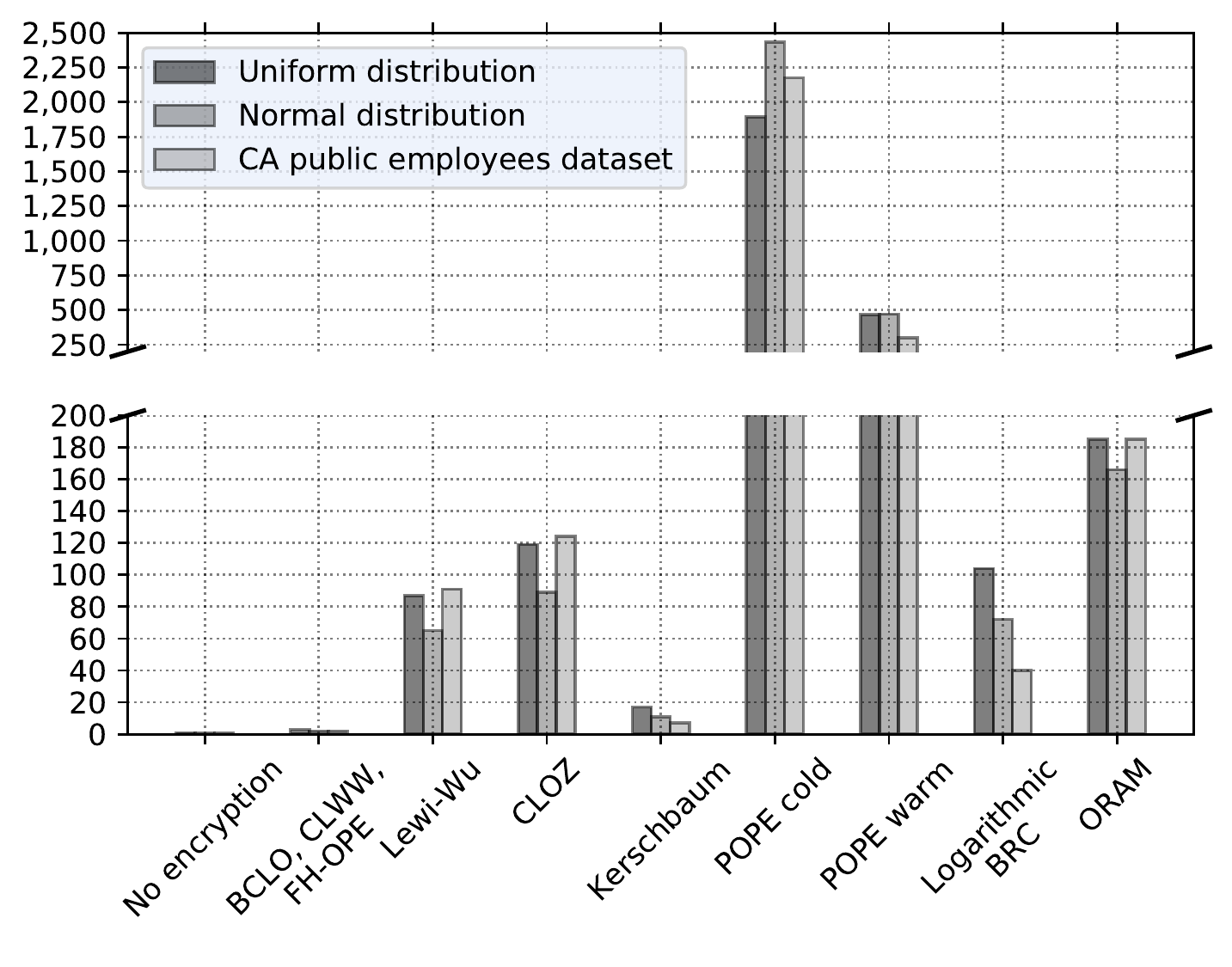}
		\caption{Queries stage number of \acrshort{io} requests}%
		\label{figure:protocols-ios:q}
	\end{subfigure}%
	\caption{Number of \acrshort{io} requests for different protocols and data distributions}%
	\label{figure:protocols-ios}
\end{figure*}

\begin{figure*}[ht!]
	\captionsetup[subfigure]{justification=centering}
	\centering
	\begin{subfigure}[t]{0.5\textwidth}
		\centering
		\includegraphics[width=\linewidth]{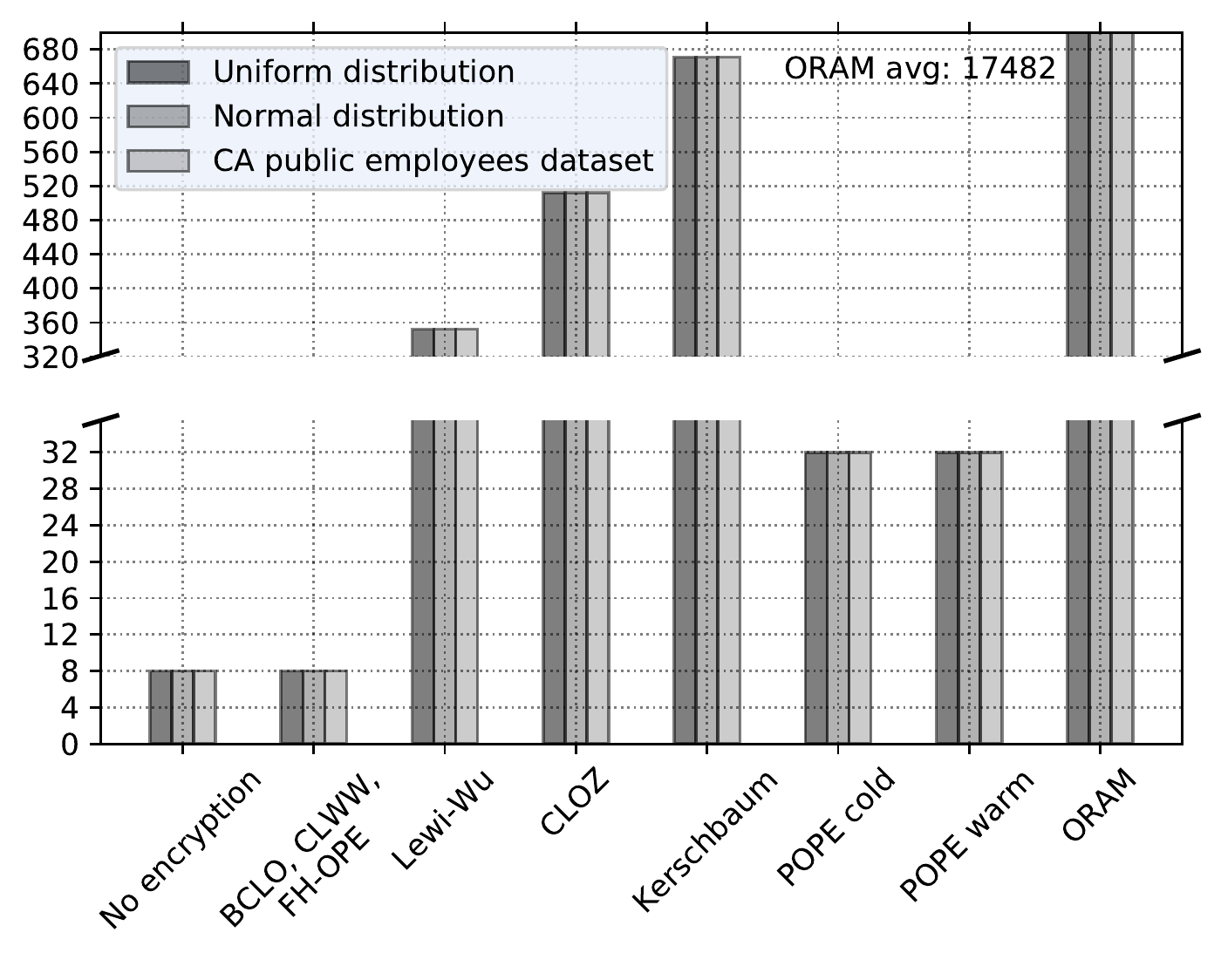}
		\caption{Construction stage communication size (bytes transferred)}%
		\label{figure:protocols-size:c}
	\end{subfigure}%
	\hfill
	\begin{subfigure}[t]{0.5\textwidth}
		\centering
		\includegraphics[width=\linewidth]{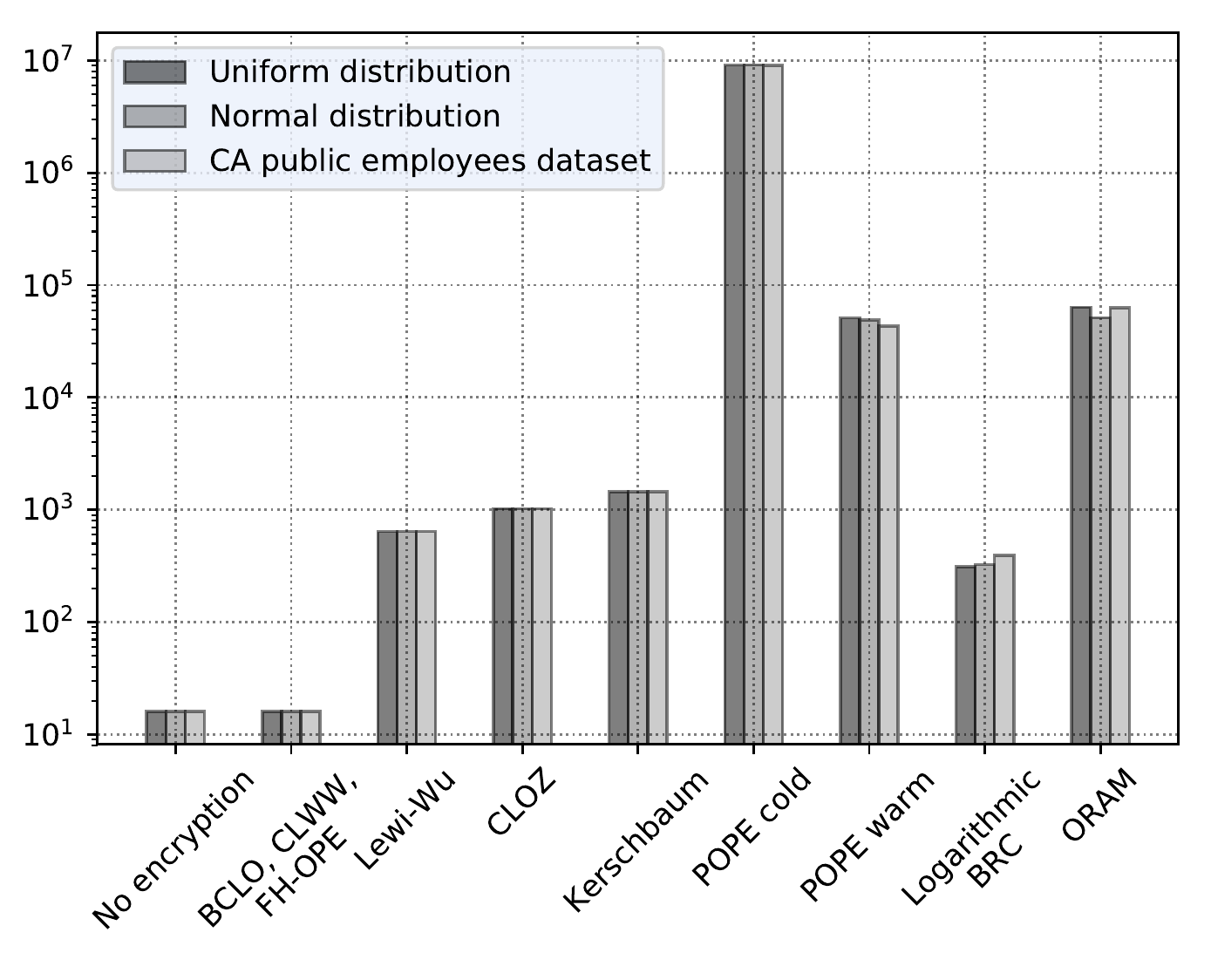}
		\caption{Queries stage communication size (transferred bytes, log scale)}%
		\label{figure:protocols-size:q}
	\end{subfigure}%
	\caption{Communication size for different protocols and data distributions}%
	\label{figure:protocols-size}
\end{figure*}

\begin{figure*}[ht!]
	\captionsetup[subfigure]{justification=centering}
	\centering
	\begin{subfigure}[t]{0.5\textwidth}
		\centering
		\includegraphics[width=\linewidth]{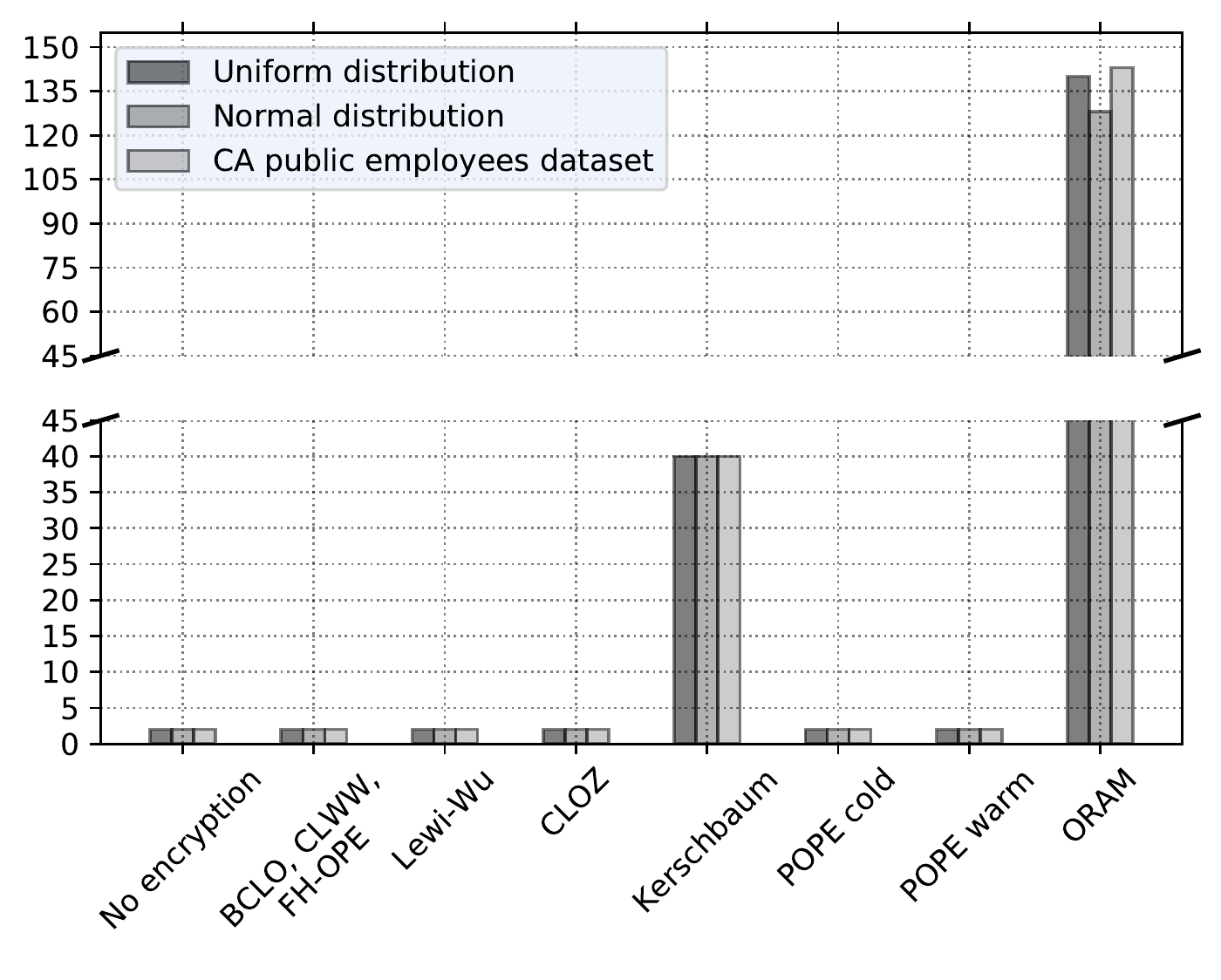}
		\caption{Construction stage communication volume (number of messages)}%
		\label{figure:protocols-vol:c}
	\end{subfigure}%
	\hfill
	\begin{subfigure}[t]{0.5\textwidth}
		\centering
		\includegraphics[width=\linewidth]{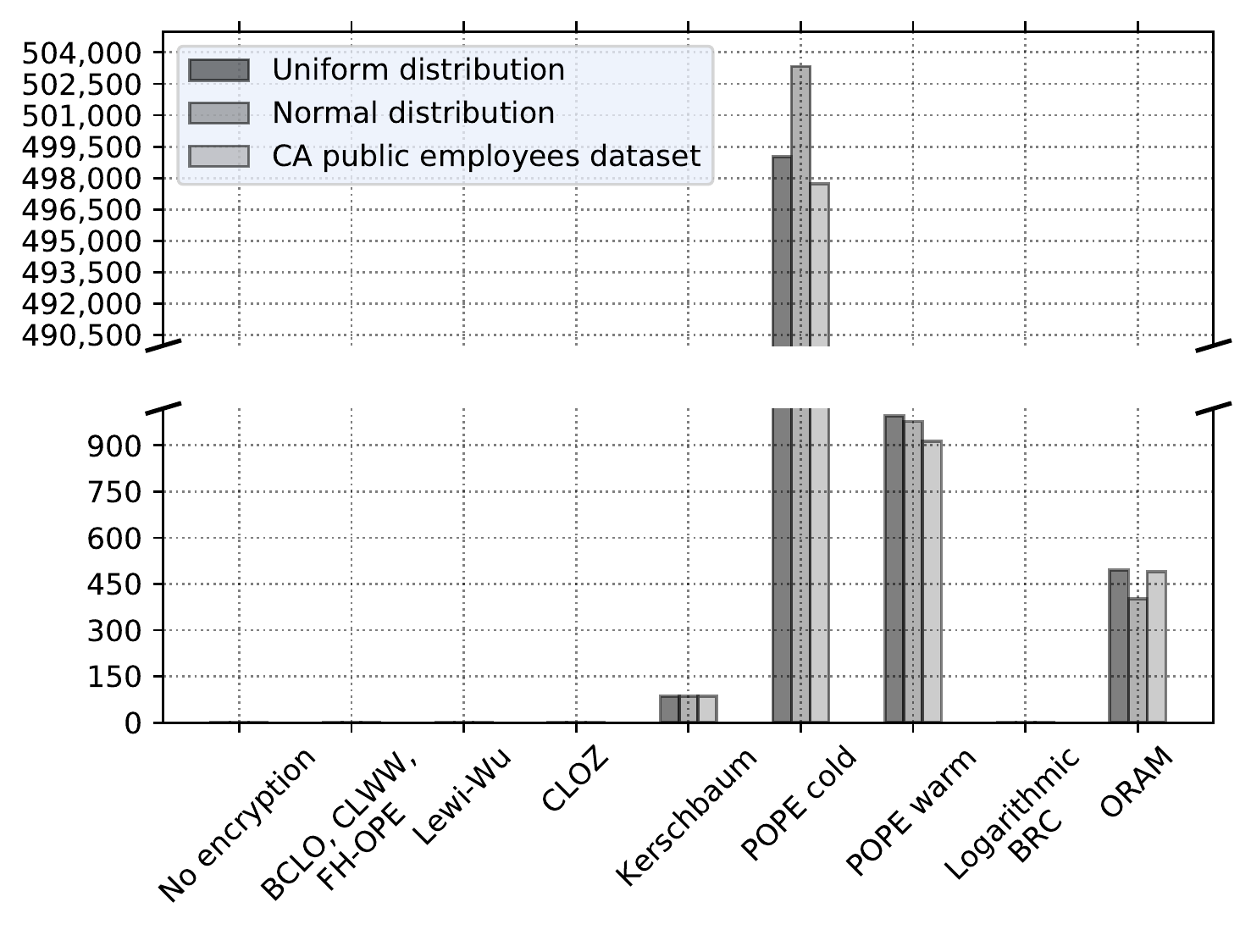}
		\caption{Queries stage communication volume (number of messages)}%
		\label{figure:protocols-vol:q}
	\end{subfigure}%
	\caption{Communication volume for different protocols and data distributions}%
	\label{figure:protocols-vol}
\end{figure*}

			Finally, primitive benchmark uses the same tool, but compares the primitives.
			We use it to compare different implementations of primitives (e.g.\ Feistel \acrshort{prp} vs pre-generated permutation) and to approximate time consumption of the schemes and protocols based on primitive usage.

	\subsection{Setup}

		For our simulations, we have used three datasets.
		Two synthetic distributions, that are uniform (range is third of data size) and normal (standard deviation is $0.1$ of data size).
		The real dataset is California public employees salaries (``total pay and benefits'' column) \cite{ca-dataset}.
		Synthetic datasets and subsets of the real dataset are generated pseudo\hyp{}randomly.
		Queries are generated uniformly at random with a range as a percentage of data size.

	\subsection{Results}

		\subsubsection{Primitive usage by schemes}\label{section:range-snapshot:schemes-primitive-usage}

			In \cref{table:primitive-usage-theory} we show the simulation-derived values of each \acrshort{ope} and \acrshort{ore} scheme's primitive usage.
			Each scheme is given \num{1000} data points of each dataset.
			First, the scheme encrypts each data point, then decrypts each ciphertext and then performs five comparisons (all possible types) pairwise.
			This micro-simulation is repeated \num{100} times.
			Resulting values for primitive usage are averaged for each scheme.
			State and ciphertext sizes are calculated after each operation and the values are averaged.
			Please note that the simulated values are consistent with the theoretical calculations.

		\subsubsection{Benchmarks of schemes and primitives}

			Using the Benchmark.NET tool \cite{benchmark-net}, we have accurately tracked the performance of the schemes and primitives running of different parameters (see \cref{figure:benchmarks:schemes,figure:benchmarks:primitives}).
			\acrshort{ore} schemes benchmark setup is the same as in primitive usage simulation \cref{section:range-snapshot:schemes-primitive-usage}.
			Primitives were given randomly generated byte inputs and keys of different sizes (e.g.\ \acrshort{prp} of $2$ to $32$ bits).
			Benchmark.NET decides how many times to run the routine to get statistically sound results.
			For example, large variance results in more runs.
			To improve the accuracy, each run is compiled in release mode as a separate project and runs in a separate process with the highest priority.

			Please note the logarithmic scale of the schemes' performances.
			FH-OPE is fast since it does not perform \acrshort{cpu}-heavy operations and works in main memory.
			Lewi-Wu performance degrades exponentially with the increase of block size mainly due to exponential number of \acrshort{prf} executions and the performance of \acrshort{prp} degrading exponentially.
			Note also that Lewi-Wu comparison takes noticeable time due to Hash primitive usage.

			In the primitives benchmark, it is clear that most primitives use \acrshort{aes} under the hood.
			\acrshort{prg} and \acrshort{prf} take less than \acrshort{aes} because they do not include the initialization vector generation needed for symmetric encryption.
			\acrshort{prp} is implemented as a Knuth shuffle \cite{knuth-shuffle} and its complexity is exponential in the input bit length.
			Input size of 2 bits is shown on \cref{figure:benchmarks:schemes}.
			\acrshort{prg} does not discard the entropy generated by \acrshort{aes} cycle, so one \acrshort{aes} cycle can supply four 32-bit integers.
			\acrshort{prp} generates the permutation table once and does not regenerate it if the same key and number of bits are supplied.

		\subsubsection{Protocols}\label{section:range-snapshot:results-protocols}

			In this experiment we have run each protocol with each of the three datasets.
			Dataset sizes are \num{247000} (bounded by California Employees dataset size) and the number of queries is \num{1000}.
			Queries are generated uniformly at random with a fixed range --- \SI{0.5}{\percent} of data size.
			The cache size is fixed to \num{128} blocks, and the {\BPlus} tree branching factor as well as block sizes for other protocols are set such that the page size is \SI{4}{\kibi\byte}.
			The values we are measuring are the number of \acrshort{io} operations, communication volume, and size for both construction and query stages.

			See \cref{table:protocols} for the snapshot for particular distribution (CA employees).
			\cref{figure:protocols-ios,figure:protocols-size,figure:protocols-vol} shows all values we tracked for all protocols and distributions.
			Values for \acrshort{ore} based protocols are averaged.
			Being ``cold'' in our simulations means executing the first query and being ``warm'' means the first query has been previously executed.
			This difference makes sense only for POPE as its first query incurs disproportionately large overhead by design.

			Note that all \acrshort{ore} based protocols behave the same except when ciphertext size matters.
			Thus, since BCLO, CLWW and FH-OPE have the same ciphertext size, they create {\BPlus} trees with the same page capacity and have the same number of \acrshortpl{io} for different operations.
			Lewi-Wu and CLOZ schemes have relatively large ciphertexts and thus induce larger traffic (see \cref{figure:protocols-size:c}) and smaller {\BPlus} tree branching factor resulting in greater number of \acrshort{io} requests (see \cref{figure:protocols-ios:q}).
			Kerschbaum protocol requires high number of \acrshort{io} requests during construction since it needs to insert an element into the arbitrary place in an array and rotate the data structure on a disk.

			POPE suffers huge penalty on the first query (see \cref{figure:protocols-ios:q,figure:protocols-vol:q,figure:protocols-size:q}) since it reads and sends all blocks to the client for sorting.
			POPE performance improves as more queries are executed.

			\begin{figure}[ht!]
	\captionsetup{justification=centering}
	\centering
	\begin{subfigure}[t]{0.5\textwidth}
		\centering
		\includegraphics[width=\linewidth]{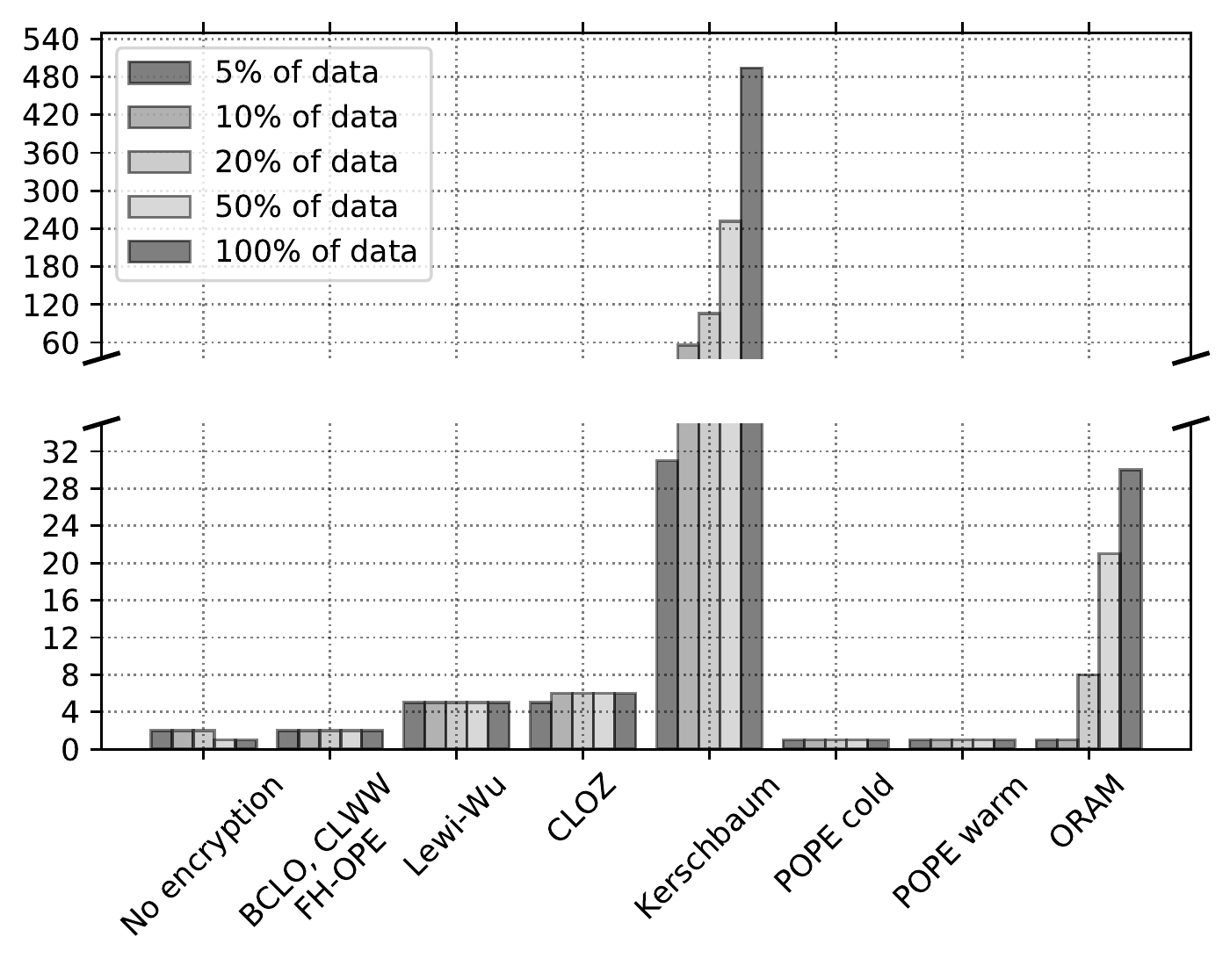}
		\caption{Construction stage \acrshort{io} requests}%
		\label{figure:protocols-data-percent-ios:c}
	\end{subfigure}%
	\hfill
	\begin{subfigure}[t]{0.5\textwidth}
		\centering
		\includegraphics[width=\linewidth]{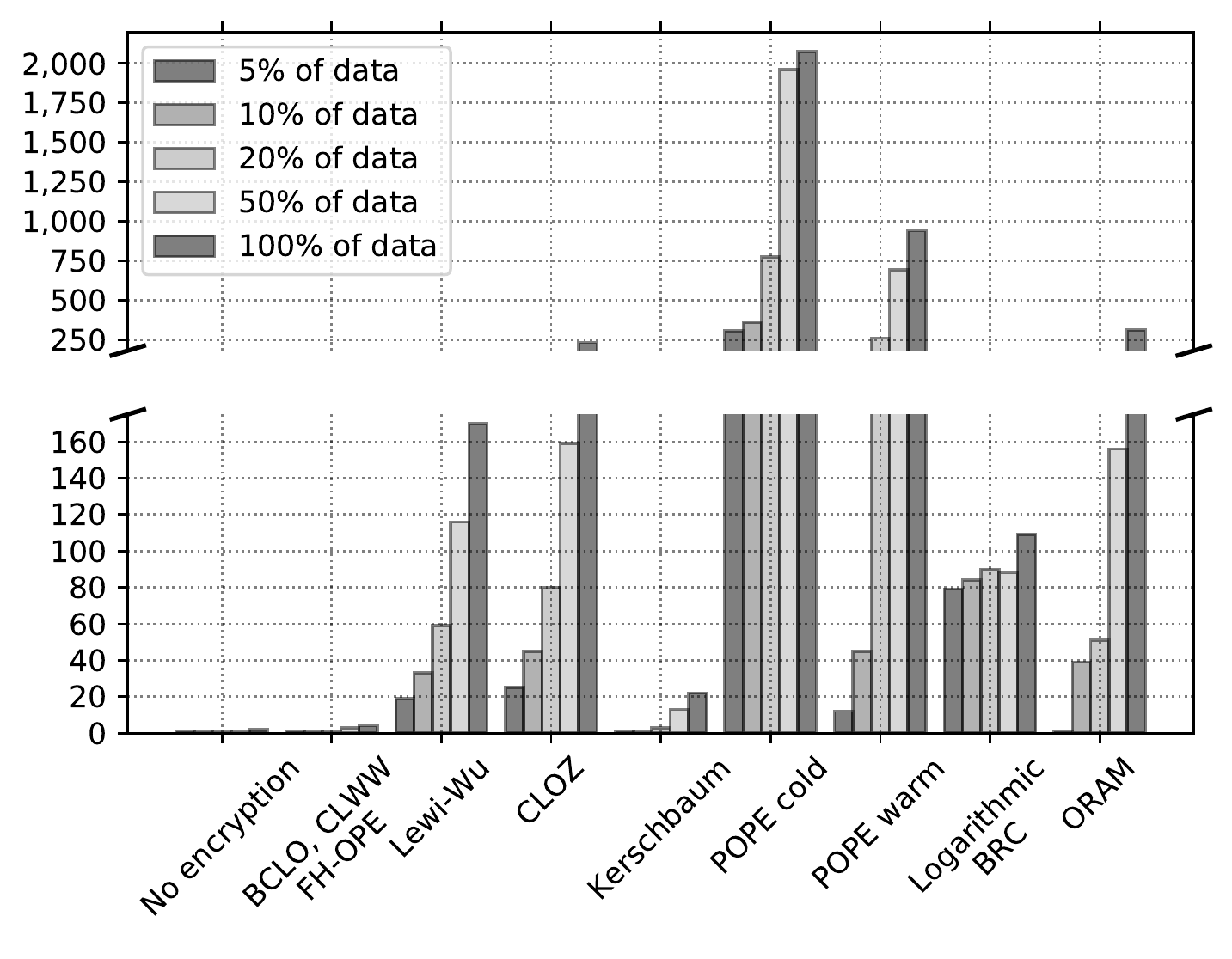}
		\caption{Queries stage \acrshort{io} requests}%
		\label{figure:protocols-data-percent-ios:q}
	\end{subfigure}%
	\caption{Scalability: number of \acrshort{io} requests}%
	\label{figure:protocols-data-percent-ios}
\end{figure}

\begin{figure}[ht!]
	\captionsetup{justification=centering}
	\centering
	\begin{subfigure}[t]{0.5\textwidth}
		\centering
		\includegraphics[width=\linewidth]{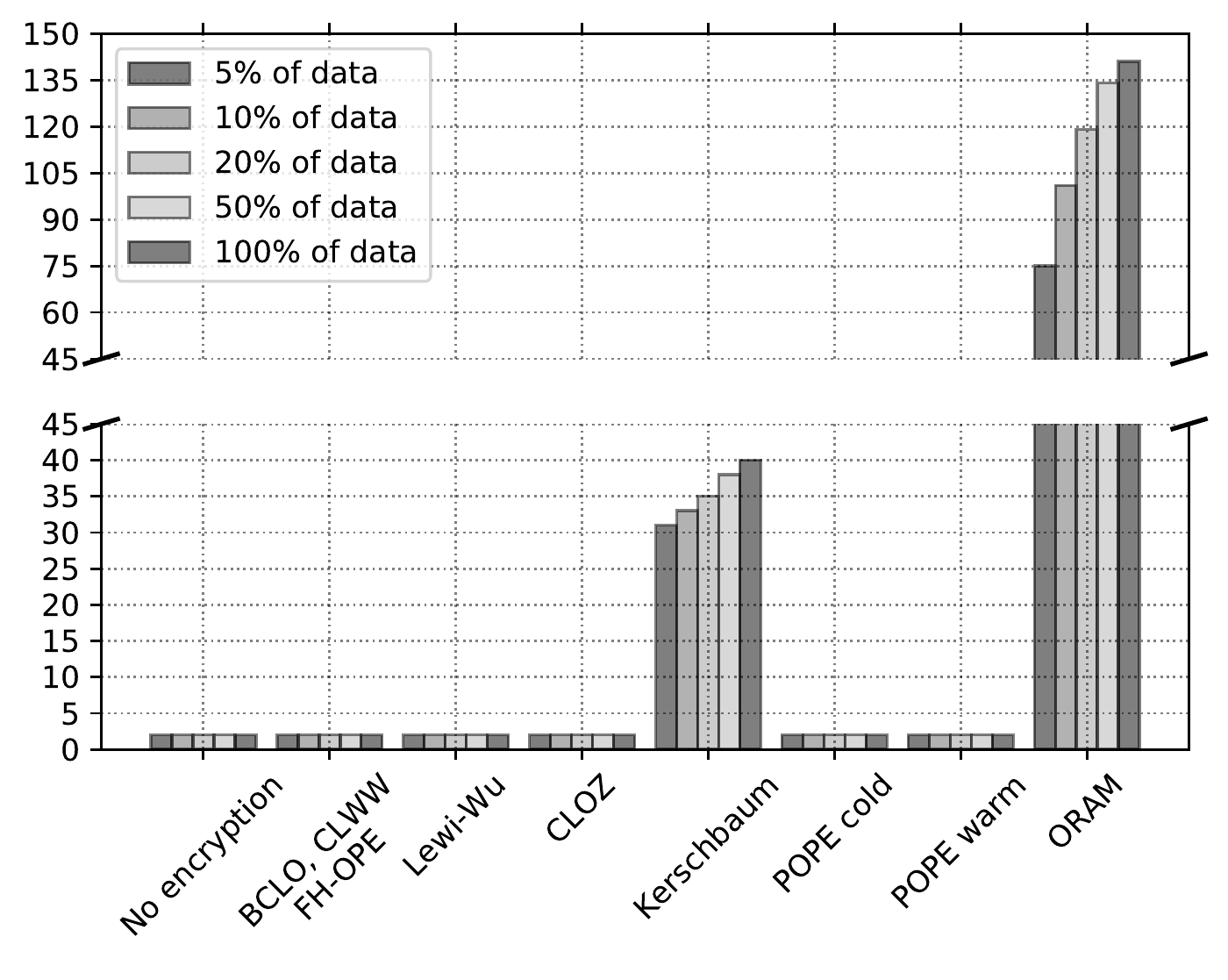}
		\caption{Construction stage number of messages}%
		\label{figure:protocols-data-percent-vol:c}
	\end{subfigure}%
	\hfill
	\begin{subfigure}[t]{0.5\textwidth}
		\centering
		\includegraphics[width=\linewidth]{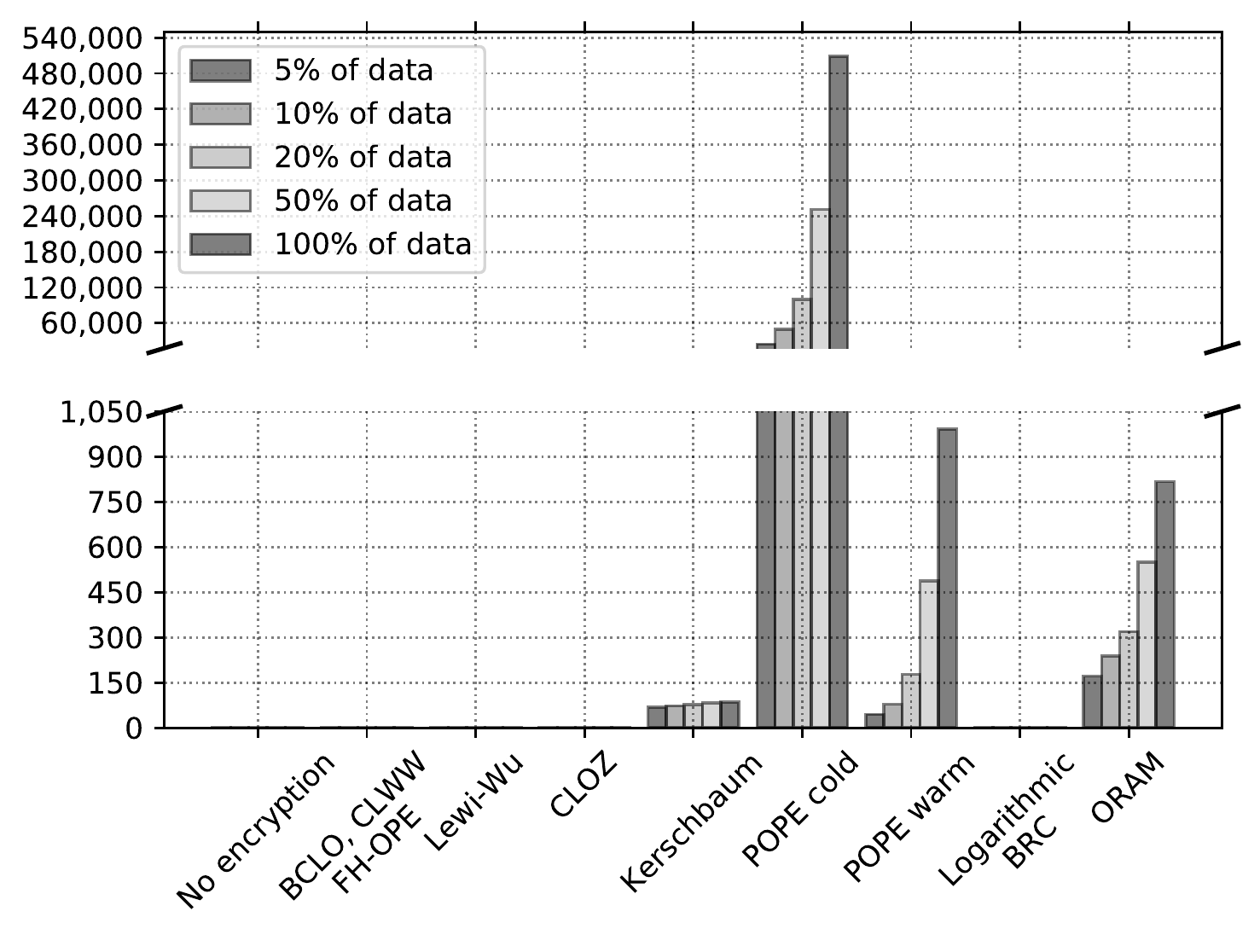}
		\caption{Queries stage number of messages}%
		\label{figure:protocols-data-percent-vol:q}
	\end{subfigure}%
	\caption{Scalability: communication volume}%
	\label{figure:protocols-data-percent-vol}
\end{figure}

\begin{figure}[ht!]
	\centering
	\begin{minipage}{.5\textwidth}
		\captionsetup[figure]{justification=centering}
		\centering
		\includegraphics[width=\linewidth]{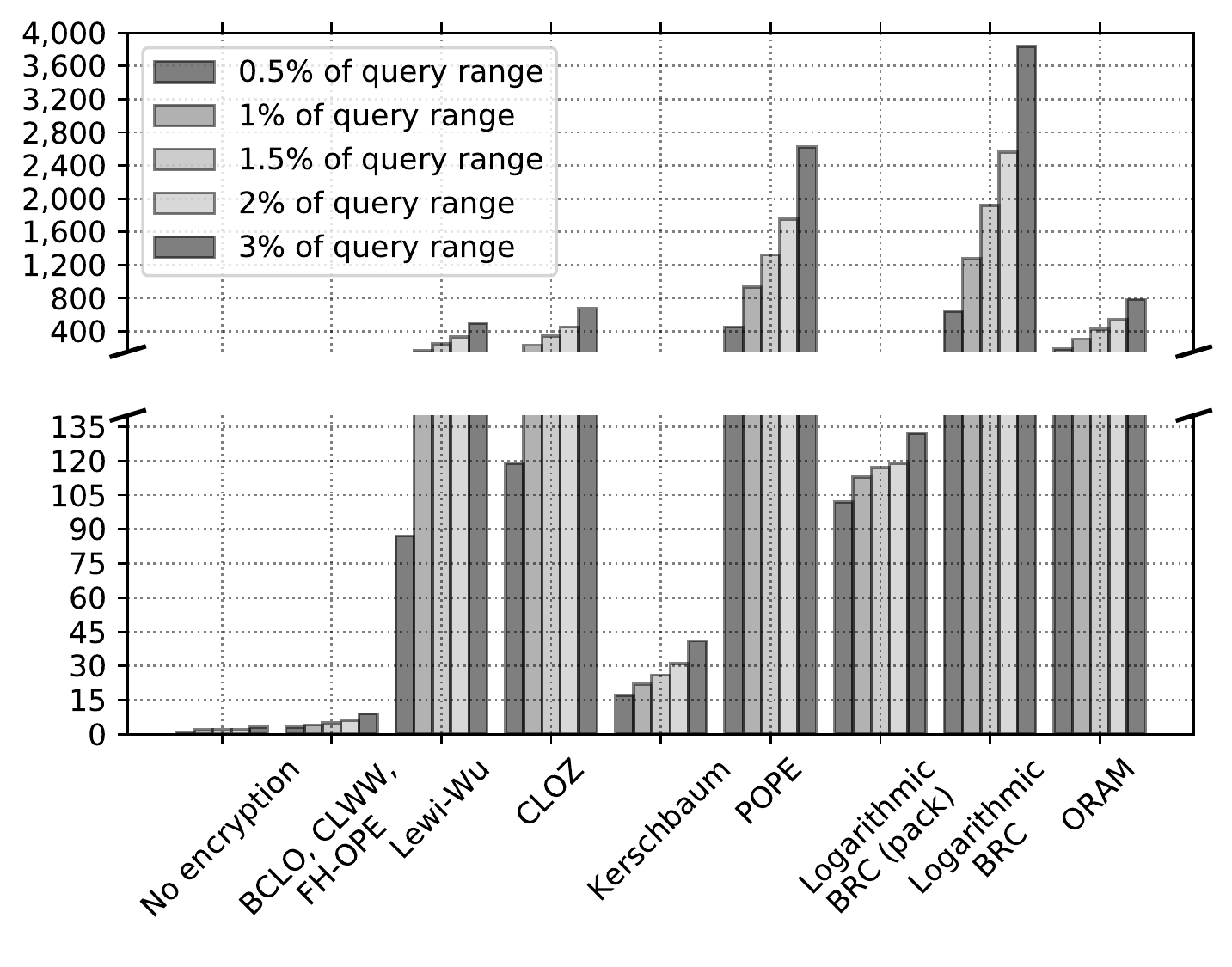}
		\captionof{figure}{Performance for different query sizes}%
		\label{figure:protocols-query-sizes}
	\end{minipage}%
	\hfill
	\begin{minipage}{.5\textwidth}
		\captionsetup[figure]{justification=centering}
		\centering
		\includegraphics[width=\linewidth]{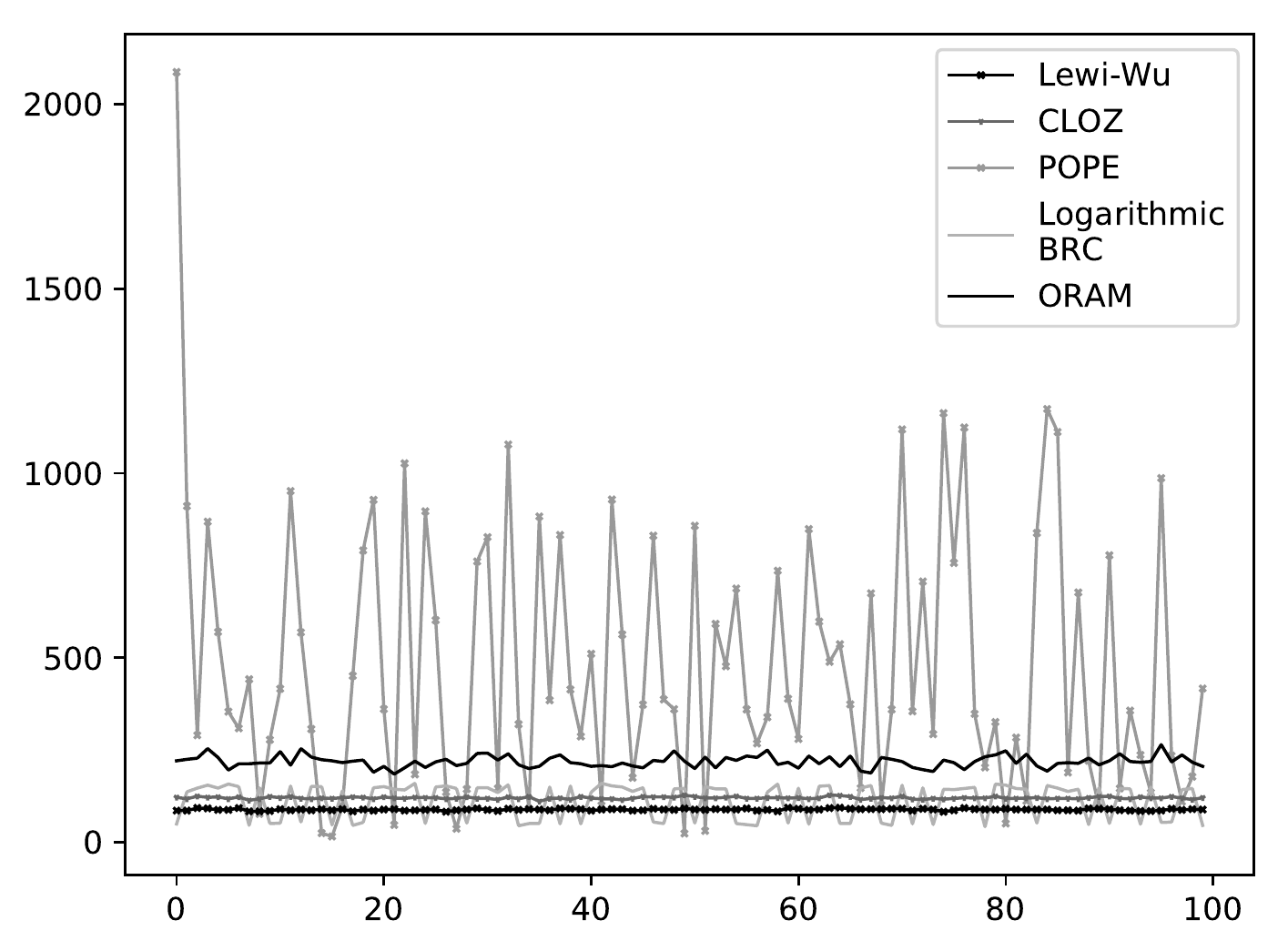}
		\captionof{figure}{Performance over time (queries)}%
		\label{figure:cold-vs-warm}
	\end{minipage}
\end{figure}

			Logarithmic-BRC does not support interactive insertions and thus its construction stage is not benchmarked.
			Otherwise it is the most performant of all non-\acrshort{ore} protocols.
			Note, however, that its performance depends on the result size, not data size.

			As expected, \acrshort{oram} performs worse than the \acrshort{ore}-based protocols, but its performance is in-line with the non-\acrshort{ore} protocols.
			It may seem that \acrshort{oram} does especially bad in construction communication (\cref{figure:protocols-vol:q,figure:protocols-size:q}), but it is only because POPE has a shortcut in construction.
			This ``debt'' is being payed off during queries (\cref{figure:protocols-size:q}).

			Note that the values do not vary a lot among different data distributions except for \acrshort{io} requests.
			\acrshort{io} performance depends on the result size for queries, and is therefore more sensitive to data distribution.

			Also note that using an \acrshort{ore} scheme with relatively small ciphertext in {\BPlus} tree does not add any substantial \acrshort{io} overhead (see ``No encryption'').

			On \cref{figure:protocols-query-sizes} it is clear that query performance does not depend substantially on the query size, except for Logarithmic-BRC, for which the relation is linear.
			Note that Logarithmic-BRC with optimally configured \texttt{pack} extension shows almost no growth.
			This is because for large ranges \acrshort{brc} will return the higher nodes (keywords matching many documents), which are optimally packed in \acrshort{io} pages.
			As query range doubles, higher nodes are involved increasing the chance that requested keywords have their documents packed.

			\cref{figure:protocols-data-percent-vol,figure:protocols-data-percent-ios} show \cref{table:protocols} asymptotic values.
			The simulation was run for uniform dataset of \num{247000} records (hundred percent), \num{1000} queries, \SI{0.5}{\percent} query range and \num{128} blocks cache size.
			Kerschbaum construction \acrshortpl{io} and cold POPE query values grow linearly with inputs, while the other protocols grow logarithmically, square-logarithmically, or do not grow.

			\cref{figure:cold-vs-warm} shows how the performance of protocols fluctuates as queries are processed.
			Note that POPE and Logarithmic-BRC fluctuate the most (which is, in general, undesirable), and POPE is the only protocol where cold versus warm makes a difference.

	\section{Remarks and conclusion}\label{section:range-snapshot:conclusion}

	Having done theoretical and practical evaluations of the protocols, we have found that primitive usage is a much better performance measure than the plain time measurements.
	When it comes to practical use, the observed time of a query execution is a mix of a number of factors and \acrshort{io} requests can slow the system down dramatically.

	\acrshort{ore}-based {\BPlus} tree protocol is provably \acrshort{io} optimal and can potentially be extended by using another data structure with \acrshort{ore}.
	Its security/performance trade off is tunable by choosing and parametrizing the underlying \acrshort{ore} scheme.
	Each scheme we considered has its own unique advantages and drawbacks.
	BCLO \cite{bclo-ope} is the least secure scheme in the benchmark, but is stateless and produces numerical ciphertexts, so it may be used in the databases without any modifications.
	Frequency-hiding \acrshort{ope} \cite{fh-ope} also has this property, hides the frequency of the ciphertexts, but is stateful and requires uniform input.
	Lewi-Wu \cite{lewi-wu-ore} is easily customizable in terms of tuning performance to security ratio, and it offers the security benefits of left / right framework --- particuarly useful for {\BPlus} tree.
	CLWW \cite{clww-ore} provides weaker security guarantees but is the fastest scheme in the benchmark.

	Kerschbaum protocol \cite{florian-protocol} offers semantically secure ciphertexts, hiding the location of the smallest and largest of them, and has a simple implementation.
	The protocol is well-suited for bulk insertions and scales well.

	POPE \cite{pope} offers a ``deferred'' {\BPlus} tree implementation.
	By deferring the sorting of its ciphertexts, POPE remains more secure for the small number of queries.
	POPE has the fastest insertion routine and does not reveal the order of most of its ciphertexts.
	It will be more performant for the systems where there are a lot more insertions than queries.
	We would also recommend to ``warm up'' the structure to avoid a substantial delay upon the first query.

	Logarithmic-BRC is a perfect choice for huge datasets where query result size is limited.
	It is the only protocol with substantial space overhead, but it offers scalability and perfect (in a snapshot setting) security,
	and a carefully chosen and configured \acrshort{sse} scheme ensures that \acrshort{io} grows slowly as a function of result size.

	\acrshort{oram} has shown the most interesting result.
	Its performance is not only adequate, but also in-line with the other even less secure protocols.
	With this empirical result, we expect more interest in \acrshort{oram} research, possibly discovering tighter bounds, faster constructions and efficient ways to use the schemes.
	The performance of \acrshort{oram} gives an upper bound on the acceptable performance level of less secure (access pattern revealing) protocols, as practitioners will choose \acrshort{oram} over both less secure and less performant solutions.

	We found our framework to be a powerful tool for analyzing the protocols, and we hope developers of new protocols will contribute implementations and evaluate them.

	An important future work is to understand better the meaning of the different leakage profiles and their implications.
	Furthermore, another direction is to try to improve the performance of the most secure schemes (e.g. \cite{parameter-hiding-ore}). 

	\cleardoublepage%
	\chapter{Range queries in the persistent model}\label{section:range-persistent}
\thispagestyle{myheadings}

	To protect the data in the outsourced database systems, various cryptographic techniques are used to ensure data privacy, while allowing efficient querying.
	A rich collection of attacks on such systems has emerged.
	Even with strong cryptography, just communication volume or access pattern is enough for an adversary to succeed.

	In this chapter we present a model for differentially private outsourced database system and a concrete construction, \epsolute{}, that provably conceals the aforementioned leakages, while remaining efficient and scalable.
	In our solution, differential privacy is preserved at the record level even against an untrusted server that controls data and queries.
	\epsolute{} combines \acrlong{oram} and differentially private sanitizers to create a generic and efficient construction.

	We go further and present a set of improvements to bring the solution to efficiency and practicality necessary for real-world adoption.
	We describe the way to parallelize the operations, minimize the amount of noise, and reduce the number of network requests, while preserving the privacy guarantees.
	We have run an extensive set of experiments, dozens of servers processing up to 10 million records, and compiled a detailed result analysis proving the efficiency and scalability of our solution.
	While providing strong security and privacy guarantees we are less than an order of magnitude slower than range query execution of a non-secure plain-text optimized \acrshort{rdbms} like MySQL and PostgreSQL\@.

	\emph{Some of the following sections were paraphrased or taken verbatim from the following published work.}

	\cite{epsolute} \printpublication{epsolute}

	\section{Introduction}

	Secure outsourced database systems aim at helping organizations outsource their data to untrusted third parties, without compromising data confidentiality or query efficiency.
	The main idea is to encrypt the data records before uploading them to an untrusted server along with an index data structure that governs which encrypted records to retrieve for each query.
	While strong cryptographic tools can be used for this task, existing implementations such as CryptDB \cite{crypt-db}, Cipherbase \cite{cipherbase-daas}, StealthDB \cite{stealth-db} and TrustedDB \cite{trusted-db} try to optimize performance but do not provide strong security guarantees when answering queries.
	Indeed, a series of works \cite{multidimensional-range-queries, inference-attack-islam-14, leakage-abuse-attacks-cash-15, inference-attacks-naveed-15, generic-attacks-kellaris, attacks-tao-of-inference, grubbs-attacks, access-pattern-disclosure, attacks-improved-reconstruction} demonstrate that these systems are vulnerable to a variety of reconstruction attacks.
	That is, an adversary can fully reconstruct the distribution of the records over the domain of the indexed attribute.
	This weakness is prominently due to the \emph{access pattern leakage}: the adversary can tell if the same encrypted record is returned on different queries.

	More recently, \cite{generic-attacks-kellaris, state-of-uniform, attacks-improved-reconstruction, pump-volume-attacks, volume-range-attacks} showed that reconstruction attacks are possible even if the systems employ heavyweight cryptographic techniques that hide the access patterns, such as homomorphic encryption \cite{arbitrary-functions-encrypted, fully-homomorphic-encryption} or \acrfull{oram} \cite{oram-theory, oram-original}, because they leak the size of the result set of a query to the server (this is referred to as \emph{communication volume leakage}). 
	Thus, even some recent systems that provide stronger security guarantees like ObliDB \cite{oblidb}, Opaque \cite{opaque} and Oblix \cite{oblix} are susceptible to these attacks. 
	This also means that no outsourced database system can be both optimally efficient and privacy-preserving: secure outsourced database systems should not return the exact number of records required to answer a query.

	We take the next step towards designing secure outsourced database systems by presenting novel constructions that strike a provable balance between efficiency and privacy.
	First, to combat the access pattern leakage, we integrate a layer of \acrshort{oram} storage in our construction.
	Then, we bound the communication volume leakage by utilizing the notion of \acrfull{dp} \cite{differential-privacy-original}.
	Specifically, instead of returning the exact number of records per query, we only reveal perturbed query answer sizes by adding random encrypted records to the result so that the communication volume leakage is bounded.
	Our construction guarantees privacy of any single record in the database which is necessary in datasets with stringent privacy requirements.
	In a medical \acrshort{hipaa}-compliant setting, for example, disclosing that a patient exists in a database with a rare diagnosis correlating with age may be enough to reveal a particular individual.

	The resulting mechanism achieves the required level of privacy, but implemented na\"{\i}vely the construction is prohibitively slow.
	We make the solution practical by limiting the amount of noise and the number of network roundtrips while preserving the privacy guarantees.
	We go further and present a way to parallelize the construction, which requires adapting noise-generation algorithms to maintain differential privacy requirements.

	Using our system, we have run an extensive set of experiments over cloud machines, utilizing large datasets --- that range up to 10 million records --- and queries of different sizes, and we report our experimental results on efficiency and scalability.
	We compare against best possible solutions in terms of efficiency (conventional non-secure outsourced database systems on unencrypted data) and against an approach that provides optimal security (retrieves the full table from the cloud or runs the entire query obliviously with maximal padding).
	We report that our solution is very competitive against both baselines.
	Our performance is comparable to that of unsecured plain-text optimized database systems (like MySQL and PostgreSQL): while providing strong security and privacy guarantees, we are only 4 to 8 times slower in a typical setting.
	Compared with the optimally secure solution, a linear scan (downloading all the records), we are 18 times faster in a typical setting and even faster as database sizes scale up.

	\smallskip

	To summarize, our contributions in this work are as follows:
	\begin{itemize}
		\item
			We present a new model for a \emph{differentially private} outsourced database system, \acrshort{cdpodb}, its security definition, query types, and efficiency measures.
			In our model, the adversarial honest-but-curious server cannot see the record values, access patterns, or exact communication volume.

		\item
			We describe a novel construction, \epsolute{}, that satisfies the proposed security definition, and provide detailed algorithms for both range and point query types.
			In particular, to conceal the access pattern and communication volume leakages, we provide a secure storage construction, utilizing a combination of \acrlong{oram} \cite{oram-theory, oram-original} and differentially private sanitization \cite{non-interactive-database-privacy}.
			Towards this, we maintain an index structure to know how many and which objects we need to retrieve.
			This index can be stored locally for better efficiency (in all our experiments this is the case), but crucially, it can also be outsourced to the adversarial server and retrieved on-the-fly for each query.

		\item
			We improve our generic construction to enable parallelization within a query.
			The core idea is to split the storage among multiple \acrshortpl{oram}, but this requires tailoring the overhead required for differential privacy proportionally to the number of \acrshortpl{oram}, in order to ensure privacy.
			We present practical improvements and optimization techniques that dramatically reduce the amount of fetched noise and the number of network roundtrips.

		\item
			Finally, we provide and open-source a high-quality C++ implementation of our system.
			We have run an extensive set of experiments on both synthetic and real datasets to empirically assess the efficiency of our construction and the impact of our improvements.
			We compare our solutions to the na\"{\i}ve approach (linear scan downloading all data every query), oblivious processing and maximal padding solution (Shrinkwrap \cite{shrinkwrap}), and to a non-secure regular \acrshort{rdbms} (PostgreSQL and MySQL), and we show that our system is very competitive.
	\end{itemize}

	\section{Differentially private outsourced database systems}\label{section:range-persistent:dpodb}

	In this section we present our model, \emph{differentially private outsourced database system}, \acrshort{cdpodb}, its security definition, query types and efficiency measures.
	It is an extension of the outsourced database model in \cref{section:introduction:model:odb}.

	\subsection{Adversarial model}\label{section:range-persistent:dpodb:adversarial-models}

		We consider an honest-but-curious polynomial time adversary that attempts to breach differential privacy with respect to the input database \database{}.
		We observe later in \cref{section:range-persistent:dpodb:adversarial-models:adaptive} that it is impossible to completely hide the number of records returned on each query without essentially returning all the database records on each query.
		This, in turn, means that different query sequences may be distinguished, and, furthermore, that differential privacy may not be preserved if the query sequence depends on the content of the database records.
		We hence, only require the protection of differential privacy with respect to every fixed query sequence.
		Furthermore, we relax to computational differential privacy (following \cite{computational-dp}).

		In the following definition, the notation \view{\protocol{}}{\database, \fromNtoM{\query}{1}{m}} denotes the view of the server \server{} in the execution of protocol \protocol{} in answering queries \fromNtoM{\query}{1}{m} with the underlying database \database{}.

		\begin{definition}
			We say that an outsourced database system \protocol{} is $(\epsilon, \delta)$\hyp{}computationally differentially private (a.k.a.~\acrshort{cdpodb}) if for every polynomial time distinguishing adversary \adversary{}, for every neighboring databases $\database \sim \database^\prime$, and for every query sequence $\fromNtoM{\query}{1}{m} \in \querySet^m$ where $m = \mathsf{poly}(\lambda)$,

			\begin{multline*}
				\probability{\adversary \left( 1^\lambda, \view{\protocol{}}{\database, \fromNtoM{\query}{1}{m}} \right) = 1 } \leq \\
				\exp{(\epsilon)} \cdot \probability{\adversary \left( 1^\lambda, \view{\protocol{}}{\database^\prime, \fromNtoM{\query}{1}{m}} \right) = 1} + \delta +\negl \; ,
			\end{multline*}
			where the probability is over the randomness of the distinguishing adversary \adversary{} and the protocol \protocol{}.
		\end{definition}

	\begin{remark}[Informal]
		We note that security and differential privacy in this model imply protection against communication volume and access pattern leakages and thus prevent a range of attacks, such as \cite{leakage-abuse-attacks-cash-15,inference-attacks-naveed-15,generic-attacks-kellaris}. 
	\end{remark}

	\subsubsection{On impossibility of adaptive queries}\label{section:range-persistent:dpodb:adversarial-models:adaptive}

		Non-adaptivity in our \acrshort{cdpodb} definition does not reflect a deficiency of our specific protocol but rather an inherent source of leakage when the queries may depend on the decrypted data.
		Consider an adaptive \acrshort{cdpodb} definition that does not fix the query sequence \fromNtoM{q}{1}{m} in advance but instead an arbitrary (efficient) user \user{} chooses them during the protocol execution with \server{}.
		As before, we ask that the \server{}'s view is \acrshort{dp} on neighboring databases for every such \user{}.
		We observe that this definition cannot possibly be satisfied by \emph{any} outsourced database system without unacceptable efficiency overhead.
		Note that non-adaptivity here does not imply that the client knows all the queries in advance, but rather can choose them at any time (e.g., depending on external circumstances) as long as they do not depend on true answers to prior queries.

		To see this, consider two neighboring databases $\database, \database^\prime$.
		Database \database{} has 1 record with $\mathsf{key} = 0$ and $\database^\prime$ has none.
		Furthermore, both have 50 records with $\mathsf{key} = 50$ and 100 records with $\mathsf{key} = 100$.
		User \user{} queries first for the records with $\mathsf{key} = 0$, and then if there is a record with $\mathsf{key} = 0$ it queries for the records with $\mathsf{key} = 50$, otherwise for the records with $\mathsf{key} = 100$.
		Clearly, an efficient outsourced database system cannot return nearly as many records when $\mathsf{key} = 50$ versus $\mathsf{key} = 100$ here.
		Hence, this allows distinguishing $\database, \database^\prime$ with probability almost 1.

		To give a concrete scenario, suppose neighboring medical databases differ in one record with a rare diagnosis ``Alzheimer's disease''.
		A medical professional queries the database for that diagnosis first (point query), and if there is a record, she queries the senior patients next (range query, \texttt{age $\ge$ 65}), otherwise she queries the general population (resulting in more records).
		We leave it open to meaningfully strengthen our definition while avoiding such impossibility results, and we defer the formal proof to future work.


\newlength{\setupLength}
\setlength{\setupLength}{16.8em}
\newlength{\queryLength}
\setlength{\queryLength}{16em}

\begin{algorithm*}[ht!]

	\begin{pcvstack}

		\procedure[linenumbering]{\protocolSetup{}}{
													\textbf{User \user}																\>																\> \textbf{Server \server}	\\
			\label{algorithm:dp-oram:setup:line-2}	\pcinput{\database}																\>																\> \pcinput{\emptyset}		\\
			\label{algorithm:dp-oram:setup:line-3}	\indexI \gets \algo{CreateIndex}{\database}										\>																\>							\\
			\label{algorithm:dp-oram:setup:line-4}	\oramProgram = \left. (\oramWrite, \recordID_i, \record_i) \right|_{i = 1}^n	\>																\>							\\
			\label{algorithm:dp-oram:setup:line-5}																					\> \sendmessageboth*[\setupLength]{\algo{ORAM}{\oramProgram}}	\>							\\
			\label{algorithm:dp-oram:setup:line-6}	\serverDS \gets \algo{A}{\fromNtoM{\searchKey}{1}{\domainSize}}					\> \sendmessageright*[\setupLength]{\serverDS}					\>							\\
			\label{algorithm:dp-oram:setup:line-7}	\pcouput{\indexI}																\>																\> \pcouput{\serverDS}
		}

		\vspace{0.5em}

		\procedure[linenumbering]{\protocolQuery{}}{
													\textbf{User \user}																				\>																											\> \textbf{Server \server}				\\
			\label{algorithm:dp-oram:query:line-2}	\pcinput{\query, \indexI}																		\>																											\> \pcinput{\serverDS}					\\
			\label{algorithm:dp-oram:query:line-3}	T \gets \algo{Lookup}{\indexI, \query}															\> \sendmessageright*[\queryLength]{\query}																	\> c \gets \algo{B}{\serverDS, \query}	\\
			\label{algorithm:dp-oram:query:line-4}	\oramProgram_\mathsf{true} = \left. (\oramRead, \recordID_i, \bot) \right|_{i \in T}			\> \sendmessageleft*[\queryLength]{c}																		\>										\\
			\label{algorithm:dp-oram:query:line-5}	\oramProgram_\mathsf{noise} = \left. (\oramRead, S \setminus T, \bot) \right|_{1}^{c - \abs{T}}	\>																											\>										\\
			\label{algorithm:dp-oram:query:line-6}	R																								\> \sendmessageboth*[\queryLength]{\algo{ORAM}{\oramProgram_\mathsf{true} \| \oramProgram_\mathsf{noise}}}	\>										\\
			\label{algorithm:dp-oram:query:line-7}	\pcouput{R}																						\>																											\>	\pcouput{\emptyset}
		}

	\end{pcvstack}

	\caption[\epsolute{} protocol]{
		\epsolute{} protocol.
		$\algo{ORAM}{\cdot}$ denotes an execution of \acrshort{oram} protocol (\cref{section:background:oram}), where \user{} plays the role of the client.
		\acrshort{oram} protocol client and server states are implicit.
		$S \setminus T$ represents a set of valid record \acrshortpl{id} $S$ that are not in the true result set $T$.
	}%
	\label{algorithm:dp-oram}
\end{algorithm*}

	\subsection{Query types}

		In this work we are concerned with the following query types:
		\begin{description}[style=unboxed, leftmargin=0em]
			\item[Range queries]
				Here we assume a total ordering on \searchKeyDomain{}.
				A query \query{a}{b} is associated with an interval $\interval{a}{b}$ for $1 \leq a \leq b \leq \domainSize$ such that $\query{a}{b}(c) = 1$ iff $c \in \interval{a}{b}$ for all $c \in \searchKeyDomain$.
				The equivalent \acrshort{sql} query is:

				\smallskip
				\indent\texttt{SELECT * FROM table WHERE attribute BETWEEN a AND b;}
				\smallskip

			\item[Point queries]
				Here \searchKeyDomain{} is arbitrary and a query predicate \query{a} is associated with an element $a \in \searchKeyDomain$ such that $\query{a}(b) = 1$ iff $a = b$.
				In an ordered domain, point queries are degenerate range queries.
				The equivalent \acrshort{sql} query is:

				\smallskip
				\indent\texttt{SELECT * FROM table WHERE attribute = a;}

		\end{description}

	\subsection{Measuring Efficiency}

		We define two basic efficiency measures for a \acrshort{cdpodb}.
		\begin{description}[style=unboxed, leftmargin=0em]
			\item[Storage efficiency]
				is defined as the sum of the bit-lengths of the records in a database relative to the bit-length of a corresponding encrypted database.
				Specifically, we say that an outsourced database system has \emph{storage efficiency} of $(\efficiencyCoefficient, \efficiencyOffset)$ if the following holds.
				Fix any \databaseDef{} and let $n_1 = \sum_{i=1}^n \abs{r_i}$.
				Let $\server_\mathsf{state}$ be an output of \server{} on a run of \protocolSetup{} where \user{} has input \database{}, and let $n_2 = \abs{ \server_\mathsf{state} }$.
				Then $n_2 \leq \efficiencyCoefficient n_1 + \efficiencyOffset$.

			\item[Communication efficiency]
				is defined as the sum of the lengths of the records in bits whose search keys satisfy the query relative to the actual number of bits sent back as the result of a query.
				Specifically, we say that an outsourced database system has \emph{communication efficiency} of $(\efficiencyCoefficient, \efficiencyOffset)$ if the following holds.
				Fix any \query{} and \serverDS{} output by \protocolSetup{}, let \user{} and \server{} execute \protocolQuery{} where \user{} has inputs \query{}, and output $R$, and \server{} has input \serverDS{}.
				Let $m_1$ be the amount of data in bits transferred between \user{} and \server{} during the execution of \protocolQuery{}, and let $m_2 = \abs{ R }$.
				Then $m_2 \leq \efficiencyCoefficient m_1 + \efficiencyOffset$.
		\end{description}

		Note that $\efficiencyCoefficient \geq 1$ and $\efficiencyOffset \geq 0$ for both measures.
		We say that an outsourced database system is \emph{optimally storage efficient} (resp., \emph{optimally communication efficient}) if it has storage (resp., communication) efficiency of $(1, 0)$.

	\section{\texorpdfstring{\epsolute{}}{Epsolute}}\label{section:range-persistent:dp-oram}

	In this section we present a construction, \epsolute{}, that satisfies the security definition in \cref{section:range-persistent:dpodb}, detailing algorithms for both range and point query types.
	We also provide efficiency guarantees for approximate and pure \acrshort{dp} versions of \epsolute{}.

	\subsection{General construction}

		Let \querySet{} be a collection of queries.
		We are interested in building a differentially private outsourced database system for \querySet{}, called \epsolute{}.
		Our solution will use these building blocks.
		\begin{itemize}
			\item
				A $(\eta_1, \eta_2)$-\acrshort{oram} protocol \algo{ORAM}{\cdot}.
			\item
				An $(\epsilon, \delta, \alpha, \beta)$-differentially private sanitizer $(\algo{A}, \algo{B})$ for \querySet{} and negligible $\beta$, which satisfies the non-negative noise guarantee from \cref{remark:dp-sanitizer-guarantees}.
			\item
				A pair of algorithms \algo{CreateIndex} and \algo{Lookup}.
				\algo{CreateIndex} consumes \database{} and produces an index data structure \indexI{} that maps a search key \searchKey{} to a list of record \acrshortpl{id} \recordID{} corresponding to the given search key.
				\algo{Lookup} consumes \indexI{} and \query{} and returns a list $T = \fromNtoM{\recordID}{1}{{\abs{T}}}$ of record \acrshortpl{id} matching the supplied query.
		\end{itemize}

\begin{figure}[!ht]
	\centering
	\includegraphics[width=\linewidth]{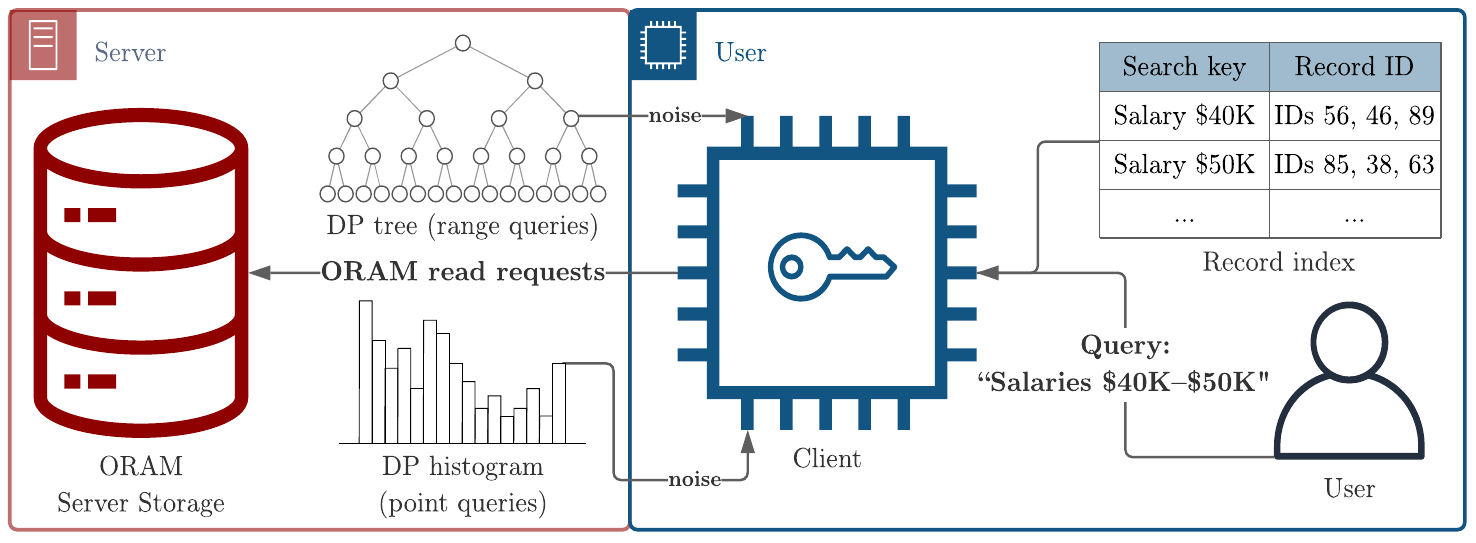}
	\caption{\epsolute{} construction}%
	\label{figure:dp-oram}
\end{figure}

		Our protocol $\protocol = (\protocolSetup, \protocolQuery)$ of \epsolute{} works as shown in \cref{algorithm:dp-oram}.
		Hereafter, we reference lines in \cref{algorithm:dp-oram}.
		See \cref{figure:dp-oram} for a schematic description of the protocol.

		\paragraph*{Setup protocol \; \texorpdfstring{\protocolSetup{}}{}}

			Let \user{}'s input be a database \databaseDef{} (\cref{algorithm:dp-oram:setup:line-2}).
			\user{} creates an index \indexI{} mapping search keys to record \acrshortpl{id} corresponding to these keys (\cref{algorithm:dp-oram:setup:line-3}).
			\user{} sends over the records to \server{} by executing the \acrshort{oram} protocol on the specified sequence (\crefrange{algorithm:dp-oram:setup:line-4}{algorithm:dp-oram:setup:line-5}).
			\user{} generates a \acrshort{dp} structure \serverDS{} over the search keys using sanitizer \algo{A}, and sends \serverDS{} over to \server{} (\cref{algorithm:dp-oram:setup:line-6}).
			The output of \user{} is \indexI{} and of \server{} is \serverDS{}; final \acrshort{oram} states of \server{} and \user{} are implicit, including encryption key \queryKey{} (\cref{algorithm:dp-oram:setup:line-7}).

		\paragraph*{Query protocol \; \texorpdfstring{\protocolQuery{}}{}}

			\user{} starts with a query \query{} and index \indexI{}, \server{} starts with a \acrshort{dp} structure \serverDS{}.
			One can think of these inputs as outputs of \protocolSetup{} (\cref{algorithm:dp-oram:query:line-2}).
			\user{} immediately sends the query to \server{}, which uses the sanitizer \algo{B} to compute the total number of requests $c$, while \user{} uses index \indexI{} to derive the true indices of the records the query \query{} targets (\cref{algorithm:dp-oram:query:line-3}).
			\user{} receives $c$ from \server{} and prepares two \acrshort{oram} sequences: $\oramProgram_\mathsf{true}$ for real records retrieval, and $\oramProgram_\mathsf{noise}$ to pad the number of requests to $c$ to perturb the communication volume.
			$\oramProgram_\mathsf{noise}$ includes valid non-repeating record \acrshortpl{id} that are not part of the true result set $T$ (\crefrange{algorithm:dp-oram:query:line-4}{algorithm:dp-oram:query:line-5}).
			\user{} fetches the records, both real and fake, from \server{} using the \acrshort{oram} protocol (\cref{algorithm:dp-oram:query:line-6}).
			The output of \user{} is the filtered set of records requested by the query $\query{}$; final \acrshort{oram} states of \server{} and \user{} are implicit (\cref{algorithm:dp-oram:query:line-7}).

		The protocols for point and range queries only differ in sanitizer implementations, see \cref{section:range-persistent:dp-oram:point,section:range-persistent:dp-oram:range}.
		Note above that in any execution of \protocolQuery{} we have $c \geq \query(\database)$ with overwhelming probability $1 - \beta$ (by using sanitizers satisfying \cref{remark:dp-sanitizer-guarantees}), and thus the protocol is well-defined and its accuracy is $1 - \beta$.
		Also note that the \acrshort{dp} parameter $\delta$ is lower-bounded by $\beta$ because sampling negative noise, however improbable, violates privacy, and therefore the final construction is $(\epsilon, \beta)$-\acrshort{dp}.

	\subsection{Security}

		\begin{theorem}
			\epsolute{} is $(\beta \cdot m)$-wrong and $(\epsilon, \delta)$-\acrshort{cdpodb} where the negligible term is $\negl = 2 \cdot \eta_2$.
		\end{theorem}

		\begin{proof}
			We consider a sequence of views
			\[
				\view{1} \to \view{2} \to \view{3} \to \view{4} \; .
			\]
			\view{1} is \view{\protocol}{\database, \fromNtoM{\query}{1}{m}}.
			\view{2} is produced only from $\serverDS \gets \algo{A}(\fromNtoM{\searchKey}{1}{\domainSize})$. 
			Namely, compute $c_i \gets \algo{A}{\serverDS, \query_i}$ for all $i$ and run \acrshort{oram} simulator on $\sum_i c_i$.
			By \acrshort{oram} security,
			\[
				\probability{\adversary(\view{1})} - \probability{\adversary(\view{2})} \leq \eta_2 \; .
			\]
			\view{3} is produced similarly but $\serverDS \gets \algo{A}{\fromNtoM{\searchKey^\prime}{1}{\domainSize}}$ instead.
			Note that the $c_i$ are simply post-processing on \serverDS{} via \algo{B} so
			\[
				\probability{\adversary(\view{2})} = \exp(\epsilon) \cdot \probability{\adversary(\view{3})} + \delta \; .
			\]
			$\view{4} = \view{\protocol}{\database^\prime, \fromNtoM{\query}{1}{m}}$.
			It follows by \acrshort{oram} security
			\[
				\probability{\adversary(\view{3})} - \probability{\adversary(\view{4})} \leq \eta_2 \; .
			\]
			Putting this all together completes the proof.
		\end{proof}

	\subsection{Efficiency}

		For an \acrshort{oram} with communication efficiency $(a_1, a_2)$ and an $(\alpha, \beta)$-differentially private sanitizer, the \epsolute{} communication efficiency is $(a_1, a_2 \cdot \alpha)$.
		The efficiency metrics demonstrate how the total storage or communication volume (the number of stored or transferred bits) changes additively and multiplicatively as the functions of data size \dataSize{} and domain \domainSize{}.
		We therefore have the following corollaries for the efficiency of the system in the cases of approximate and pure differential privacy.
		\begin{corollary}\label{corollary:comm-efficiency-approximate-dp}
			\epsolute{} is an outsourced database system with storage efficiency \efficiency{1}{0}.
			Depending on the query type, assume it offers the following communication efficiency.
			\begin{description}
				\item[Range queries] $\efficiency{\log \dataSize}{2^{\log^* \domainSize} \log \dataSize}$
				\item[Point queries] $\efficiency{\log \dataSize}{\log \dataSize}$
			\end{description}
			Then, there is a negligible $\delta$ such that \epsolute{} satisfies $(\epsilon, \delta)$\hyp{}differential privacy for some $\epsilon$.\footnote{
				Note that the existence of $\epsilon$ in this setting implies that the probability of an adversary breaking the \acrshort{dp} guarantees is bounded by it.
			}
		\end{corollary}

		\begin{proof}
			By using \acrshort{oram}, we store only the original data once and hence, we get optimal storage efficiency.

			The communication efficiency depends on the upper bound of the error for each sanitizer when $\delta > 0$, as described in \cref{section:background:dp-sanitizers} and \cref{remark:dp-sanitizer-guarantees}.
			The most efficient \acrshort{oram} protocol to date has $\bigO{\log \dataSize}$ communication overhead (see \cref{section:background:oram}).
		\end{proof}

		\begin{corollary}\label{corollary:comm-efficiency-pure-dp}
			\epsolute{} is an outsourced database system with storage efficiency \efficiency{1}{0}.
			Depending on the query type, assume it offers the following communication efficiency.
			\begin{description}
				\item[Range queries] $\efficiency{\log \dataSize}{\log \domainSize \log \dataSize}$
				\item[Point queries] $\efficiency{\log \dataSize}{\log \domainSize \log \dataSize}$
			\end{description}
			Then, \epsolute{} satisfies $\epsilon$-differential privacy for some $\epsilon$.
		\end{corollary}

		\begin{proof}
			Similarly, we derive the proof by considering the use of \acrshort{oram} and the upper bound of the error for each sanitizer when $\delta = 0$ in \cref{section:background:dp-sanitizers}.
		\end{proof}

	\subsection{Extending to multiple attributes}\label{section:range-persistent:dp-oram:multiple-attributes}

		We will now describe how \epsolute{} supports multiple indexed attributes and what the privacy and performance implications are.
		The na\"{\i}ve way is to simply duplicate the entire stack of states of \user{} and \server{}, and during the query use the states whose attribute the query targets.
		However, \epsolute{} design allows to keep the most expensive part of the state --- the \acrshort{oram} state --- shared for all attributes and both types of queries.
		Specifically, the index \indexI{} and \acrshort{dp} structure \serverDS{} are generated per attribute and query type, while \user{} and \server{} \acrshort{oram} states are generated once.
		This design is practical since \serverDS{} is tiny and index \indexI{} is relatively small compared to \acrshort{oram} states, see \cref{section:range-persistent:experiments}.

		We note that in case the indices grow large in number, it is practical to outsource them to the adversarial server using \acrshort{oram} and download only the ones needed for each query.
		In terms of privacy, the solution is equivalent to operating different \epsolute{} instances because \acrshort{oram} hides the values of records and access patterns entirely.
		Due to \cref{theorem:composition} for non-disjoint datasets, the total privacy budget of the multi-attribute system will be the sum of individual budgets for each attribute / index.

		Next, we choose two \acrshort{dp} sanitizers for our system, for point and for range queries, and calculate the $\alpha$ values to make them output positive values with high probability, consistent with \cref{remark:dp-sanitizer-guarantees}.

	\subsection{\texorpdfstring{\epsolute{}}{Epsolute} for point queries}\label{section:range-persistent:dp-oram:point}

		For point queries, we use the \acrshort{lpa} method as the sanitizer to ensure pure differential privacy.
		Specifically, for every histogram bin, we draw noise from the Laplace distribution with mean $\alpha_p$ and scale $\lambda = \nicefrac{1}{\epsilon}$.
		To satisfy \cref{remark:dp-sanitizer-guarantees}, we have to set $\alpha_p$ such that if values are drawn from $\algo{Laplace}{ \alpha_p, \nicefrac{1}{\epsilon} }$ at least as many times as the number of bins \domainSize{}, they are all positive with high probability $1 - \beta$, for negligible $\beta$.

		We can compute the exact minimum required value of $\alpha_p$ in order to ensure drawing positive values with high probability by using the \acrshort{cdf} of the Laplace distribution.
		Specifically, $\alpha_p$ should be equal to the minimum value that satisfies the following inequality.

		\[
			\left( 1 - \frac{1}{2} e^{- \alpha_p \cdot \epsilon} \right)^\domainSize \leq 1 - \beta
		\]
		which is equivalent to
		\[
			\alpha_p = \ceil{ -\frac{ \ln \left( 2 - 2 \sqrt[\domainSize]{1 - \beta} \right) }{ \epsilon } }
		\]


\setlength{\setupLength}{15.8em}
\setlength{\queryLength}{15em}

\newcommand{\SetupGamma}{
	\procedure[linenumbering]{\protocolSetup{} of \protocolGamma{}}{
																\textbf{User \user}																							\>																\> \textbf{Server \server}	\\
		\label{algorithm:dp-oram-parallel:gamma:setup:line-2}	\pcinput{\database{}}																						\>																\> \pcinput{\emptyset}		\\
		\label{algorithm:dp-oram-parallel:gamma:setup:line-3}	\indexI \gets \algo{CreateIndex}{\database, \oramsNumber}													\>																\>							\pclb
		\pcintertext[dotted]{$\pcfor j \in \set{1, \ldots, \oramsNumber} \pcdo$ \; \text{(in parallel)}}
		\label{algorithm:dp-oram-parallel:gamma:setup:line-4}	\left\langle \overline{\record}, \overline{\recordID} \right\rangle\ \text{s.t.}\ \algo{H}{\recordID} = j	\>																\>							\\
		\label{algorithm:dp-oram-parallel:gamma:setup:line-5}	\oramProgram = \left\langle (\oramWrite, \overline{\recordID}, \overline{\record}) \right\rangle			\> \sendmessageboth*[\setupLength]{\algo{ORAM}_j(\oramProgram)}	\>							\pclb
		\pcintertext[dotted]{$\pcendfor$}
		\label{algorithm:dp-oram-parallel:gamma:setup:line-6}	\serverDS \gets \algo{A}{\fromNtoM{\searchKey}{1}{\domainSize}}												\> \sendmessageright*[\setupLength]{\serverDS}					\>							\\
		\label{algorithm:dp-oram-parallel:gamma:setup:line-7}	\pcouput{\indexI}																							\>																\> \pcouput{ \serverDS }
	}
}

\newcommand{\QueryGamma}{
	\procedure[linenumbering]{\protocolQuery{} of \protocolGamma{}}{
																\textbf{User \user}																					\>																												\> \textbf{Server \server}									\\
		\label{algorithm:dp-oram-parallel:gamma:query:line-2}	\pcinput{\query, \indexI}																			\>																												\> \pcinput{ \serverDS }									\\
		\label{algorithm:dp-oram-parallel:gamma:query:line-3}	\fromNtoM{T}{1}{\oramsNumber} \gets \algo{Lookup}{I, \query}										\> \sendmessageright*[\queryLength]{\query}																		\> k \gets \algo{B}{\serverDS, \query}						\\
		\label{algorithm:dp-oram-parallel:gamma:query:line-4}																										\> \sendmessageleft*[\queryLength]{c}																			\> c \gets (1 + \gamma) \frac{\tilde{k}_0}{\oramsNumber}	\pclb
		\pcintertext[dotted]{$\pcfor j \in \set{1, \ldots, \oramsNumber} \pcdo$ \; \text{(in parallel)}}
		\label{algorithm:dp-oram-parallel:gamma:query:line-5}	\oramProgram_\mathsf{true} = \left. (\oramRead, \recordID_i, \bot) \right|_{i \in T_j}				\>																												\>															\\
		\label{algorithm:dp-oram-parallel:gamma:query:line-6}	\oramProgram_\mathsf{noise} = \left. (\oramRead, S \setminus T_j, \bot) \right|_{1}^{c - \abs{T_j}}	\> \sendmessageboth*[\queryLength]{\algo{ORAM}_j(\oramProgram_\mathsf{true} \| \oramProgram_\mathsf{noise})}	\> R_j														\pclb
		\pcintertext[dotted]{$\pcendfor$}
		\label{algorithm:dp-oram-parallel:gamma:query:line-7}	\pcouput{ \left. R_j \right|_{j = 1}^\oramsNumber }													\>																												\>	\pcouput{\emptyset}
	}
}

\begin{algorithm*}[ht!]

	\begin{pcvstack}

		\begin{pcvstack}

			\SetupGamma{}

			\vspace{0.5em}

			\QueryGamma{}

		\end{pcvstack}

	\end{pcvstack}

	\caption[Parallel \epsolute{} for \protocolGamma{}]{
		Parallel \epsolute{} for \protocolGamma{}, extends \cref{algorithm:dp-oram}.
		\algo{H} is a random hash function $\algo{H} : \bin^* \to \set{1, \ldots, \oramsNumber}$.
		$\gamma$ and $\tilde{k}_0$ are computed as in \cref{section:range-persistent:prallel-dp-oram:gamma}.
	}%
	\label{algorithm:dp-oram-parallel}
\end{algorithm*}

	\subsection{\texorpdfstring{\epsolute{}}{Epsolute} for range queries}\label{section:range-persistent:dp-oram:range}

		For range queries, we implement the aggregate tree method as the sanitizer.
		Specifically, we build a complete \fanout{}-ary tree on the domain, for a given \fanout{}.
		A leaf node holds the number of records falling into each bin plus some noise.
		A parent node holds sum of the leaf values in the range covered by this node, plus noise.
		Every time a query is issued, we find the minimum number of nodes that cover the range, and determine the required number of returned records by summing these node values.
		Then, we ask the server to retrieve the records in the range, plus to retrieve multiple random records so that the total number of retrieved records matches the required number of returned records.

		The noise per node is drawn from the Laplace distribution with mean $\alpha_h$ and scale $\lambda = \frac{\log_{\fanout} \domainSize}{\epsilon}$.
		Consistent with \cref{remark:dp-sanitizer-guarantees}, we determine the mean value $\alpha_h$ in order to avoid drawing negative values with high probability.
		We have to set $\alpha_h$ such that if values are drawn from $\algo{Laplace}{ \alpha_h, \frac{\log_{\fanout} \domainSize}{\epsilon} }$ at least as many times as the number of nodes in the tree, they are all positive with high probability $1 - \beta$, for negligible $\beta$.

		Again, we can compute the exact minimum required value of $\alpha_h$ in order to ensure drawing positive values with high probability by using the \acrshort{cdf} of the Laplace distribution.
		Specifically, $\alpha_h$ should be equal to the minimum value that satisfies the following inequality.
		\[
			\left( 1 - \frac{1}{2} e^{- \frac{\alpha_h \cdot \epsilon}{\log_{\fanout} \domainSize}} \right)^\mathsf{nodes} \leq 1 - \beta
		\]
		which is equivalent to
		\begin{equation}\label{equation:min-mu-for-range}
			\alpha_h = \ceil{ -\frac{ \ln{ (2 - 2 \sqrt[\mathsf{nodes}]{ 1 - \beta } ) } \cdot \log_{\fanout} \domainSize }{\epsilon} }
		\end{equation}
		where $\mathsf{nodes} = \frac{ \fanout^{ \ceil{ \log_{\fanout} (\fanout - 1) + \log_{\fanout} \domainSize - 1 }} - 1}{ \fanout - 1 } + \domainSize$ is the total number of tree nodes.

	\section{An efficient Parallel \texorpdfstring{\epsolute{}}{Epsolute}}\label{section:range-persistent:prallel-dp-oram}

	While the previously described scheme is a secure and correct \acrshort{cdpodb}, a single-threaded implementation may be prohibitively slow in practice.
	To bring the performance closer to real-world requirements, we need to be able to scale the algorithm horizontally.
	In this section, we describe an upgrade of \epsolute{} --- a scalable parallel solution.

	We suggest two variants of parallel \epsolute{} protocol.
	Both of them work by operating \oramsNumber{} \acrshortpl{oram} and randomly assigning to each of them $\nicefrac{n}{\oramsNumber}$ database records.
	For each query, we utilize the index \indexI{} to find the required records from the corresponding \acrshortpl{oram}.
	For each \acrshort{oram}, we execute a separate thread to retrieve the records.
	The threads work in parallel and there is no need for locking, since each \acrshort{oram} works independently from the rest.
	We present two methods that differ in the way they build and store \acrshort{dp} structure \serverDS{}, and hence the number of \acrshort{oram} requests they make.

	\subsection{\texorpdfstring{No-$\gamma$-method}{No-gamma-method}: \acrshort{dp} structure per \acrshort{oram}}

		In \protocolNoGamma{}, for each \acrshort{oram} / subset of the dataset, we build a \acrshort{dp} index the same way as described in \cref{section:range-persistent:dp-oram}.
		We note that \cref{theorem:composition} for disjoint datasets applies to this construction: the privacy budget $\epsilon$ for the construction is the largest (least private) among the $\epsilon$'s of the \acrshort{dp} indices for each \acrshort{oram} / subset of the dataset.

		The communication efficiency changes because
		\begin{enumerate*}[label={(\roman*)}]
			\item
				we essentially add \oramsNumber{} record subsets in order to answer a query, each having at most $\alpha$ extra random records, and
			\item
				each \acrshort{oram} holds fewer records than before, resulting in a tree of height $\log \frac{\dataSize}{\oramsNumber}$.
		\end{enumerate*}

		However, we cannot expect that the records required for each query are equally distributed among the different \acrshortpl{oram} in order to reduce the multiplicative communication cost from $\log \dataSize$ to $\frac{\log \dataSize}{\oramsNumber}$.
		Instead, we need to bound the worst case scenario which is represented by the maximum number of records from any \acrshort{oram} that is required to answer a query.
		This can be computed as follows.

		Let $X_j$ be $1$ if a record for answering query \query{} is in a specific $\algo{ORAM}_j$, and $0$ otherwise.
		Due to the random assignment of records to \acrshortpl{oram}, $\probability{X_j = 1} = \nicefrac{1}{\oramsNumber}$.
		Assume that we need $k_0$ records in order to answer query \query{}.
		The maximum number of records from $\algo{ORAM}_j$ in order to answer \query{} is bounded as follows.

		\begin{equation}\label{equation:gamma}
			\probability{ \sum_{i=1}^{k_0} X_i > ( 1 + \gamma ) \frac{k_0}{\oramsNumber} } \leq \exp{ \left( - \frac{ k_0 \gamma^2 }{ 3 \oramsNumber } \right) }
		\end{equation}

		Finally, we need to determine the value of $\gamma$ such that $\exp{ \left( - \frac{ k_0 \gamma^2 }{ 3 \oramsNumber } \right) }$ is smaller than the value $\beta$.
		Thus, $\gamma = \sqrt{ \frac{-3 \oramsNumber \log \beta}{ k_0 } }$.
		The communication efficiency for each query type is described in the following corollary.

		\begin{corollary}\label{corollary:no-gamma}
			Let \protocolNoGamma{} be an outsourced database system with storage efficiency \efficiency{1}{0}.
			Depending on the query type, \protocolNoGamma{} offers the following communication efficiency.
			\begin{description}
				\item[Range queries] $\efficiency{\left( 1 + \sqrt{ \frac{- 3 \oramsNumber \log \beta}{k_0} } \right) \log \frac{\dataSize}{\oramsNumber} }{ \frac{ \log^{1.5} \domainSize}{ \epsilon } \oramsNumber \log \dataSize }$
				\item[Point queries] $\efficiency{\left( 1 + \sqrt{ \frac{- 3 \oramsNumber \log \beta}{k_0} } \right) \log \frac{\dataSize}{\oramsNumber} }{ \frac{ \log \domainSize}{ \epsilon } \oramsNumber \log \dataSize }$
			\end{description}

			Then, \protocolNoGamma{} satisfies $\epsilon$-differential privacy for some $\epsilon$.
		\end{corollary}

		In our experiments, we set \oramsNumber{} as a constant depending on the infrastructure.
		However, if \oramsNumber{} is set as $\bigO{\log n}$, the total communication overhead of the construction will still exceed the lower-bound presented in \cite{multi-server-orams}.

	\subsection{\texorpdfstring{$\gamma$-method}{Gamma-method}: shared \acrshort{dp} structure}\label{section:range-persistent:prallel-dp-oram:gamma}

		In \protocolGamma{}, we maintain a single shared \acrshort{dp} structure \serverDS{}.
		When a query is issued, we must ensure that the number of records retrieved from every \acrshort{oram} is the same.
		As such, depending on the required noisy number of records $\tilde{k}_0$, we need to retrieve at most $( 1 + \gamma ) \frac{ \tilde{k}_0 }{\oramsNumber}$ records from each \acrshort{oram}, see \cref{equation:gamma}, for $\gamma = \sqrt{ \frac{-3 \oramsNumber \log \beta}{ \tilde{k}_0 } }$.
		Setting $\tilde{k}_0 = k_0 + \frac{\log^{1.5} \domainSize}{\epsilon}$ for range queries and $\tilde{k}_0 = k_0 + \frac{\log \domainSize}{\epsilon}$ for point queries, the communication efficiency is as follows.

		\begin{corollary}\label{corollary:gamma}
			Let \protocolGamma{} be an outsourced database system with storage efficiency \efficiency{1}{0}.
			Depending on the query type, \protocolGamma{} offers the following communication efficiency.
			\begin{description}
				\item[Range queries] $\efficiency{ \left( 1 + \sqrt{ \frac{-3 \oramsNumber \log \beta}{k_0 + \frac{ \log^{1.5} \domainSize }{ \epsilon }} }\right) \log \frac{\dataSize}{\oramsNumber} \left( 1 + \frac{\log^{1.5} \domainSize}{\epsilon} \right) }{0}$
				\item[Point queries] $\efficiency{ \left( 1 + \sqrt{ \frac{-3 \oramsNumber \log \beta}{k_0 + \frac{ \log \domainSize }{ \epsilon }} }\right) \log \frac{\dataSize}{\oramsNumber} \left( 1 + \frac{\log \domainSize}{\epsilon} \right) }{0}$
			\end{description}
			Then, \protocolGamma{} satisfies $\epsilon$-differential privacy for some $\epsilon$.
		\end{corollary}

		\protocolGamma{} is depicted in \cref{algorithm:dp-oram-parallel}.
		There are a few extensions to the subroutines and notation from \cref{algorithm:dp-oram}.
		\algo{CreateIndex} and \algo{Lookup} now build and query the index which maps a search key to a pair --- the record \acrshort{id} and the \acrshort{oram} \acrshort{id} (1 to \oramsNumber{}) which stores the record.
		\Crefrange{algorithm:dp-oram-parallel:gamma:setup:line-4}{algorithm:dp-oram-parallel:gamma:setup:line-5} of \cref{algorithm:dp-oram-parallel} \protocolSetup{} repeat for each \acrshort{oram} and operate on the records partitioned for the given \acrshort{oram} using hash function \algo{H} on the record \acrshort{id}.
		A shared \acrshort{dp} structure is created with the sanitizer \algo{A} (\cref{algorithm:dp-oram-parallel:gamma:setup:line-6}).
		In \cref{algorithm:dp-oram-parallel} \protocolQuery{}, the total number of \acrshort{oram} requests is computed once (\cref{algorithm:dp-oram-parallel:gamma:query:line-4}).
		\Crefrange{algorithm:dp-oram-parallel:gamma:query:line-5}{algorithm:dp-oram-parallel:gamma:query:line-6} repeat for each \acrshort{oram} and operate on the subset of records stored in the given \acrshort{oram}.
		Note that \user{} and \server{} implicitly maintain \oramsNumber{} \acrshort{oram} states, and the algorithm uses the $(\algo{A}, \algo{B})$ sanitizer defined in \cref{section:range-persistent:dp-oram}.

		Note that we guarantee privacy and access pattern protection on a record level.
		Each \acrshort{oram} gets accessed at least once (much more than once for a typical query) thus the existence of a particular result record in a particular \acrshort{oram} is hidden.

	\subsection{Practical improvements}\label{section:range-persistent:dp-improvements}

		Here we describe the optimizations aimed at bringing the construction's performance to the real-world demands.

		\subsubsection{\texorpdfstring{\acrshort{oram}}{ORAM} request batching}\label{section:range-persistent:dp-improvements:oram-batching}

			We have noticed that although the entire set of \acrshort{oram} requests for each query is known in advance, the requests are still executed sequentially.
			To address this inefficiency, we have designed a way to combine the requests in a batch and reduce the number of network requests to the bare minimum.
			We have implemented this method over PathORAM, which we use for the $(\eta_1, \eta_2)$-\acrshort{oram} protocol, but the idea applies to most tree-based \acrshortpl{oram} (similar to \cite{parallel-oram-improved}).

			Our optimization utilizes the fact that all PathORAM leaf \acrshortpl{id} are known in advance and paths in a tree-based storage share the buckets close to the root.
			The core idea is to read all paths first, processes the requests and and then write all paths back.
			This way the client makes a single \texttt{read} request, which is executed much faster than many small requests.
			Requests are then processed in main memory, including re-encryptions.
			Finally, the client executes the \texttt{write} requests using remapped leaves as a single operation, saving again compared to sequential execution.

			This optimization provides up to \textbf{8 times} performance boost in our experiments.
			We note that the gains in speed and \acrshort{io} overhead are achieved at the expense of main memory, which is not an issue given that the memory is released after a batch, and our experiments confirm that.
			The security guarantees of PathORAM are maintained with this optimization, since the security proof in \cite[Section 3.6]{path-oram} still holds. 
			Randomized encryption, statistically independent remapping of leaves, and stash processing do not change.

		\subsubsection{Lightweight \texorpdfstring{\acrshort{oram}}{ORAM} servers}\label{section:range-persistent:dp-improvements:three-tier}

			We have found in our experiments that na\"{\i}ve increase of the number of \acrshort{cpu} cores and gigabytes of memory does not translate into linear performance improvement after some threshold.
			Investigating the observation we have found that the \epsolute{} protocol, executing parallel \acrshort{oram} protocols, is highly intensive with respect to main memory access, cryptographic operations and network usage.
			The bottleneck is the hardware --- we have confirmed that on a single machine the memory and network are saturated quickly preventing the linear scaling.

			\begin{figure}[ht!]
	\centering
	\includegraphics[width=\linewidth]{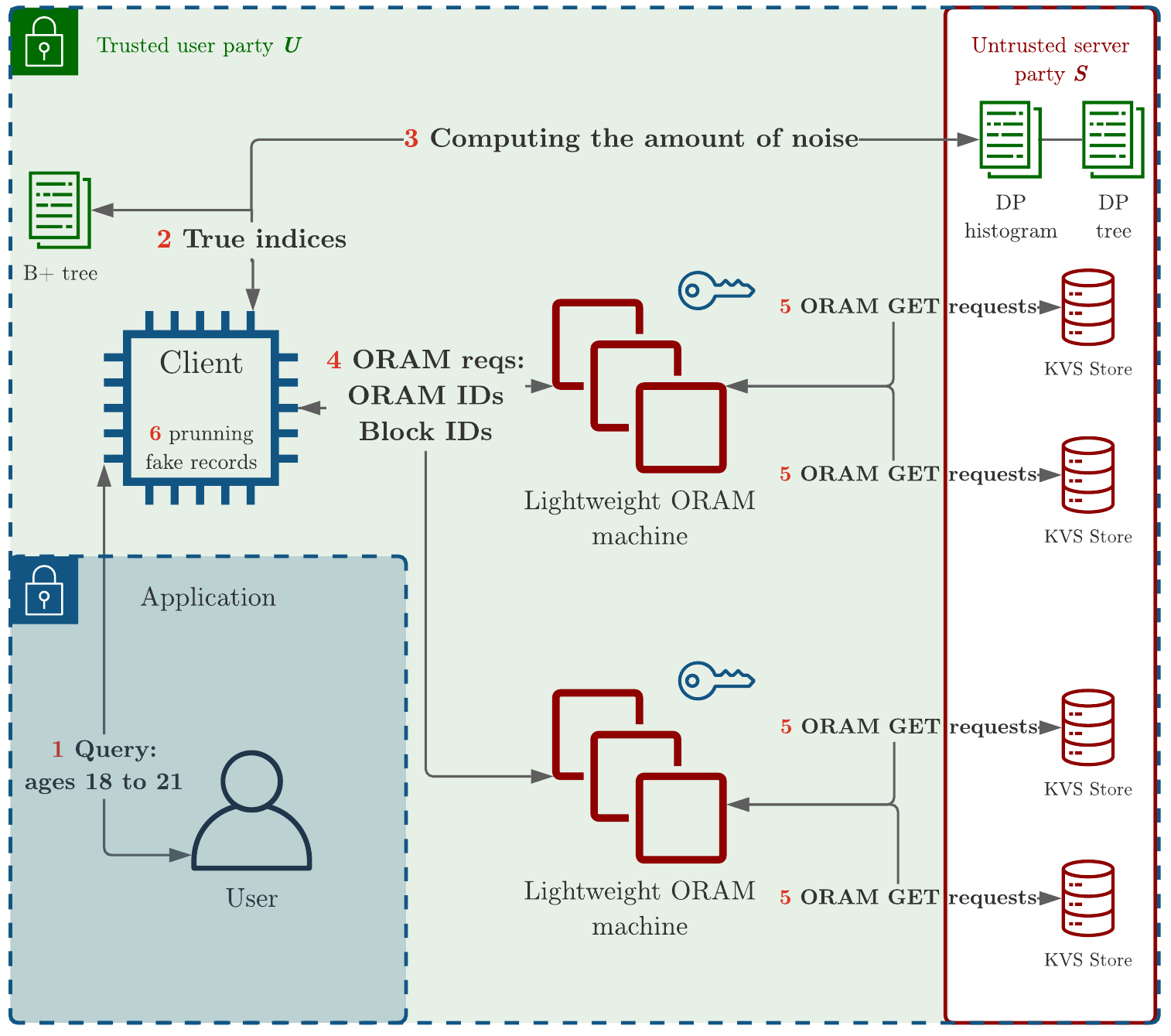}
	\caption[Parallel \epsolute{} construction]{
		Parallel \epsolute{} construction.
		A \emph{user} sends a query to \user{} modeled as the \emph{client} machine, which uses local \emph{data index} and \emph{\acrshort{dp} structures} to prepare a set of \acrshort{oram} requests, which are sent to respective \emph{\acrshort{oram} machines}.
		These machines execute the \acrshort{oram} protocol against the \emph{untrusted storage} of \server{}.
	}%
	\label{figure:three-tier}
\end{figure}

			To address the problem, we split the user party \user{} into multiple lightweight machines that are connected locally to each other and reside in a single trust domain (e.g., same data center).
			Specifically, we maintain a \emph{client machine} that receives user requests and prepares \acrshort{oram} \texttt{read} requests, and up to \oramsNumber{} lightweight \emph{\acrshort{oram} machines}, whose only job is to run the \acrshort{oram} protocols in parallel.
			See \cref{figure:three-tier} for the schematic representation of the architecture.
			We emphasize that \user{} is still a single party, therefore, the security and correctness guarantees remain valid.

			The benefit of this approach is that each of the lightweight machines has its own hardware stack.
			Communication overhead among \user{} machines is negligible compared to the one between \user{} and \server{}.
			The approach is also flexible: it is possible to use up to \oramsNumber{} \acrshort{oram} machines and the machines do not have to be identical.
			Our experiments show that when the same number of \acrshort{cpu} cores and amount of memory are consumed the efficiency gain is up to \textbf{5 times}.

	\section{Experimental Evaluation}\label{section:range-persistent:experiments}

	We have implemented our solution as a modular client-server application in C++.
	We open-sourced all components of the software set: PathORAM\footnote{\url{https://github.com/epsolute/path-oram}} and B+ tree\footnote{\url{https://github.com/epsolute/b-plus-tree}} implementations and the main query executor\footnote{\url{https://github.com/epsolute/epsolute}}.
	We provide PathORAM and B+ tree components as C++ libraries to be used in other projects; the code is documented, benchmarked and tested (228 tests covering \SI{100}{\percent} of the code).
	We have also published our datasets and query sets.\footnote{\url{http://csr.bu.edu/dp-oram/}}

	For cryptographic primitives, we used OpenSSL library (version 1.1.1i).
	For symmetric encryption in \acrshort{oram} we have used \acrshort{aes} in \acrshort{cbc} mode \cite{aes-nist,nist-modes} with a 256-bits key (i.e., $\eta_2 = 2^{-256}$), for the hash algorithm \algo{H} used to partition records among \acrshortpl{oram} we have used \acrshort{sha}-256 algorithm \cite{nist-hash}.
	Aggregate tree fanout \fanout{} is 16, proven to be optimal in \cite{hierarchical-methods-for-dp}.

	We designed our experiments to answer the following questions:
	\newlength{\questionLength}
	\settowidth{\questionLength}{Question-5}
	\begin{description}[
		font=\bfseries,
		leftmargin=\dimexpr\questionLength+1.0em\relax,
		labelindent=0pt,
		labelwidth=\questionLength%
	]
		\item[Question-1\label{item:question-practicality}] How practical is our system compared to the most efficient and most private real-world solutions?
		\item[Question-2\label{item:question-storage}] How practical is the storage overhead?
		\item[Question-3\label{item:question-parameters}] How different inputs and parameters of the system affect its performance?
		\item[Question-4\label{item:question-scalability}] How well does the system scale?
		\item[Question-5\label{item:question-optimizations}] What improvements do our optimizations provide?
		\item[Question-6\label{item:question-attributes}] What is the impact of supporting multiple attributes?
	\end{description}

	For \ref{item:question-practicality} we have run the default setting using conventional \acrshort{rdbms} (MySQL and PostgreSQL), Linear Scan approach and Shrinkwrap \cite{shrinkwrap}. 
	To target \ref{item:question-storage}, we measured the exact storage used by the client and the server for different data, record and domain sizes. 
	To answer \ref{item:question-parameters}, we ran a default setting and then varied all parameters and inputs, one at a time. 
	For \ref{item:question-scalability} we gradually added \acrshortpl{cpu}, \acrshort{oram} servers and \acrshort{kvs} instances and observed the rate of improvement in performance. 
	To target \ref{item:question-optimizations} we have run the default setting with our optimizations toggled. 
	Lastly, for \ref{item:question-attributes} we have used two datasets to construct two indices and then queried each of the attributes. 

	\subsection{Data sets}\label{section:range-persistent:experiments:data-sets}

		We used two real and one synthetic datasets --- California public pay pension database 2019 \cite{ca-employees-dataset} (referred to as ``CA employees''), Public Use Microdata Sample from US Census 2018 \cite{pums-dataset} (referred to as ``PUMS'') and synthetic uniform dataset.
		We have used salary / wages columns of the real datasets, and the numbers in the uniform set also represent salaries.
		The \texttt{NULL} and empty values were dropped.

		We created three versions of each dataset --- $10^5$, $10^6$ and $10^7$ records each.
		For uniform dataset, we simply generated the target number of entries.
		For PUMS dataset, we picked the states whose number of records most closely matches the target sizes (Louisiana for $10^5$, California for $10^6$ and the entire US for $10^7$).
		Uniform dataset was also generated for different domain sizes --- number of distinct values for the record.
		For CA employees dataset, the set contains \num{260 277} records, so we contracted it and expanded in the following way.
		For contraction we uniformly randomly sampled $10^5$ records.
		For expansion, we computed the histogram of the original dataset and sampled values uniformly within the bins.

		Each of the datasets has a number of corresponding query sets.
		Each query set has a selectivity or range size, and is sampled either uniformly or following the dataset distribution (using its \acrshort{cdf}).

	\subsection{Default setting}\label{section:range-persistent:experiments:default-setting}

		The default setting uses the \protocolGamma{} from \cref{section:range-persistent:prallel-dp-oram} and lightweight \acrshort{oram} machines from \cref{section:range-persistent:dp-improvements:three-tier,figure:three-tier}.
		We choose the \protocolGamma{} because it outperforms \protocolNoGamma{} in all experiments (see \textbf{\ref{item:question-scalability}} in \cref{section:range-persistent:experiments:results}).
		In the setting, there are 64 Redis services (8 services per one Redis server \acrshort{vm}), 8 \acrshort{oram} machines communicating with 8 Redis services each, and the client, which communicates with these 8 \acrshort{oram} machines.
		We have empirically found this configuration optimal for the compute nodes and network that we used in the experiments.
		\acrshort{oram} and Redis servers run on \acrshort{gcp} \texttt{n1-standard-16} \acrshortpl{vm} (Ubuntu 18.04), in regions \texttt{us-east4} and \texttt{us-east1} respectively.
		Client machine runs \texttt{n1-highmem-16} \acrshort{vm} in the same region as \acrshort{oram} machines.
		The ping time between the regions (i.e.\ between trusted and untrusted zones) is \SI{12}{\milli\second} and the effective bandwidth is \SI{150}{\mega\byte\per\second}.
		Ping within a region is negligible.

		Default \acrshort{dp} parameters are $\epsilon = \ln(2) \approx \num{0.693}$ and $\beta = 2^{-20}$, which are consistent with the other \acrshort{dp} applications in the literature \cite{choosing-epsilon}.
		Buckets number is set as the largest power of $\fanout = 16$ that is no greater than the domain of the dataset \domainSize{}.

		Default dataset is a uniform dataset of $10^6$ records with domain size $10^4$, and uniformly sampled queries with selectivity \SI{0.5}{\percent}.
		Default record size is \SI{4}{\kibi\byte}.

	\subsection{Experiment stages}

		Each experiment includes running 100 queries such that the overhead is measured from loading query endpoints into memory to receiving the exact and whole query response from all \acrshort{oram} machines.
		The output of an experiment is, among other things, the overhead (in milliseconds), the number of real and noisy records fetched and communication volume averaged per query.

	\subsection{\texorpdfstring{\acrshort{rdbms}}{RDBMS}, Linear Scan and Shrinkwrap}

		On top of varying the parameters, we have run similar workloads using alternative mechanisms --- extremes representing highest performance or highest privacy.
		Unless stated otherwise, the client and the server are in the trusted and untrusted regions respectively, with the network configuration as in \cref{section:range-persistent:experiments:default-setting}.

		\subsubsection*{Relational databases}

			Conventional \acrshort{rdbms} represents the most efficient and least private and secure solution in our set.
			While MySQL and PostgreSQL offer some encryption options and no differential privacy, for our experiments we turned off security features for maximal performance.
			We have run queries against MySQL and PostgreSQL varying data and record sizes.
			We used \texttt{n1-standard-32} \acrshort{gcp} \acrshortpl{vm} in \texttt{us-east1} region, running MySQL version 14.14 and PostgreSQL version 10.14.

		\subsubsection*{Linear Scan}

			Linear scan is a primitive mechanism that keeps all records encrypted on the server then downloads, decrypts and scans the entire database to answer every query.
			This method is trivially correct, private and secure, albeit not very efficient.
			There are \acrshort{rdbms} solutions, which, when configured for maximum privacy, exhibit linear scan behavior (e.g., MS-SQL Always Encrypted with Randomized Encryption\footnote{\url{https://docs.microsoft.com/sql/relational-databases/security/encryption/always-encrypted-database-engine}} and Oracle Column Transparent Data Encryption\footnote{\url{https://docs.oracle.com/database/121/ASOAG/introduction-to-transparent-data-encryption.htm}}).
			For a fair comparison we make the linear scan even more efficient by allowing it to download data via parallel threads matching the number of threads and bytes per request to that of our solution.
			Although linear scan is wasteful in the amount of data it downloads and processes, compared to our solution it has a benefit of not executing an \acrshort{oram} protocol with its logarithmic overhead and network communication in both directions.

		\subsubsection*{Shrinkwrap}

			Shrinkwrap \cite{shrinkwrap} is a construction that answers federated \acrshort{sql} queries hiding both access pattern and communication volume.
			Using the \acrshort{emp} \cite{emp-toolkit} and the code Shrinkwrap authors shared with us, we implemented a prototype that only answers range queries.
			This part of Shrinkwrap amounts to making a scan over the input marking the records satisfying the range, sorting the input, and then revealing the result set plus \acrshort{dp} noise to the client.
			For the latter part we have adapted Shrinkwrap's Truncated Laplace Mechanism \cite[Definition 4]{shrinkwrap} to hierarchical method \cite{hierarchical-methods-for-dp} in order to be able to answer an unbounded number of all possible range queries.
			We have emulated the outsourced database setting by using two \texttt{n1-standard-32} servers in different regions (\SI{12}{\milli\second} ping and \SI{150}{\mega\byte\per\second} bandwidth) executing the algorithm in a circuit model (the faster option per Shrinkwrap experiments) and then revealing the result to the trusted client.
			We note that although the complexity of a Shrinkwrap query is $\bigO{n \log n}$ due to the sorting step, its functionality is richer as it supports more relational operators, like \texttt{JOIN}, \texttt{GROUP BY} and aggregation.
			We also note that since MySQL, PostgreSQL and Shrinkwrap are not parallelized within the query, experiments using more \acrshortpl{cpu} do not yield higher performance.

	\subsection{Results and Observations}\label{section:range-persistent:experiments:results}

		After running the experiments, we have made the following observations.
		Note that we report results based on the default setting.
		\begin{itemize}[leftmargin=*]
			\item
				\epsolute{} is efficient compared to a strawman approach, \acrshort{rdbms} and Shrinkwrap: it is three orders of magnitude faster than Shrinkwrap, 18 times faster than the scan and only 4--8 times slower than a conventional database.
				In fact, for different queries, datasets, and record sizes, our system is much faster than the linear scan, as we show next.
			\item
				\epsolute{}'s client storage requirements are very practical: client size is just below \SI{30}{\mega\byte} while the size of the offloaded data is over 400 times larger.
			\item
				\epsolute{} scales predictably with the change in its parameters: data size affects performance logarithmically, record size --- linearly, and privacy budget $\epsilon$ --- exponentially.
			\item
				\epsolute{} is scalable: using \protocolGamma{} with the lightweight \acrshort{oram} machines, the increase in the number of threads translates into linear performance boost.
			\item
				The optimizations proposed in \cref{section:range-persistent:dp-improvements} provide up to an order of magnitude performance gain.
			\item
				\epsolute{} efficiently supports multiple indexed attributes.
				The overhead and the client storage increase slightly due to a lower privacy budget and extra local indices.
		\end{itemize}

		For the purposes of reproducibility we have put the log traces of all our experiments along with the instructions on how to run them on a publicly available page \href{https://epsolute.org}{epsolute.org}.
		Unless stated otherwise, the scale in the figures is linear and the $x$-axis is categorical.

		\subsubsection*{\textbf{\texorpdfstring{\ref{item:question-practicality}:}{} against \acrshort{rdbms}, Linear Scan and Shrinkwrap}}

			\begin{figure}[!ht]
	\centering
	\includegraphics[width=\linewidth]{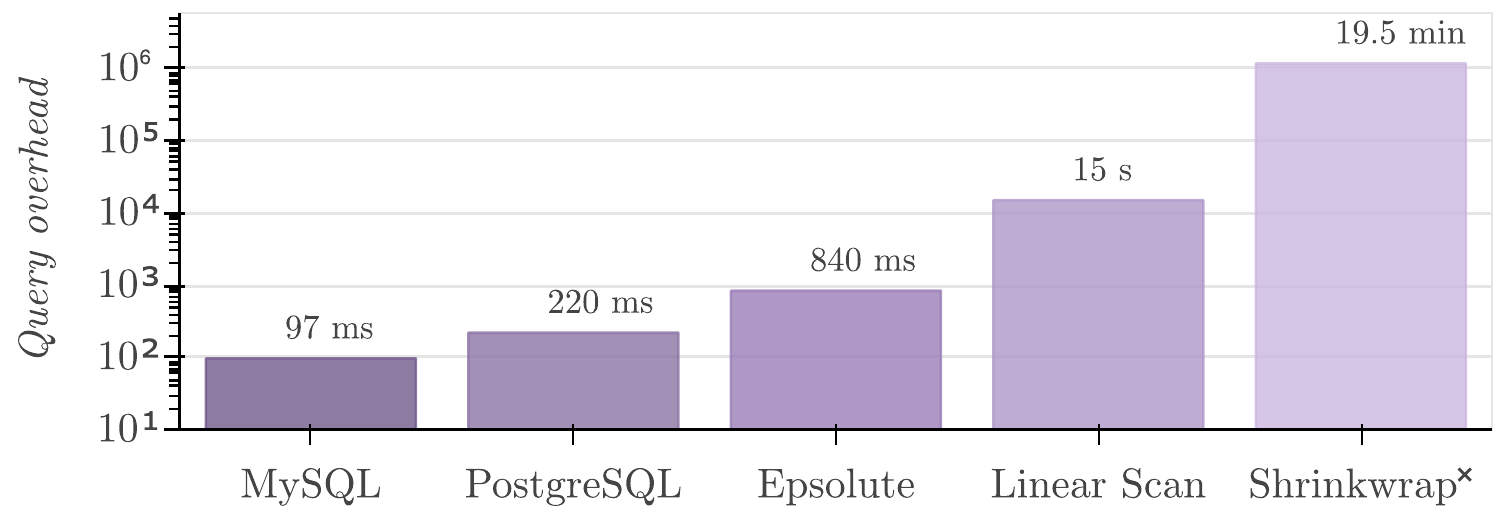}
	\caption[Different range-query mechanisms]{
		Different range-query mechanisms, logarithmic scale.
		Default setting: $10^6$ \SI{4}{\kibi\byte} uniformly-sampled records with the range $10^4$.
	}%
	\label{figure:mechanism}
\end{figure}

			The first experiment we have run using \epsolute{} is the default setting in which we observed the query overhead of \textbf{\SI[detect-all=true]{840}{\milli\second}}.
			To put this number in perspective, we compare \epsolute{} to conventional relational databases, the linear scan and Shrinkwrap.

			For the default setting, MySQL and PostgreSQL, configured for no privacy and maximum performance, complete in \SI{97}{\milli\second} and \SI{220}{\milli\second} respectively, which is just \textbf{8 to 4 times} faster than \epsolute{}, see \cref{figure:mechanism}.
			Conventional \acrshort{rdbms} uses efficient indices (B+ trees) to locate requested records and sends them over without noise and encryption, and it does so using less hardware resources.
			In our experiments \acrshort{rdbms} performance is linearly correlated with the result and record sizes.

			\begin{figure}[!ht]
	\centering
	\includegraphics[width=\linewidth]{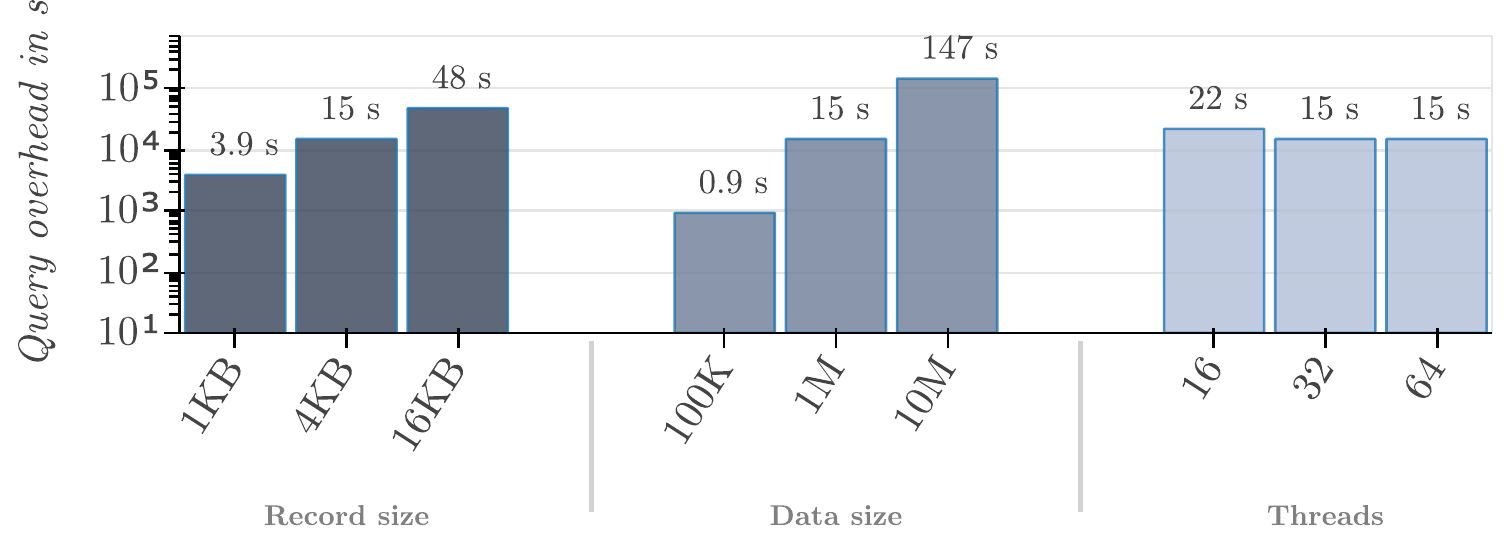}
	\caption[Linear scan performance]{
		Linear scan performance, logarithmic scale.
		The experiments are run for the default setting of $10^6$ records of size \SI{4}{\kibi\byte} and 64 threads, with one of the three parameters varying.
	}%
	\label{figure:linear-scan}
\end{figure}

			Linear scan experiments demonstrate the practicality of \epsolute{} compared to a trivial ``download everything every time'' approach, see \cref{figure:linear-scan}.
			Linear scan's overhead is $\bigO{n}$ regardless of the queries, while \epsolute{}'s overhead is $\bigO{\log{n}}$ times the result size.
			According to our experiments, \epsolute{} eclipses the linear scan at \SI{4}{\kibi\byte}, 64 threads and only \emph{ten thousand records} (both mechanisms complete in about \SI{120}{\milli\second}).
			For a default setting (at a million records), the difference is \textbf{18 times}, see \cref{figure:linear-scan}.

			Because Shrinkwrap sorts the input obliviously in a circuit model, it then incurs $\bigO{n \log n}$ comparisons, each resulting in multiple circuit gates, which is much more expensive than the linear scan.
			Unlike linear scan, however, Shrinkwrap does not require much client memory as the client merely coordinates the query.
			While Shrinkwrap supports richer set of relational operators, for range queries alone \epsolute{} is \textbf{three orders of magnitude} faster.

		\subsubsection*{\textbf{\texorpdfstring{\ref{item:question-storage}:}{} storage}}


\newcommand{\lightvbar}{\color{lightGrey}\vrule}
\newcommand{\lightcline}[1]{\arrayrulecolor{lightGrey}\cline{#1}\arrayrulecolor{black}}

\begin{table}[!ht]
	\renewcommand{\arraystretch}{1.2}
	\sisetup{detect-all = true}
	\begin{tabular*}{\linewidth}{ !{\extracolsep\fill} c | r !{\lightvbar{}} r | r !{\lightvbar{}} r | r !{\lightvbar{}} r} 
		\toprule
			\diagbox{\dataSize}{\scriptsize{Record}}			& \multicolumn{2}{c !{\lightvbar{}}}{\SI{1}{\kibi\byte}}				& \multicolumn{2}{c !{\lightvbar{}}}{\SI{4}{\kibi\byte}}				& \multicolumn{2}{c}{\SI{16}{\kibi\byte}}												\\
		\midrule
			\multirow{2}{*}{$10^5$}								& \small\SI{400}{\kibi\byte}			& \small\SI{400}{\byte}			& \small\SI{400}{\kibi\byte}			& \small\SI{102}{\kibi\byte}	& \small\SI{400}{\kibi\byte}					& \small\SI{1.6}{\mega\byte}			\\ \lightcline{2-3} \lightcline{4-5} \lightcline{6-7}
																& \small\bfseries\SI{396}{\mega\byte}	& \small\SI{4.6}{\mega\byte}	& \small\bfseries\SI{1.5}{\giga\byte}	& \small\SI{14}{\mega\byte}		& \small\bfseries\SI{6.2}{\giga\byte}			& \small\SI{51}{\mega\byte}				\\

		\midrule
			\multirow{2}{*}{$10^6$}								& \small\SI{3.9}{\mega\byte}			& \small\SI{400}{\byte}			& \small\SI{3.9}{\mega\byte}			& \small\SI{102}{\kibi\byte}	& \small\SI{3.9}{\mega\byte}					& \small\SI{1.6}{\mega\byte}			\\ \lightcline{2-3} \lightcline{4-5} \lightcline{6-7}
																& \small\bfseries\SI{3.2}{\giga\byte}	& \small\SI{15}{\mega\byte}		& \small\bfseries\SI{12}{\giga\byte}	& \small\SI{25}{\mega\byte}		& \small\bfseries\SI{48}{\giga\byte}			& \small\SI{62}{\mega\byte}				\\

		\midrule
			\multirow{2}{*}{$10^7$}								& \small\SI{40}{\mega\byte}				& \small\SI{400}{\byte}			& \small\SI{40}{\mega\byte}				& \small\SI{102}{\kibi\byte}	& \small\itshape\SI{40}{\mega\byte}				& \small\itshape\SI{1.6}{\mega\byte}	\\ \lightcline{2-3} \lightcline{4-5} \lightcline{6-7}
																& \small\bfseries\SI{24}{\giga\byte}	& \small\SI{99}{\mega\byte}		& \small\bfseries\SI{96}{\giga\byte}	& \small\SI{109}{\mega\byte}	& \small\itshape\bfseries\SI{384}{\giga\byte}	& \small\itshape\SI{146}{\mega\byte}	\\

		\midrule
			\diagbox[dir=SW, width=6em]{\dataSize}{\domainSize}	& \multicolumn{2}{c !{\lightvbar{}}}{$100$}								& \multicolumn{2}{c !{\lightvbar{}}}{$10^4$}							& \multicolumn{2}{c}{$10^6$}															\\
		\toprule
	\end{tabular*}
	\sisetup{detect-none = true}
	\caption[\epsolute{} storage usage for varying data, record and domain sizes]{
		\epsolute{} storage usage for varying data, record and domain sizes.
		The values are as follows.
		Left top: index \indexI{} (B+ tree), right top: aggregate tree \serverDS{}, right bottom: \acrshort{oram} \user{} state and left bottom (bold): \acrshort{oram} \server{} state.
		\textit{Italic} indicates that the value is estimated.
	}%
	\label{table:storage}
\end{table}

			While \epsolute{} storage efficiency is near-optimal \efficiency{1}{0}, it is important to observe the absolute values.
			Index \indexI{} is implemented as a B+ tree with fanout 200 and occupancy \SI{70}{\percent}, and its size, therefore, is roughly $5.7 \dataSize$ bytes.
			Most of the \acrshort{oram} client storage is the PathORAM stash with its size chosen in a way to bound failure probability to about $\eta_1 = 2^{-32}$ (see \cite[Theorem 1]{path-oram}). 
			In \cref{table:storage}, we present \epsolute{} storage usage for the parameters that affect it --- data, record and domain sizes.
			We measured the sizes of the index \indexI{}, \acrshort{dp} structure \serverDS{}, and \acrshort{oram} client and server states.
			Our observations are:
			\begin{enumerate*}[label={(\roman*)}]
				\item index size expectedly grows only with the data size,
				\item \serverDS{} is negligibly small in practice,
				\item small \indexI{} and \serverDS{} sizes imply the efficiency of supporting multiple indexed attributes,
				\item \server{} to \user{} storage size ratio varies from \textbf{85} in the smallest setting to more than \textbf{\num[detect-all=true]{2000}} in the largest, and
				\item one can trade client storage for \acrshort{oram} failure probability.
			\end{enumerate*}
			We conclude that the storage requirements of \epsolute{} are practical.



		\subsubsection*{\textbf{\texorpdfstring{\ref{item:question-parameters}:}{} varying parameters}}

			\begin{figure}[!ht]
	\centering
	\begin{minipage}{0.48\textwidth}
		\centering
		\includegraphics[width=\linewidth]{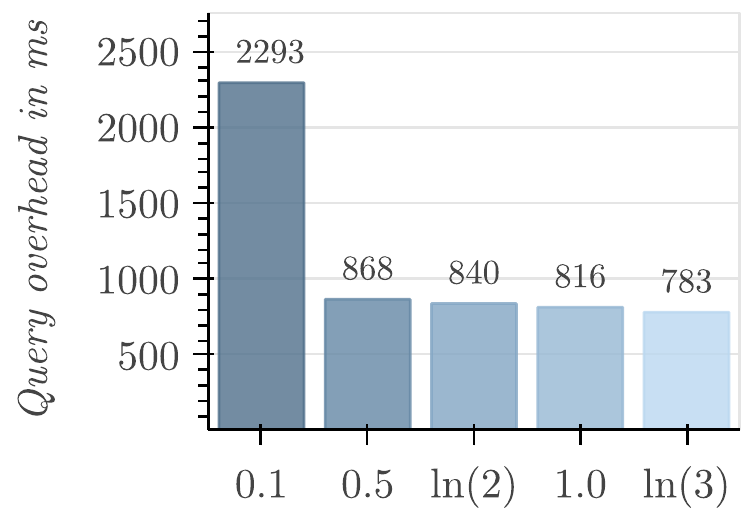}
		\captionof{figure}{Privacy budget $\epsilon$}%
		\label{figure:epsilon}
	\end{minipage}
	~ 
	\begin{minipage}{0.48\textwidth}
		\centering
		\includegraphics[width=\linewidth]{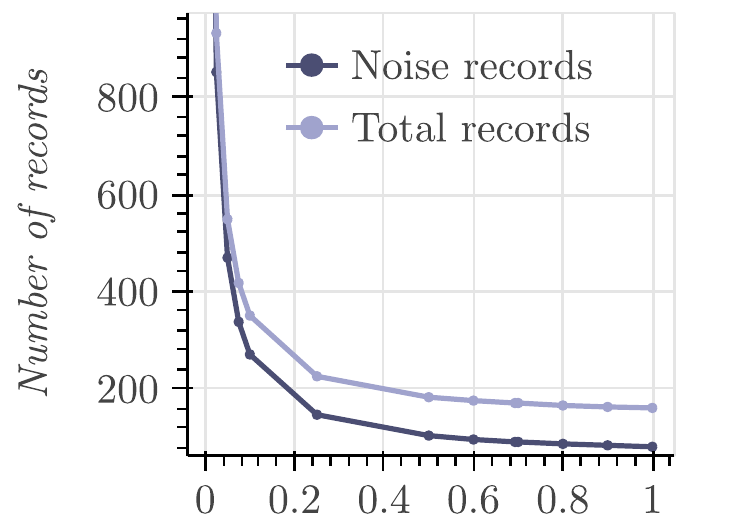}
		\captionof{figure}{Effect of $\epsilon$}%
		\label{figure:epsilon-effect}
	\end{minipage}
\end{figure}

			To measure and understand the impact of configuration parameters on the performance of our solution we have varied $\epsilon$, record size, data size \dataSize{}, domain size \domainSize{}, selectivities, as well as data and query distributions.
			The relation that is persistent throughout the experiments is that for given data and record sizes, the performance (the time to completely execute a query) is strictly proportional to the total number of records, fake and real, that are being accessed per query.
			Each record access goes through the \acrshort{oram} protocol, which, in turn, downloads, re-encrypts and uploads $\bigO{\log{\dataSize}}$ blocks.
			These accesses contribute the most to the overhead and all other stages (e.g., traversing index or aggregate tree) are negligible.

			\paragraph*{Privacy budget \texorpdfstring{$\epsilon$}{epsilon} and its effect}

				We have run the default setting for $\epsilon = \{ 0.1, \allowbreak 0.5, \allowbreak \ln{2}, \allowbreak 1.0, \allowbreak \ln{3} \}$.
				$\epsilon$ strictly contributes to the amount of noise, which grows exponentially as $\epsilon$ decreases, see \cref{figure:epsilon}, observe sharp drop.
				As visualized on \cref{figure:epsilon-effect}, at high $\epsilon$ values the noise contributes a fraction of total overhead, while at low values the noise dominates the overhead entirely.

			\begin{figure}[!ht]
	\centering
	\includegraphics[width=\linewidth]{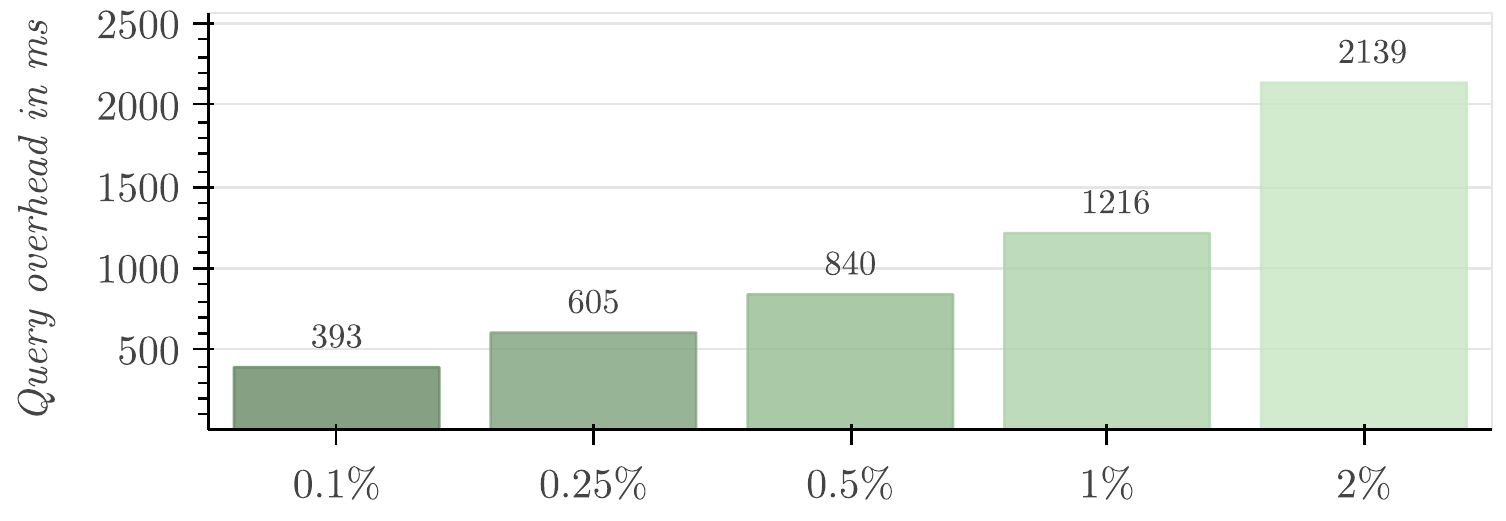}
	\caption{Selectivity}%
	\label{figure:selectivity}
\end{figure}

			\paragraph*{Selectivity}

				We have ranged the selectivity from \SI{0.1}{\percent} to \SI{2}{\percent} of the total number of records, see \cref{figure:selectivity}.
				Overhead expectedly grows with the result size.
				For smaller queries, and thus for lower overhead, the relation is positive, but not strictly proportional.
				This phenomena, observed for the experiments with low resulting per-query time, is explained by the variance among parallel threads.
				During each query the work is parallelized over \oramsNumber{} \acrshortpl{oram} and the query is completed when the \emph{last} thread finishes.
				The problem, in distributed systems known as ``the curse of the last reducer'' \cite{curse-of-last-reducer}, is when one thread takes disproportionally long to finish.
				In our case, we run 64 threads in default setting, and the delay is usually caused by a variety of factors --- blocking \acrshort{io}, network delay or something else running on a shared virtual \acrshort{cpu}.
				This effect is noticeable when a single thread does relatively little work and small disruptions actually matter; the effect is negligible for large queries.

			\begin{figure}[!ht]
	\centering
	\begin{minipage}{0.31\columnwidth}
		\centering
		\includegraphics[width=\linewidth]{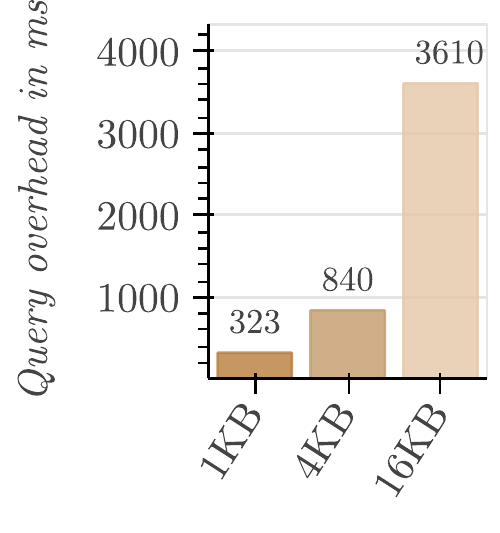}
		\captionof{figure}{Record size}%
		\label{figure:record-size}
	\end{minipage}
	~ 
	\begin{minipage}{0.31\columnwidth}
		\centering
		\includegraphics[width=\linewidth]{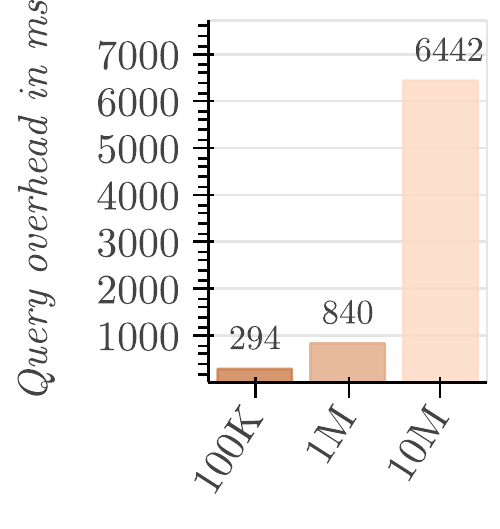}
		\captionof{figure}{Data size}%
		\label{figure:data-size}
	\end{minipage}
	~ 
	\begin{minipage}{0.31\columnwidth}
		\centering
		\includegraphics[width=\linewidth]{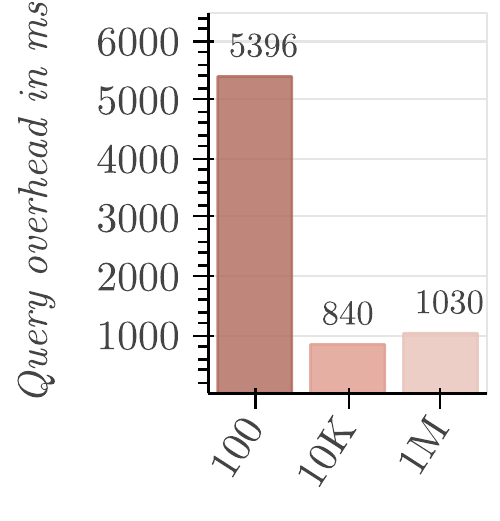}
		\captionof{figure}{Domain size}%
		\label{figure:domain-size}
	\end{minipage}
\end{figure}

			\paragraph*{Record, data and domain sizes}

				We have tried \SI{1}{\kibi\byte}, \SI{4}{\kibi\byte} and \SI{16}{\kibi\byte} records, see \cref{figure:record-size}.
				Trivially, the elapsed time is directly proportional to the record size.

				We set \dataSize{} to $10^5$, $10^6$ and $10^7$, see \cref{figure:data-size}.
				The observed correlation of overhead against the data size is positive but non-linear, 10 times increment in \dataSize{} results in less than 10 times increase in time.
				This is explained by the \acrshort{oram} overhead --- when \dataSize{} changes, the \acrshort{oram} storage gets bigger and its overhead is logarithmic.

				For synthetic datasets we have set \domainSize{} to $100$, $10^4$ and $10^6$, see \cref{figure:domain-size}.
				The results for domain size correlation are more interesting: low and high values deliver worse performance than the middle value.
				Small domain for a large data set means that a query often results in a high number of real records, which implies significant latency regardless of noise parameters.
				A sparse dataset, on the other hand, means that for a given selectivity wider domain is covered per query, resulting in more nodes in the aggregate tree contributing to the total noise value.

			\begin{figure}[!ht]
	\centering
	\begin{minipage}{0.48\columnwidth}
		\centering
		\includegraphics[width=\linewidth]{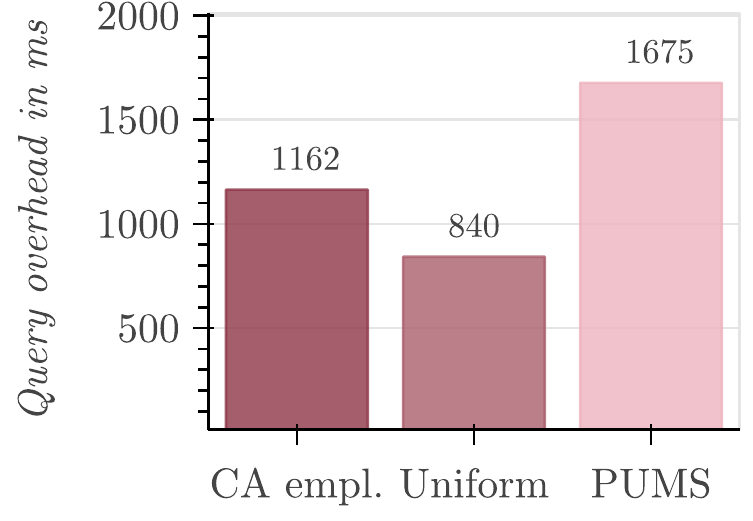}
		\captionof{figure}{Data distribution}%
		\label{figure:data-distribution}
	\end{minipage}
	~ 
	\begin{minipage}{0.48\columnwidth}
		\centering
		\includegraphics[width=\linewidth]{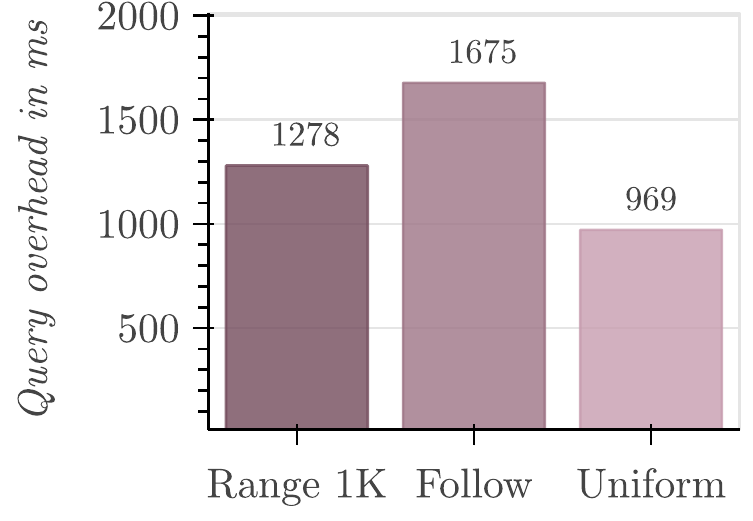}
		\captionof{figure}{Query distribution}%
		\label{figure:query-distribution}
	\end{minipage}
\end{figure}

			\paragraph*{Data and query distributions}

				Our solution performs best on the uniform data and uniform ranges, see \cref{figure:data-distribution,figure:query-distribution}.
				Once a skew of any kind is introduced, there appear sparse and dense regions that contribute more overhead than uniform regions.
				Sparse regions span over wider range for a given selectivity, which results in more noise.
				Dense regions are likely to include more records for a given range size, which again results in more fetched records.
				Both real datasets are heavily skewed towards smaller values as few people have ultra-high salaries.

		\subsubsection*{\textbf{\texorpdfstring{\ref{item:question-scalability}:}{} scalability}}

			\begin{figure}[!ht]
	\centering
	\includegraphics[width=\linewidth]{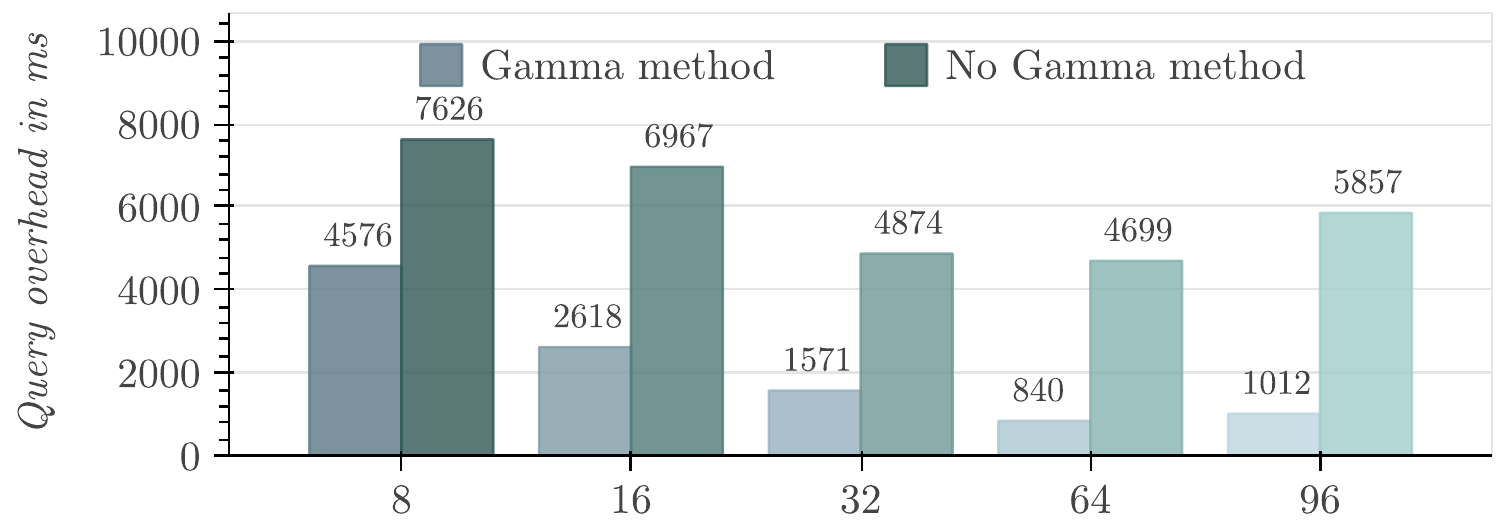}
	\caption{Scalability measurements for \protocolGamma{} and \protocolNoGamma{}}%
	\label{figure:scalability}
\end{figure}

			Horizontal scaling is a necessity for a practical system, this is the motivation for the parallelization in the first place.
			Ideally, performance should improve proportionally to the parallelization factor, number of \acrshortpl{oram} in our case, \oramsNumber{}.

			For scalability experiments we run the default setting for both \protocolNoGamma{} and \protocolGamma{} (\emph{no-$\gamma$-method} and \emph{$\gamma$-method} respectively) varying the number of \acrshortpl{oram} \oramsNumber{}, from 8 to 96 (maximum virtual \acrshortpl{cpu} on a \acrshort{gcp} \acrshort{vm}).
			The results are visualized on \cref{figure:scalability}.
			We report two positive observations:
			\begin{enumerate*}[label={(\roman*)}]
				\item the $\gamma$-method provides substantially better performance and storage efficiency, and
				\item when using this method the system scales linearly with the number of \acrshortpl{oram}.
			\end{enumerate*}
			($\oramsNumber = 96$ is a special case because some \acrshortpl{oram} had to share a single \acrshort{kvs}.)

		\subsubsection*{\textbf{\texorpdfstring{\ref{item:question-optimizations}:}{} optimizations benefits}}

			\begin{table}[!ht]
	\begin{tabular*}{\linewidth}{ !{\extracolsep\fill} l c c >{\bfseries}c } 
		\toprule
			Improvement (section)													& Enabled					& Disabled					& Boost			\\
		\midrule
			\acrshort{oram} batching (\ref{section:range-persistent:dp-improvements:oram-batching})				& \SI{840}{\milli\second}	& \SI{6978}{\milli\second}	& 8.3x			\\
			Lightweight \acrshort{oram} machines (\ref{section:range-persistent:dp-improvements:three-tier})	& \SI{840}{\milli\second}	& \SI{4484}{\milli\second}	& 5.3x			\\
			Both improvements														& \SI{840}{\milli\second}	& \SI{8417}{\milli\second}	& \emph{10.0x}	\\
		\bottomrule
	\end{tabular*}
	\caption{Improvements over parallel \epsolute{}}%
	\label{table:optimizations}
\end{table}

			\cref{table:optimizations} demonstrates the boosts our improvements provide; when combined, the speedup is up to an order of magnitude.

			\acrshort{oram} request batching (\cref{section:range-persistent:dp-improvements:oram-batching}) makes the biggest difference.
			We have run the default setting with and without the batching.
			The overhead is substantially smaller because far fewer \acrshort{io} requests are being made, which implies benefits across the full stack: download, re-encryption and upload.

			Using lightweight \acrshort{oram} machines (\cref{section:range-persistent:dp-improvements:three-tier}) makes a difference when scaling.
			In the default setting, 64 parallel threads quickly saturate the memory access and network channel, while spreading computation among nodes removes the bottleneck.

		\subsubsection*{\textbf{\texorpdfstring{\ref{item:question-attributes}:}{} multiple attributes}}

			\begin{figure}[!ht]
	\centering
	\includegraphics[width=\columnwidth]{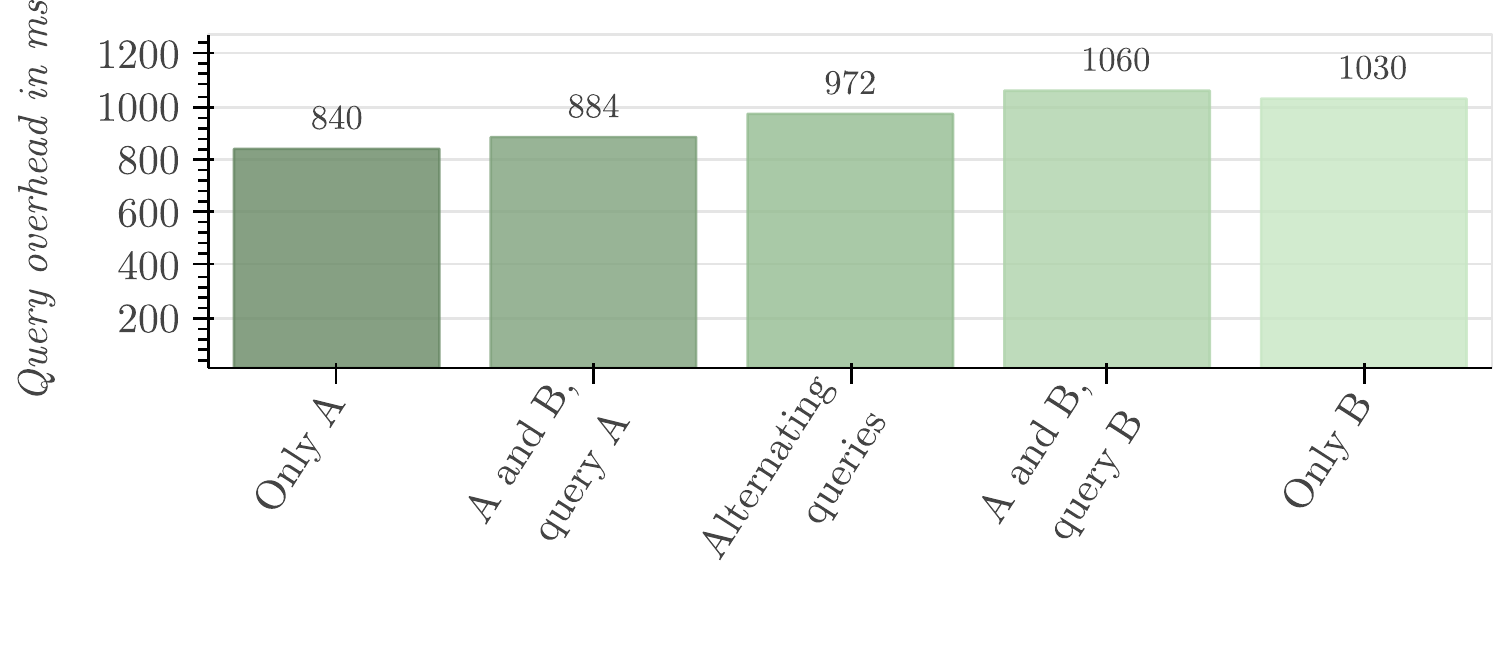}
	\caption[Query overhead when using multiple attributes]{
		Query overhead when using multiple attributes.
		\emph{Only A} and \emph{Only B} index one attribute.
		\emph{A and B} indexes both attributes and then queries one of them.
		\emph{Alternating} indexes both attributes and runs half of the queries against \emph{A} and another half against \emph{B}.
	}%
	\label{figure:attributes}
\end{figure}

			\epsolute{} supports multiple indexed attributes.
			In \cref{section:range-persistent:dp-oram:multiple-attributes} we described that the performance implications amount to having an index \indexI{} and a \acrshort{dp} structure \serverDS{} per attribute and sharing the privacy budget $\epsilon$ among all attributes.
			As shown in \cref{table:storage}, \indexI{} and \serverDS{} are the smallest components of the client storage.
			To observe the query performance impact, we have used the default dataset with domains $10^4$ and $10^6$ as indexed attributes \emph{A} and \emph{B} respectively.
			We ran queries against only \emph{A}, only \emph{B} and against both attributes in alternating fashion.
			Each of the attributes used $\epsilon = \frac{\ln{2}}{2}$ to match the default privacy budget of $\ln(2)$.

			\cref{figure:attributes} demonstrates the query overhead of supporting multiple attributes.
			The principal observation is that the overhead increases only slightly due to a lower privacy budget.
			The client storage went up by just \SI{9}{\mega\byte}, and still constitutes only \SI{3.3}{\percent} of the server storage, which is not affected by the number of indexed attributes.


	\section{Conclusion and Future Work}

	In this work, we present a system called \epsolute{} that can be used to store and retrieve encrypted records in the cloud while providing strong and provable security guarantees, and that exhibits excellent query performance for range and point queries.
	We use an optimized \acrlong{oram} protocol that has been parallelized together with very efficient \acrshort{dp}-sanitizers that hide both the access patterns and the exact communication volume sizes and can withstand advanced attacks that have been recently developed.
	We provide a prototype of the system and present an extensive evaluation over very large and diverse datasets and workloads that show excellent performance for the given security guarantees.

	In our future work, we plan to investigate methods to extend our approaches to use a \acrfull{tee}, like \acrshort{sgx}, in order to improve the performance even further.
	We will also explore a multi-user setting without the need for a shared stateful client, and enabling dynamic workloads with insertions and updates.
	We will also consider how adaptive and non-adaptive security models would change in the case of dynamic environments.
	Lastly, we plan to explore other relational operations like \texttt{JOIN} and \texttt{GROUP BY}.

	\cleardoublepage%
	\chapter{\texorpdfstring{\acrshort{knn}}{kNN} queries in the snapshot model}\label{section:knn-snapshot}
\thispagestyle{myheadings}

	In this chapter we describe and analyze \kanon{}, a system that executes secure \acrshort{knn} queries in the snapshot adversary model.
	We study the effect of protecting the records with a type of property-preserving encryption on quality of search and efficiency of certain attacks.
	Specifically, we examine theoretically and practically how accuracy of both \acrshort{knn} search and \acrshort{ml}-based inversion attack degrade with added security.

	\section{Introduction}

		Nearest-neighbor search is a type of optimization problem that, given a set of objects and a distance metric, requires finding the object closest to a given point according to the distance metric.
		A \acrfull{knn} search is a subtype of a general nearest-neighbor problem where $k$ closest objects are requested.
		Applications that use \acrshort{knn} search only need to define the objects and the metric.
		For example, a street map application would define the 2D coordinates of the buildings as objects and Euclidean distance as a metric, then the query could be ``give 5 restaurants closest to the current user position''.
		A document search application would define the keyword vector for a document as an object and an inner product distance as a metric, then the query could be ``give 3 documents most similar to the given text'' (similar applications may search images, videos and sounds).

		In this chapter we propose a method and an analysis of running secure \acrshort{knn} queries in an outsourced database model.
		We model our application as a generic document similarity search, where the server stores the embeddings of the documents and returns the $k$ closest records to the query embedding.
		We propose to apply a type of property-preserving encryption over the embeddings on the server while retaining its ability to do nearest-neighbor search.
		Finally, we simulate an attack against the records --- an \acrshort{ml}-based inversion attack that aims to recover the set of words of a document from its embedding.
		Our goal is to observe and study the correlation of the security parameter (an approximation term) with search accuracy and attack efficiency.

		To summarize, our contributions in this work are as follows:
		\begin{itemize}
			\item
				We construct \kanon{} --- a secure similarity search system in snapshot adversary setting.
				We analyze the correlation of system's security with search accuracy and attack efficiency, and conclude on practicality of the \kanon{} system.

			\item
				We implement a \acrfull{dcpe} scheme from \cite{dcpe} and adapt it to encrypt text embeddings.
				We analyze the practical aspects of \acrshort{dcpe} scheme's security (e.g., the effects of floating point representation) and benchmark the construction.

			\item
				We conduct a set of experiments to study the effect of the security parameter on search accuracy.
				We use a fine-tuned \acrshort{bert} model to produce embeddings for the \acrshort{trec} 2020 collection.
				Given the \acrshort{trec} validation set (i.e., ``correct answers'') we run the search for varying security levels and report a number of ranking quality measures.

			\item
				We adapt a recent \acrshort{ml}-based inversion attack by \textcite{embedding-attacks} against embeddings in our setting.
				The attack works by training an \acrshort{lstm} model on pairs of sentences and their embeddings.
				We run this attack for varying security levels, and training on both plaintext and encrypted records.

		\end{itemize}

	\section{\texorpdfstring{\acrlong{dcpe}}{Distance Comparison Preserving Encryption}}

		A promising approach in secure \acrshort{knn} evaluation is using a property-preserving encryption scheme to allow the existing search algorithms to work with minimal alterations.
		\acrshort{aspe} scheme by \textcite{knn-aspe} is a step in this direction, but their scheme has been shown insecure under a type of chosen plaintext attack in \cite{secure-nn-revisited-break-aspe}.
		Using a common \acrshort{ope} scheme over vector values to encrypt the objects for the purpose of running \acrshort{knn} queries on them has been explored in \cite{quick-n}, but this approach incurs high overhead linear in dimensionality.
		See more detailed related work analysis in \cref{section:related-work:knn}.
		We, therefore, need a different method --- a scheme that operates over high-dimensional vectors and preserves a property that is required to answer the nearest-neighbor queries.

		A classical nearest-neighbor search \cite{knn-wong,knn-cunningham} simply orders the objects according to their distances from the target.
		It is important to note that knowing the exact distance is not required, merely the knowledge of \emph{distance comparison} suffices (i.e., $x$ is closer to $y$ than $z$ is).
		An encryption scheme that preserves the distance comparison would satisfy the \acrshort{knn} search correctness, but not necessarily security or even practicality.
		First, a fully deterministic \acrfull{dcpe} would reveal at least the frequency of data points (i.e., how many times a point appears in the dataset).
		Second, even in the plaintext world the use of approximate nearest-neighbor search \cite{scalable-nn,approximate-nn-fixed-d} may be preferred due to the curse of dimensionality \cite{nn-meaningful,nn-curse-of-d} (the exact distance is less important in higher dimensions).

		\subsection{\texorpdfstring{\acrshort{dcpe}}{DCPE} construction}

			A candidate \emph{approximate} \acrshort{dcpe} scheme that we adapt to our solution has been recently proposed by \textcite{dcpe}.
			The scheme provides the following guarantee on its ciphertexts
			\begin{multline*}
				\forall x, y, z \in \mathbb{X} : \algo{Dist}{x, y} < \algo{Dist}{x, z} - \beta \\
				\implies \algo{Dist}{f(x), f(y)} < \algo{Dist}{f(x), f(z)}
			\end{multline*}
			where $\mathbb{X} \subseteq \mathbb{R}^d$ is the set of $d$-dimensional vectors of real numbers, \algo{Dist} is the $\text{L}_2$ distance over elements in $\mathbb{X}$, and $\beta$ is the approximation term.
			Parameter $\beta$ partially defines the security of the encrypted set --- the larger $\beta$, the fewer distance comparisons are preserved, the less accurate the search and the reconstruction attacks would be.
			\textcite{dcpe} prove protection against membership inference attacks \cite{memebership-inference-attacks-knn} (whether an individual is in the database or not), and against the approximate frequency-finding attacks (how many times the element appears in the set, see \cite{leakage-abuse-grubs-2017} for \acrshort{ore} frequency attacks).
			As for the choice of $\beta$, \textcite{dcpe} prove that $\beta \approx \sqrt{\max N}$ would hide about half of the input bits, for $\max N$ being the maximum vector length in the dataset.


\newlength{\dcpeAlgInterLength}
\setlength{\dcpeAlgInterLength}{0.18ex}

\begin{algorithm*}[ht!]

	\begin{pchstack}

		\procedure[linenumbering]{\algo{KeyGen}{\secparam, \mathbb{S}}}{
			s \sample \mathbb{S}		\\
			\key \sample \bin^\secpar	\\
			\pcreturn (s, \key)
		}

		\hspace{\dcpeAlgInterLength}

		\procedure[linenumbering]{\algo{Enc}{ (s, \key), \vec{m} }}{
			n \sample \bin^\secpar														\\
			\mathsf{coins}_n || \mathsf{coins}_u \gets \algo{\acrshort{prf}}{\key, n}	\\
			\vec{n} \sample \algo{Normal}{0, I_d; \mathsf{coins}_n}						\\
			u \sample \algo{Uniform}{0, 1; \mathsf{coins}_u}							\\
			x \gets \frac{s \beta}{4} \cdot \sqrt[d]{u}									\\
			\vec{\delta} \gets \frac{\vec{n}}{\|\vec{n}\|} \cdot x						\\
			\vec{c} \gets s \cdot \vec{m} + \vec{\delta}								\\
			\pcreturn \vec{c}
		}

		\hspace{\dcpeAlgInterLength}

		\procedure[linenumbering]{\algo{Dec}{ (s, \key), (\vec{c}, n) }}{
			\mathsf{coins}_n || \mathsf{coins}_u \gets \algo{\acrshort{prf}}{\key, n}	\\
			\vec{n} \sample \algo{Normal}{0, I_d; \mathsf{coins}_n}						\\
			u \sample \algo{Uniform}{0, 1; \mathsf{coins}_u}							\\
			x \gets \frac{s \beta}{4} \cdot \sqrt[d]{u}									\\
			\vec{\delta} \gets \frac{\vec{n}}{\|\vec{n}\|} \cdot x						\\
			\vec{m} \gets \frac{\vec{c} - \vec{\delta}}{s}								\\
			\pcreturn \vec{m}
		}

	\end{pchstack}

	\caption[\acrshort{dcpe} scheme]{
		\acrlong{dcpe} scheme, adapted from \cite[Algorithm 2]{dcpe}.
	}%
	\label{algorithm:dcpe}
\end{algorithm*}

			\textcite{dcpe} offer an instantiation of the $\beta$-\acrshort{dcpe} scheme (though not an implementation) that we have adapted to our needs and show on \cref{algorithm:dcpe}.

			The \algo{KeyGen} procedure generates a key \key{} and an amplification factor $s$.
			The key participates in generating the random coins needed to produce deterministic execution, and the amplification factor controls the magnitude of projection of a plaintext object into the ciphertext.

\begin{figure}[!ht]
	\centering
	\includegraphics[width=1.0\textwidth]{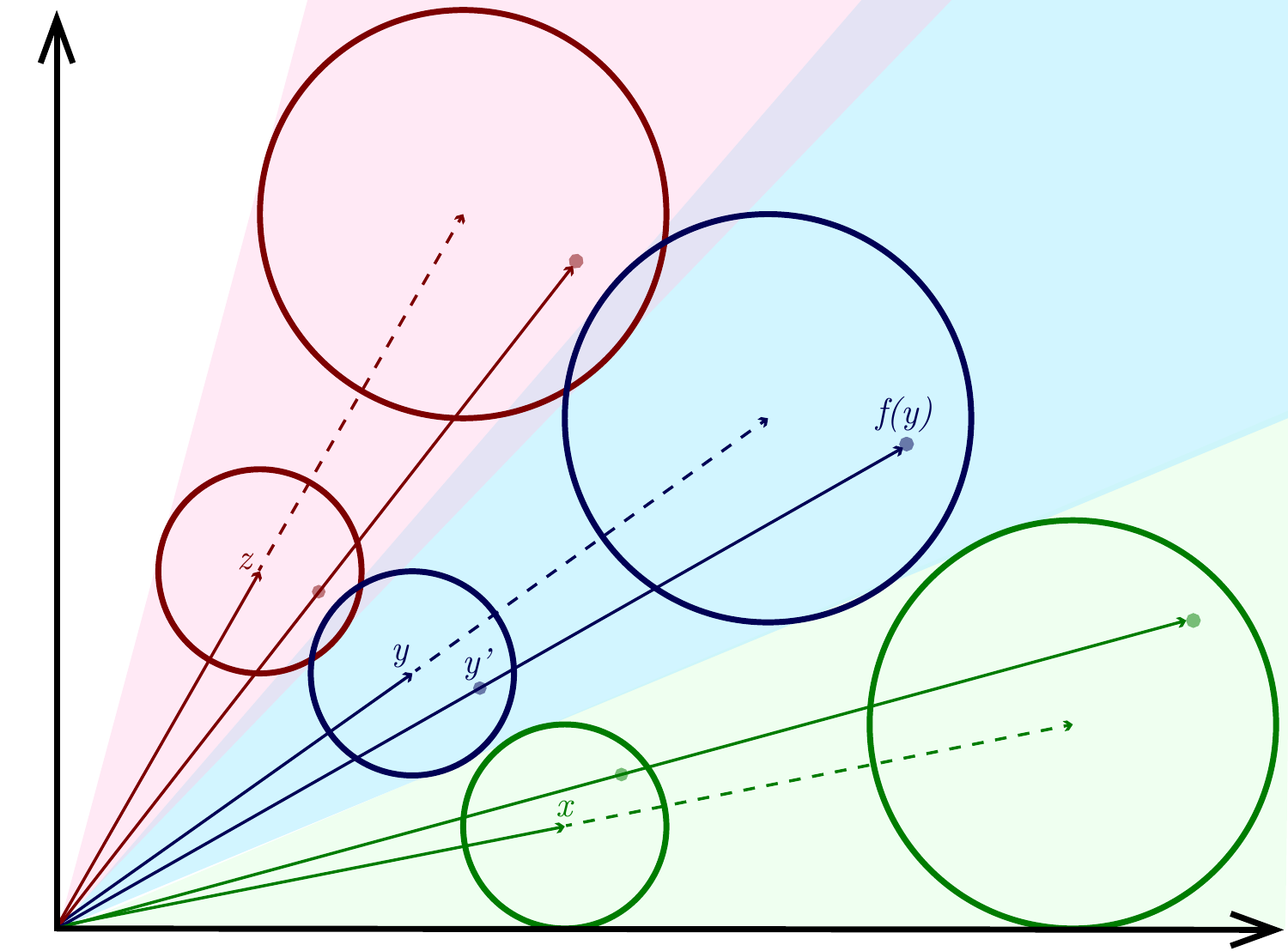}
	\caption[Schematic description of \acrshort{dcpe}]{
		Schematic description of encryption process of \acrshort{dcpe}, drawn to scale.
		In this diagram, there are two dimensions ($d = 2$), $\beta$ (the radius of a circle) is 2 units, and $s$ (the projection magnitude, the length from the origin to the larger circle over the length to the smaller one) is 2.
		The encrypted point is uniformly sampled inside a $\beta$-sphere, then projected $s$ times further from the origin.
		If two points are too close, their circles intersect, and their encryptions can be sampled in a way that breaks distance comparison.
		Intuitively, larger $\beta$ implies greater ciphertext space for a point and greater security.
	}\label{figure:dcpe}
\end{figure}

			The \algo{Enc} procedure ``encrypts'' an object by moving it in space in a way that makes it hard to recover its original position while its distance-comparison respective to other encrypted points is preserved.
			The algorithm first constructs a hypersphere of radius $\beta$, the approximation term, around the input point.
			The routine then samples a new point uniformly inside that hypersphere.
			Finally, that new point is projected into the ciphertext point according to the amplification factor $s$.
			Note that for each encryption the scheme generates a fresh nonce $n$ and uses it along with the key \key{} to generate the coins for the samplers.
			That is, the point in the $\beta$-hypersphere is deterministically set from the nonce (``\textbf{n}umber used only \textbf{once}'', unique per point) and the key (one for all points), and the final ciphertext is projected the same way for all points.
			The \algo{Dec} procedure makes the same steps in reverse, correctly setting the point in the hypersphere using the nonce and the key.
			See \cref{figure:dcpe} for a visual example of \acrshort{dcpe} encryption.

		\subsection{\texorpdfstring{\acrshort{dcpe}}{DCPE} security}

			The security of the scheme thus depends on
			\begin{enumerate*}[label={(\roman*)}]
				\item the maximum amount of amplification,
				\item the radius of the hypersphere $\beta$, and
				\item the entropy of the samplers.
			\end{enumerate*}
			\textcite{dcpe} show that the amplification, $s$ parameter, affects one-wayness bounds \cite[Section 7.2]{dcpe}.
			The approximation term $\beta$ affects bit-security with $\beta \approx \sqrt{\max N}$ protecting about half of the bits.
			Finally, the key \key{} and nonce $n$ sizes, the security parameter \secparam{}, and the samplers used to generate normal multivariate and uniform samples affect the specific amount of entropy used to generate a point in the hypersphere.

			As the construction operates on real numbers, an open question remains on how to avoid negative side-effects of floating point numbers bit representation.
			Unlike integers, floating point numbers are represented in memory in a way that their precision is different depending on their value, see the IEEE 754-2019 standard \cite{ieee-floating-point}. 
			Simply put, the closer the value is to zero, the smaller the difference between two consecutive representable values is.
			For example, while the representable 32-bit IEEE 754 floating point values range from about $1.18 \cdot 10^{-38}$ to $3.4 \cdot 10^{38}$, there are only $2^{32} \approx 4 \cdot 10^9$, 4 billion representable discrete numbers.
			This, along with the rounding errors, puts some limits on how large $s$ and $\beta$ can be.

		\subsection{\texorpdfstring{\acrshort{dcpe}}{DCPE} implementation and benchmarks}

			We offer the first implementation of \cite{dcpe} $\beta$-\acrshort{dcpe} for 32-bit and 64-bit IEEE 754 numbers in C++.\footnote{
				\url{https://github.com/private-knn/dcpe}
			}
			The code is documented, tested and benchmarked, see \cref{table:dcpe-benchmarks}.
			Observe that the difference in performance between encryption and decryption is predictably minimal, and the overhead of encryption grows slower-than-linearly with dimensionality.

			{
	\def\arraystretch{1.0}
	\begin{table}[!ht]
		\begin{tabular*}{\linewidth}{ !{\extracolsep\fill} l l c l } 
			\toprule
				Operation						& Input size								& Dimensions $d$	& Wall-clock time			\\
			\midrule
				\algo{KeyGen}					& N/A										& N/A				& \SI{1.81}{\milli\second}	\\
			\midrule
				\multirow{6}{*}{\algo{Enc}}		& \multirow{3}{*}{32-bit (\texttt{float})}	& 1					& \SI{4.12}{\milli\second}	\\
												& 											& 100				& \SI{12.2}{\milli\second}	\\
												& 											& 768				& \SI{62.0}{\milli\second}	\\ \cline{2-4}
												& \multirow{3}{*}{64-bit (\texttt{double})}	& 1					& \SI{3.96}{\milli\second}	\\
												& 											& 100				& \SI{11.4}{\milli\second}	\\
												& 											& 768				& \SI{59.3}{\milli\second}	\\
			\midrule
				\multirow{6}{*}{\algo{Dec}}		& \multirow{3}{*}{32-bit (\texttt{float})}	& 1					& \SI{3.94}{\milli\second}	\\
												& 											& 100				& \SI{11.6}{\milli\second}	\\
												& 											& 768				& \SI{62.1}{\milli\second}	\\ \cline{2-4}
												& \multirow{3}{*}{64-bit (\texttt{double})}	& 1					& \SI{3.96}{\milli\second}	\\
												& 											& 100				& \SI{11.3}{\milli\second}	\\
												& 											& 768				& \SI{59.6}{\milli\second}	\\
			\bottomrule
		\end{tabular*}
		\caption{\acrshort{dcpe} implementation benchmarks}%
		\label{table:dcpe-benchmarks}
	\end{table}
}

	\section{\texorpdfstring{\acrshort{knn}}{kNN} search accuracy}\label{section:knn-snapshot:search}

		The first part of \kanon{} is the search accuracy.
		In this set of experiments, we embed the documents and apply $\beta$-\acrshort{dcpe} to the embeddings.
		We use existing efficient \acrshort{knn} search algorithms and report ranking quality metrics for different $\beta$.

		\subsection{Secure \texorpdfstring{\acrshort{knn}}{kNN} protocol}

			With the $\beta$-\acrshort{dcpe} as a component, we can model the \kanon{} protocols similar to \acrshort{ore} with \BPlus{} tree ones.
			In the setup protocol \protocolSetup{}, \user{} simply encrypts the entire input, one vector at a time, and sends the encrypted data over to \server{}.
			In the query protocol \protocolQuery{}, \user{} encrypts the query with \acrshort{dcpe}, sends the ciphertext to \server{}, while \server{} runs a standard \acrshort{knn} search against the ciphertext.
			$k$ encrypted vectors are then returned to \user{}, which decrypts them as the last step.
			These protocols are executed for a single set of secrets and \acrshort{dcpe} parameters, including $\beta$.

			For the choice of the dataset, we use the established information retrieval \acrshort{trec} 2020 test collections (MS MARCO passage retrieval collection \cite{ms-marco}).
			A \acrshort{trec} collection consists of a set of documents, a set of topics (questions) and a corresponding set of relevance judgments (correct answers).
			The benefit of using a \acrshort{trec} dataset is being able to evaluate relevant metrics over the produced results, for example, \acrshort{mrr} \cite{mrr} and \acrshort{ndcg} \cite{dcg}.
			We can then track how these metrics, along with the simpler edit distance and set difference over the result, degrade with higher security.

			As the embedding mechanism, we use a custom fine-tuned \acrshort{bert} model.
			\acrfull{bert} is a transformer-based \acrshort{ml} technique for \acrlong{nlp}.
			An original \acrshort{bert} model, published by \textcite{bert}, has been trained on a BookCorpus \cite{bookcorpus} (800 million words) and English Wikipedia (2.5 billion words) using 24 encoders with 16 bidirectional self-attention heads (for the larger of two versions, $\text{\acrshort{bert}}_\text{LARGE}$).
			\acrshort{bert}'s main novelty is its bidirectional nature --- it processes all words in relation to each other and not one-by-one.
			The technology is now prevalent \cite{bert-is-prevalent}, and Google uses \acrshort{bert} in almost every English query\footnote{
				\url{https://searchengineland.com/google-bert-used-on-almost-every-english-query-342193}
			}.

			It is common, however, to use the original \acrshort{bert} model as a base and do training on top.
			We trained a \acrshort{bert}-based dense retrieval model that uses \acrshort{bert} for representing both queries and documents and inner product for computing their similarity \cite{hamed-custom-bert}.
			The parameters between both \acrshort{bert} models (for queries and documents) are shared.
			We used cross entropy loss function for training and used the standard MS MARCO \cite{ms-marco} training set.
			The produced vectors by the \acrshort{bert} model are used in our experiments for approximate nearest neighbor indexing of documents and retrieval for queries.
			We used the test queries produced by the Deep Learning Track of the \acrfull{trec} in 2020 in our experiments.

		\subsection{Experimental evaluation}

			The actual experiment is conducted as follows.
			First, we produce a set of embeddings for 8.8 million documents and 200 queries from \acrshort{trec} 2020 dataset.
			We observed that the maximum length of an embedding vector is about $\sqrt{\max N} \approx 11$ units.
			Second, we encrypt the record and query embeddings using \acrshort{dcpe} and range $\beta$ from 0 (meaning exact distance-comparison) to 50, with $\beta = \sqrt{11} \approx 3.3$ hiding about half of the bits of input embeddings.
			Third, we run the nearest neighbor search on these pairs of data and queries sets using \acrshort{faiss} \cite{faiss}, a \acrshort{gpu}-enabled library for efficient similarity search and clustering of dense vectors.
			Finally, we report a range of ranking quality metrics and generic \acrshort{knn} result metrics.

			\subsubsection{Ranking quality metrics}

				We report recall, \acrfull{mrr} \cite{mrr} and \acrfull{ndcg} \cite{dcg} to assess the ranking quality with respect to \acrshort{trec} relevance judgments.

				\emph{Recall} is the fraction of relevant documents that the query retrieved over all relevant documents.

				\acrfull{mrr} is the average of reciprocal ranks of a query response, which is a multiplicative inverse of the rank of the first correct answer.
				\acrshort{mrr} is defined as
				\[
					\mathsf{\acrshort{mrr}} = \frac{1}{| \mathbb{Q} |} \sum_{i = 1}^{| \mathbb{Q} |} \frac{1}{\mathsf{rank}_i} \; ,
				\]
				where $\mathbb{Q}$ is the sample of queries and $\mathsf{rank}_i$ is the rank position of the first relevant document for the $i_\text{th}$ query.

				\acrfull{ndcg} is another measure of ranking quality most used in web search engine algorithms.
				Its main goal is to produce a metric that would promote the following two assumptions.
				First, a document's relevance implies its usefulness, and second, highly relevant documents are more useful if they have higher rank (appear earlier in the result list).
				\acrshort{ndcg} is a normalized version of discounted cumulative gain, and is defined as the actual over ideal gain up to a position $p$
				\[
					\mathsf{\acrshort{ndcg}}_p = \frac{\mathsf{DCG}_p}{\mathsf{IDCG}_p} = \frac{ \sum_{i=1}^p \frac{ \mathsf{relevance}_i }{ \log_2 (i + 1) } }{ \sum_{i=1}^{ \algo{Rel}{p} } \frac{ \mathsf{relevance}_i }{ \log_2 (i + 1) } } \; ,
				\]
				where $\mathsf{relevance}_i$ is the graded relevance of the result at position $i$ and $\algo{Rel}{p}$ is the number of relevant documents in the corpus up to position $p$.
				Note that we use the classic definition by \textcite{dcg}, and not the one by \textcite{dcg-updated} that puts even greater emphasis on relevance.

				We measure all three metrics at certain cutoffs, meaning that if the number of returned documents is smaller than the cutoff, the missing documents are assumed to be irrelevant.
				The cutoff for recall is \num{1000} and for \acrshort{mrr} and \acrshort{ndcg} is 10, as is common in the information retrieval literature.

				To keep track of the isolated effect of the approximation factor on the \acrshort{knn} results, we also report the set difference and Damerau-Levenshtein distance \cite{levenshtein-distance,damerau-distance} of actual and expected \acrshort{knn} results.
				Result distance is measured as the minimal number of insert, delete and swap operations to get one set from the other.
				The inclusion of swap operation makes metric penalize less the case when the search returned the correct set with only a few documents transposed.
				This property is especially useful for us since the approximate distance comparison preservation results exactly in this kind of small error in the output.
				Result difference is simply a set difference of two outputs.
				This metric does not penalize the wrong order of documents as long as all $k$ relevant documents are present.
				For ease of exposition, we report these two metrics as the fraction of total number of returned documents.
				That is, for \num{1000} returned documents the distance of \num{853} would be reported as \SI{85.3}{\percent}.

			\subsection{Results for varying $\beta$}

				\begin{figure}[h]
	\centering
	\includegraphics[width=1.0\textwidth]{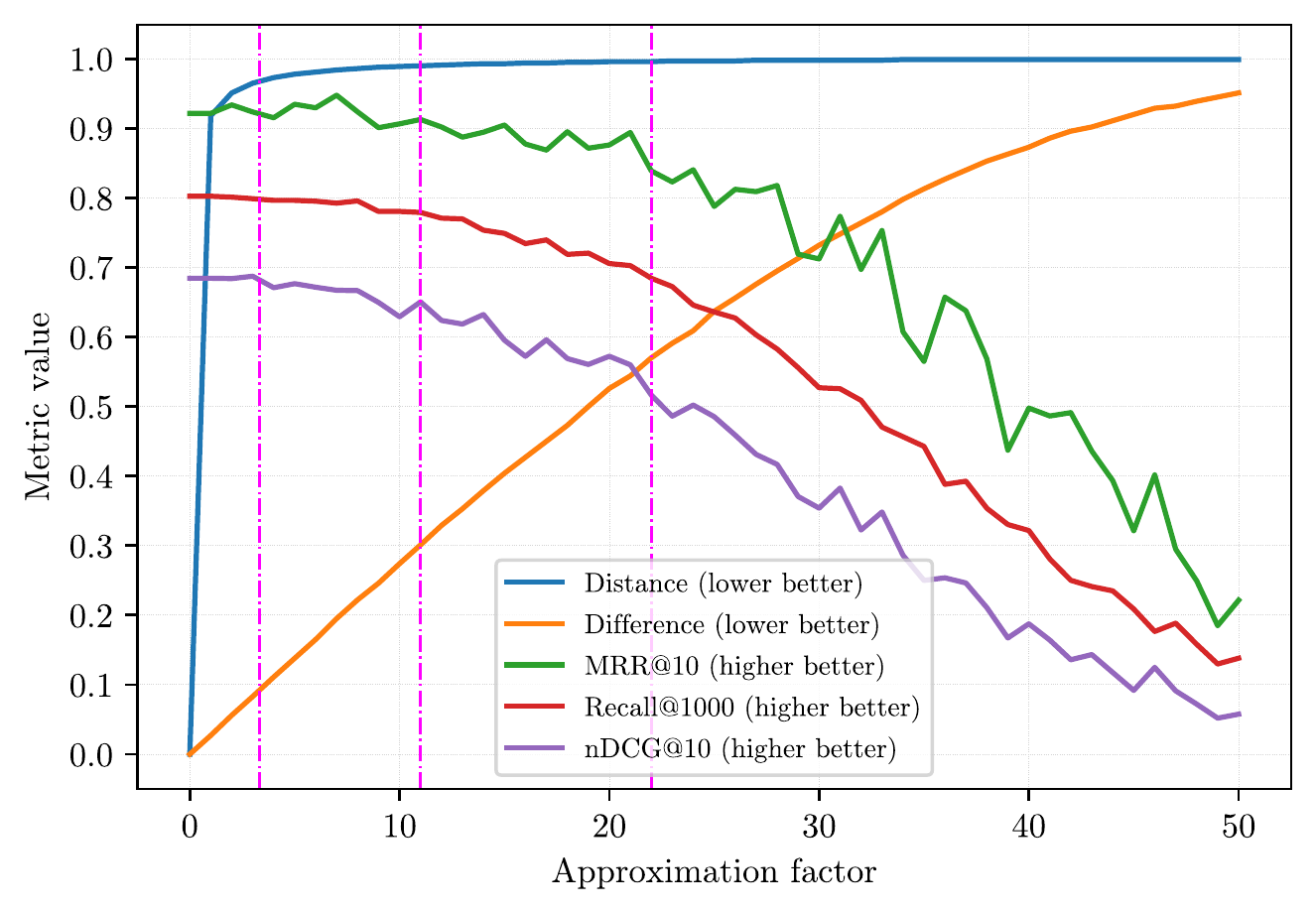} 
	\caption{Search accuracy for $\beta \in \{ 0.0, \ldots, 50.0 \}$}\label{figure:knn-search-coarse}
\end{figure}

				We run two sets of experiments to see how these metrics change with varying the security parameter $\beta$.
				For the first set of $\beta$ values, we ranged the parameter from 0 to 50 with increments of $1.0$, see \cref{figure:knn-search-coarse}.
				Higher values of $\beta$ predictably degraded the search accuracy, but we wanted to see how quickly and after which values of $\beta$ the accuracy starts to fall noticeably.

				First, we notice that the result difference grows close to linearly with $\beta$, which means that each new security level knocks out a few correct responses from the set proportionally.
				Second, we see that the result distance jumps immediately to almost \SI{100}{\percent}, which means that even a tiny approximation term significantly perturbs the order of the responses, but given low difference between the answer sets, not the content of the result.
				Finally, we observe that at about $\beta = 9$, the \acrshort{trec} metrics start to fall sharply and throughout the entire range of $\beta$ they fall in accord.

\begin{figure}[h]
	\centering
	\includegraphics[width=1.0\textwidth]{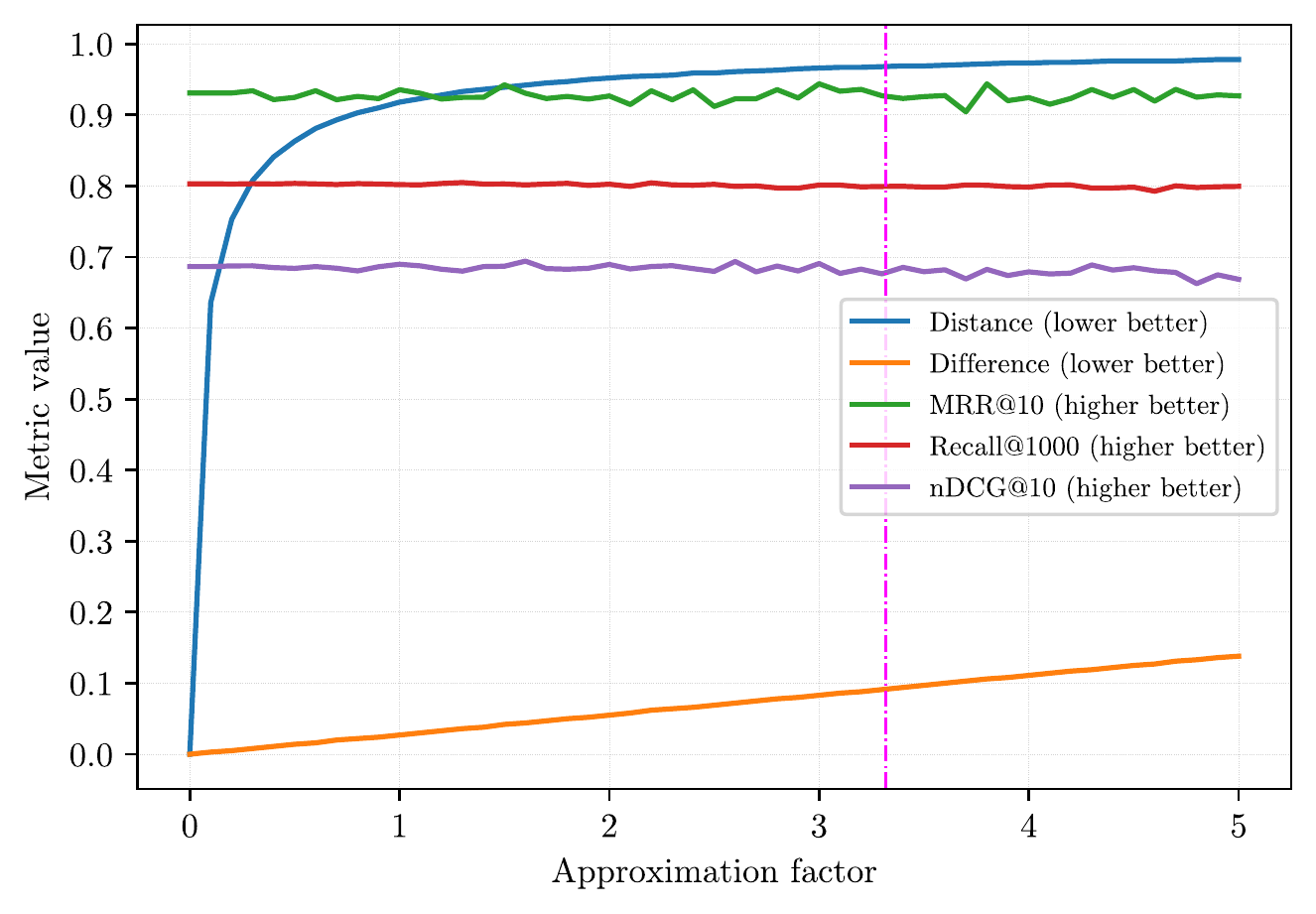}  
	\caption[Search accuracy for $\beta \in \{ 0.0, \ldots, 5.0 \}$]{
		Search accuracy for $\beta \in \{ 0.0, \ldots, 5.0 \}$.
		{\color{magenta}Highlighted} is $\beta = \sqrt{\max N}$ for $\max N \approx 11$ being the longest vector in the dataset.
	}\label{figure:knn-search-fine}
\end{figure}

				For the maximum vector length of $\sqrt{\max N} \approx 11$, the $\beta$ value that hides half of the input bits is $\beta = \sqrt{11} \approx 3.3$.
				To see the metrics behavior at around this point, we ran the experiments again, now ranging $\beta$ in finer manner, from $0.0$ to $5.0$ with increments of $0.1$, see \cref{figure:knn-search-fine}.
				We confirm that for $\beta = \sqrt{\max N} \approx 3.3$, the values of the ranking quality metrics are sufficiently close to the plaintext values.
				\emph{We therefore conclude that the bit-security \acrshort{dcpe} offers comes with a low search accuracy penalty.}

	\section{Security against attacks}\label{section:knn-snapshot:attacks}

		The second part of \kanon{} is the protection against attacks.
		In this set of experiments, we adapt a recent \acrshort{ml}-based attack by \textcite{embedding-attacks}.
		\textcite{embedding-attacks} offer three attacks against text embeddings:
		\begin{enumerate*}[label={(\roman*)}]
			\item \emph{the inversion attack}, which recovers a set of words from a document embedding,
			\item \emph{the attribute inference attack} that recovers some property of the document by its embedding, such as the gender or age of its author, and
			\item \emph{the membership inference attack}, which reveals whether a given document was or was not in the training set for the embedding model.
		\end{enumerate*}
		The inversion attack can be run in two modes: white-box and black-box.
		The white-box attack assumes the access to an embedding model and therefore can directly use its architecture and parameters to invert the inputs.
		The black-box attack relies only on being able to use the model to produce an embedding from a document, similar to how a generic embedding \acrshort{api} would work.
		We use this latter black-box inversion attack in our experiments since it most closely matches the adversary capabilities in our outsourced database setting.

		\subsection{Black-box model inversion attack \texorpdfstring{\cite{embedding-attacks}}{}}

			The black-box inversion attack assumes no knowledge of the embedding model; it can only use it to produce the embeddings.
			In this section we follow the original notation by \textcite{embedding-attacks}.

			The attack works by training another model $\Upsilon$ to recognize the correlation between the set of words $\mathcal{W}(x)$ of a document $x$ and its embedding $\Phi(x)$.
			The attacker uses some auxillary dataset $\database_\text{aux}$ (same domain works predictably better than cross domain), and produces a collection of $ ( \Phi(x), \mathcal{W}(x) ) $ for all $x \in \database_\text{aux}$.
			The adversary then trains the attack model $\Upsilon$ to maximize $ \log \Pr_\Upsilon ( \mathcal{W}(x) | \Phi(x) ) $ over dataset $ ( \Phi(x), \mathcal{W}(x) ) $.
			Finally, the adversary simply queries the model with the embedding $ \Upsilon( \Phi( x ) ) $ and recovers the original words $\mathcal{W}(x)$.

			\textcite{embedding-attacks} offer two models for the inversion attack.
			The \acrfull{mlc} model assigns a binary label of whether a word is in the set for each word in the dictionary.
			The objective function is
			\[
				\mathcal{L}_{\text{\acrshort{mlc}}} = -\sum_{ w \in \mathcal{V} } \left[ y_w \log( \hat{y}_w ) + (1 - y_w) \log(1 - \hat{y}_w) \right]
			\]
			where
			\begin{itemize}
				\item $\mathcal{V}$ is a set of possible words, a dictionary,
				\item $\hat{y}_w = \Pr_\Upsilon( y_w | \Phi( x ) )$ is the predicted probability of word $w$ given $\Upsilon$ conditioned on $\Phi( x )$, and
				\item $y_w = 1$ if word $w$ is in $x$ and 0 otherwise.
			\end{itemize}

			A disadvantage of \acrshort{mlc} model is that it predicts all words independently.
			\textcite{embedding-attacks} therefore offer a more sophisticated approach based on \cite{msp}.
			A \acrfull{msp} recurrent neural network predicts the next word in the set conditioned on the embedding $\Phi( x )$ and the up-to-date predicted set of words.
			The objective function is
			\[
				\mathcal{L}_{\text{\acrshort{msp}}} = \sum_{ i = 1 }^\ell \frac{1}{| \mathcal{W}_i |} \sum_{w \in \mathcal{W}_i } - \log \Pr_\Upsilon( w | \mathcal{W}_{ < i }, \Phi( x ) )
			\]
			where $\mathcal{W}_{ < i }$ is the set of already predicted words up to a timestamp $i$ and $\mathcal{W}_i$ is the set of words left to predict.
			According to experiments by \textcite{embedding-attacks}, \acrshort{msp} outperforms \acrshort{mlc}.

		\subsection{Experimental evaluation}\label{section:knn-snapshot:attacks:experiments}

			We have contacted \textcite{embedding-attacks} and obtained a prototype code they used to run the attack.\footnote{
				\url{https://github.com/google/embedding-tests}
			}
			The model $\Upsilon$ is implemented as a one-layer \acrshort{lstm} with \num{300} hidden units.
			The model is trained for 30 epochs with Adam optimizer \cite{adam-optimizer}, learning rate of $10^{-3}$ and batch size of 256.
			The actual implementation uses TensorFlow \cite{tensorflow} (version 1) and is naturally optimized for \acrshort{gpu} training.

			We run two sets of experiments corresponding to two adversarial settings.
			First, similar to the work of \textcite{embedding-attacks}, we assume that the adversary has a black-box access to the embedding model $\Phi$ and can use it to train $\Upsilon$, and we call this experiment a \emph{public model}.
			This setting corresponds to a scenario when the system uses some open embedding model with little to no alterations (i.e., Google AI Platform Training\footnote{
				\url{https://cloud.google.com/ai-platform/training/docs/algorithms/bert}
			}).
			Second, we assume a more realistic scenario where the embedding model $\Phi$ is not available, and the adversary can only use the system as a whole.
			That is, for a given document $x$, the adversary can only produce the encrypted embedding $\algo{Enc}{\Phi(x)}$.
			We call this experiment a \emph{private model}.
			In both experiments the actual outsourced database is encrypted and the difference is in the dataset $\database_\text{aux}$ on which the adversary can train the attack model $\Upsilon$.
			In the public model experiment, the adversary trains on plaintext embeddings while in the private model environment she trains on the already encrypted embeddings.

			We have used two datasets for both experiments.
			The first dataset is a BookCorpus \cite{bookcorpus} that \textcite{embedding-attacks} used.
			With this dataset we have been able to verify that our adapted attack implementation produces similar results to original \cite{embedding-attacks}.
			The second dataset is the \acrshort{trec} 2020, same as in \cref{section:knn-snapshot:search}.
			With this dataset we can link together the search accuracy and attack efficiency for the same levels of security.

			\subsubsection{Attack efficiency metrics}

				In line with the work of \textcite{embedding-attacks}, we define the attack efficiency metrics similar to a generic \acrshort{ml} measurers of accuracy --- precision, recall and \FOne{} score.
				Precision in our setting is defined as the number of words that the model predicted and that are part of an embedded sentence (i.e., true positives) over the total number of predicted words (all positives).
				Recall is defined as the number of correctly predicted words (true positives) over the number of all words in the sentence (true positives plus false negatives).
				\FOne{} score is then the harmonic mean of these two:
				\[
					\FOne = 2 \cdot \frac{ \mathsf{precision} \cdot \mathsf{recall} }{ \mathsf{precision} + \mathsf{recall} }
				\]

				We go a step further in this direction and given the context of the model --- recovering a set of words from a sentence embedding --- also track the fraction of common words (a.k.a.\ stop-words) in the predicted set.
				\acrshort{bert} explicitly encourages including the stop-words in the input because their relative position matters for the context and thus embedding.
				However, the attack only produces an unordered set of words and not their relative position.
				Therefore, the \emph{model prediction quality}, the \FOne{} score, may seem high, but it may not imply high \emph{attack efficiency}, because a huge fraction of predicted words are common and thus contribute little to added adversary knowledge.
				We define the list of stop words as pronouns, verb forms of ``be'', ``have'' and ``do'', some modal verbs, compound forms (e.g., ``you'll''), negations, articles, certain prepositions, conjunctions, adverbs and some more high-frequency words.\footnote{
					The list is adapted from the Snowball processing language: \url{http://snowball.tartarus.org/algorithms/english/stop.txt}.
				}
				We also include punctuation and digits in the list, as \acrshort{bert} tokenizes these along with the rest of the words.

			\subsubsection{Baselines}\label{section:knn-snapshot:attacks:experiments:baselines}

				\begin{figure}[h]
	\centering
	\includegraphics[width=1.0\textwidth]{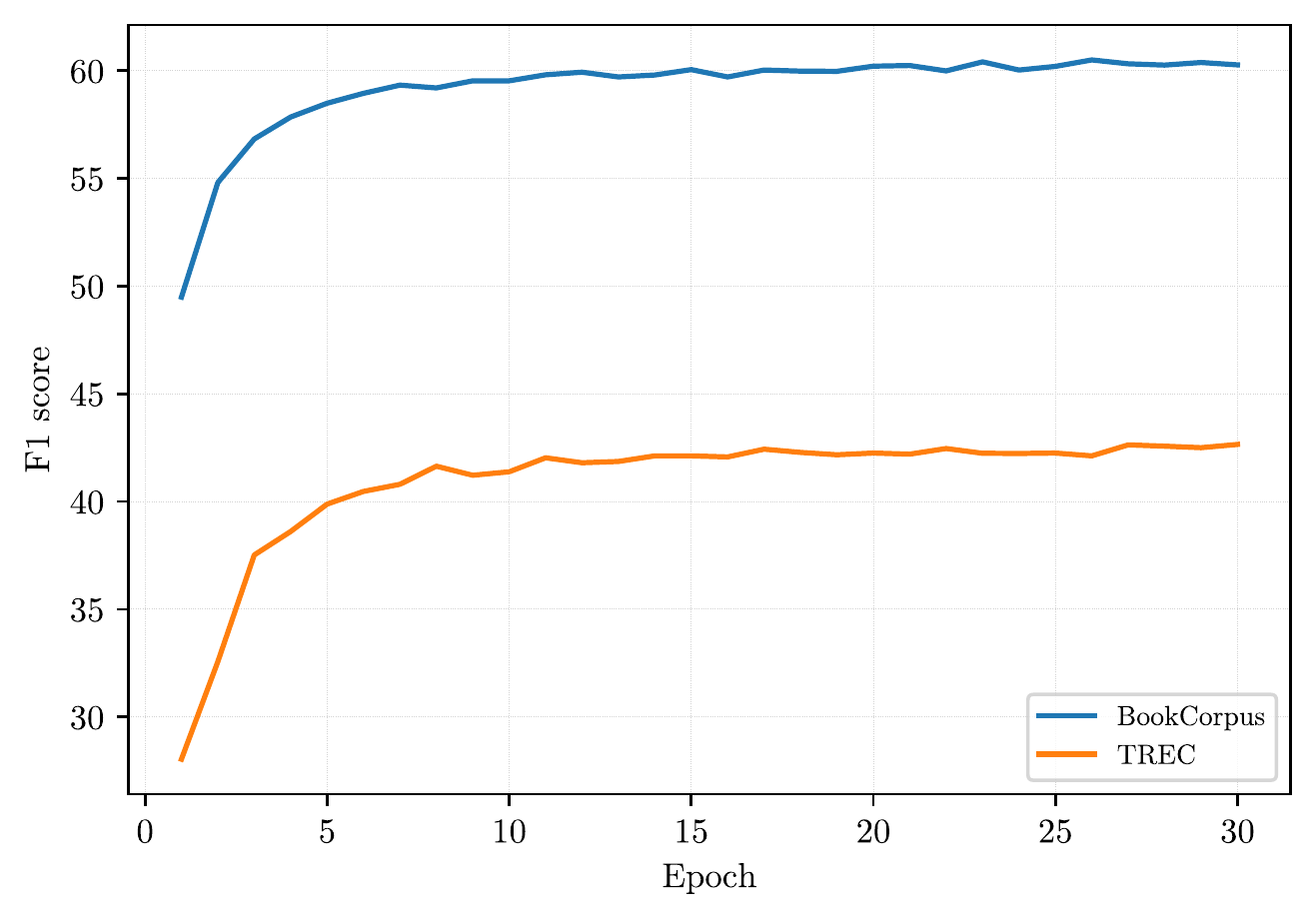} 
	\renewcommand*{\glstextformat}[1]{\textcolor{MatplotlibTwo}{#1}}
	\caption[Inversion attack \FOne{} score for different epochs]{
		Inversion attack \FOne{} score for different epochs and for the {\color{MatplotlibOne}BookCorpus} and \acrshort{trec} datasets.
	}\label{figure:knn-epochs}
	\renewcommand*{\glstextformat}[1]{\textcolor{DarkerRed}{#1}}
\end{figure}

				We have established two baselines between which the attack performance over encrypted inputs is assumed to lie.
				The first baseline is the attack on the plaintext inputs (referred to as \emph{plaintext attack}), a replica of the original attack by \textcite{embedding-attacks}.
				The second baseline is the attack on the random embeddings (referred to as \emph{random attack}), which, counterintuitively, does not produce close to zero \FOne{} score.

				In both cases, we have trained the attack model for 30 epochs (see \FOne{} score on \cref{figure:knn-epochs}).
				We note a number of observations:
				\begin{enumerate*}[label={(\roman*)}]
					\item we have replicated the efficiency result of the black-box model inversion attack from \cite[Table 2, \FOne{} score, same domain, $\mathcal{L}_{\text{\acrshort{msp}}}$, \acrshort{bert}]{embedding-attacks},
					\item for the plaintext attack, the \FOne{} score stops growing after about $10_\text{th}$ epoch,
					\item the random attack produces a far-from-negligible \FOne{} score,
					\item \acrshort{trec} dataset is much less susceptible to the attack than the originally used BookCorpus.
				\end{enumerate*}

				On this latter observation, we speculate that the reason is in the size of the input document, and to the lesser extent different embedding mechanism.
				BookCorpus input document, as used in \cite{embedding-attacks}, is merely a sentence, while \acrshort{trec} document is a larger paragraph.
				We have tuned the maximum token sequence length (shorter sequences are padded, larger ones are truncated), but it yielded no improvement.
				From a purely combinatorial perspective, we conclude that the larger input loses more information while being embedded in the same-sized vector.

				The case of a random attack is puzzling, as one would expect that there is no information to recover from an a-priori information-less (random) inputs.
				We have dived deeper than \FOne{} score and inspected the actual words that the attack recovered and observed that almost all of them are stop-words and punctuation.
				That is, the attack merely established that the input document contained a period, coma, ``the'' and ``a'', and it happened to be right some of the time.
				While such inversion technically results in an \FOne{} score as high as about \SI{22}{\percent}, it does not necessarily translate into an information leakage or a privacy violation.
				We therefore include an evaluation of the recovered words as part of our larger analysis.

				We have run all experiments for both BookCorpus and \acrshort{trec} datasets, and we have noticed that although the absolute values of attack efficiency are higher for BookCorpus, all relations and correlations are the same for both datasets.
				We therefore report on the more relevant \acrshort{trec} dataset in the rest of the chapter.

			\subsubsection{Public model}

\begin{figure}[h]
	\centering
	\includegraphics[width=1.0\textwidth]{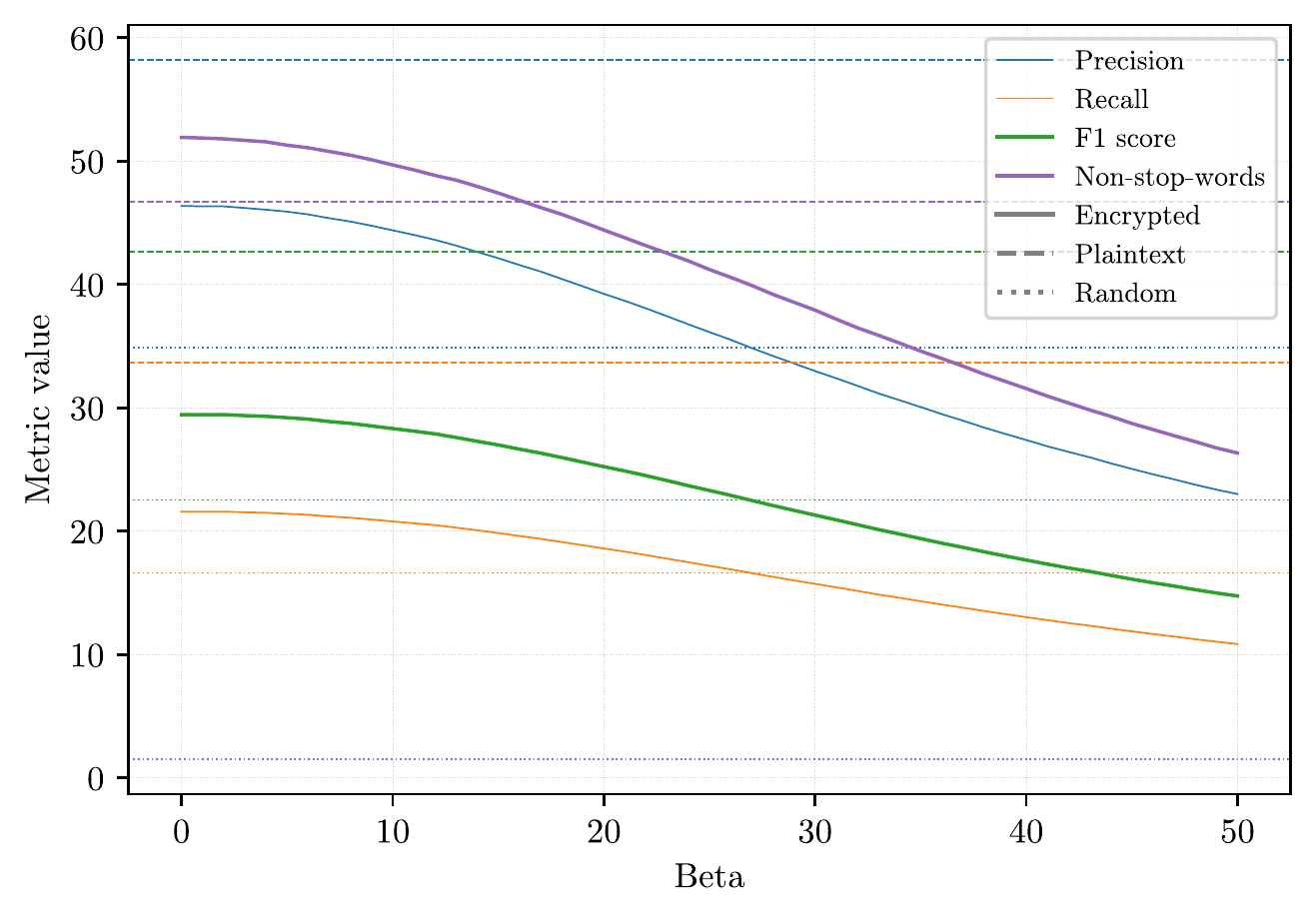} 
	\caption[Inversion attack accuracy metrics for different $\beta$ for \acrshort{trec} dataset]{
		Inversion attack {\color{MatplotlibOne}precision}, {\color{MatplotlibTwo}recall}, {\color{MatplotlibThree}\FOne{} score} and {\color{MatplotlibFive}percentage of non-stop-words among recovered words} for plaintext attack (dashed), random attack (dotted) and different values of $\beta$ (solid) for the \acrshort{trec} dataset.
		Horizontal bars depict the baselines.
	}\label{figure:knn-public}
\end{figure}

				For the public model setting we have used the trained model $\Upsilon$ from the baseline experiments and ran it against the encrypted embeddings.
				That is, for all documents $x$ in $\database_\text{aux}$, we recovered a set of words from the encrypted embeddings $\Upsilon( \algo{Enc}{ \Phi(x) } )$ and compared it to the set of words in the original document $\mathcal{W}(x)$.
				We have repeated the process for different values of $\beta \in \{ 1.0, \ldots, 50.0 \}$, see \cref{figure:knn-public}.

				We observe that even for $\beta = 0$, the attack efficiency drops sharply compared to the plaintext baseline.
				We also see that the metric values drop further as $\beta$ increases, starting slowly for small $\beta$, then accelerating for $\beta$ 10 to 30 and then slowing down again.
				Interestingly, we observe that the attack efficiency dives below the lower baseline, the random embeddings.
				We conclude that $\beta \approx 27$, which is about 2.5 times the maximum length of the input vectors, produces the security equivalent to symmetric encryption.
				Finally, we note that the fraction of stop-words grows fast as the security increases.

			\subsubsection{Private model}

				For the private model experiments, we have trained the model $\Upsilon$ on the already encrypted embeddings for 30 epochs.
				That is, the model $\Upsilon$ is trained on $ ( \algo{Enc}{ \Phi(x) }, \allowbreak \mathcal{W}(x) ) $ pairs from the training dataset and then we run predictions for the encrypted auxillary embeddings, similar to public model.
				The training and validation datasets are encrypted with the same set of public and private parameters (i.e., same key \key{}, $s$ and $\beta$).

				\begin{table}[!ht]
	\renewcommand{\arraystretch}{1.2}
	\sisetup{detect-all = true}
	\begin{tabular*}{\linewidth}{ !{\extracolsep\fill} l c c >{\bfseries}c c } 
		\toprule
			Dataset (encrypted with $\beta$)	& Precision				& Recall				& \FOne{} score 		& Non-stop-words		\\
		\midrule
			$\beta = 0$							& \SI{41.59}{\percent}	& \SI{23.36}{\percent}	& \SI{29.91}{\percent}	& \SI{3.67}{\percent}	\\
			$\beta = \ceil{\sqrt{\max N}} = 4$	& \SI{41.91}{\percent}	& \SI{23.75}{\percent}	& \SI{30.32}{\percent}	& \SI{4.96}{\percent}	\\
			$\beta \approx \max N = 11$			& \SI{40.82}{\percent} 	& \SI{24.20}{\percent} 	& \SI{30.39}{\percent}	& \SI{5.18}{\percent}	\\
			$\beta \approx 2 \cdot \max N = 22$	& \SI{40.44}{\percent} 	& \SI{23.75}{\percent} 	& \SI{29.92}{\percent}	& \SI{5.76}{\percent}	\\
		\midrule
			Random embeddings					& \SI{35.91}{\percent}	& \SI{26.49}{\percent}	& \SI{30.49}{\percent}	& \SI{0}{\percent}		\\
		\bottomrule
	\end{tabular*}
	\sisetup{detect-none = true}
	\caption[Inversion attack performance for the private model experiments]{
		Black-box inversion attack performance for the private model experiments.
		The attack model $\Upsilon$ is both trained and validated on the specified datasets.
	}%
	\label{table:knn-private}
\end{table}

				To produce the datasets, we have chosen $\beta$ values of $0.0$, $\ceil{\sqrt{\max N}}$, $\max N$ and $2 \cdot \max N$ (for $\max N \approx 11$ being the longest input length), as well as random embeddings, see \cref{table:knn-private}.
				We immediately notice that the attack model accuracy metrics are similar among all five datasets, with the metrics for encrypted and random embeddings being almost equal.
				This result implies that for the case of private model --- when the adversary has only black-box access to the system as a whole --- the protection by the \acrshort{dcpe} is absolute.

	\section{Search accuracy against security tradeoff}

		Equipped with the empirical data from \kanon{} experiments on search accuracy and attack efficiency, we can correlate these values with the security parameter, the approximation term $\beta$.

		\begin{figure}[h]
	\centering
	\includegraphics[width=1.0\textwidth]{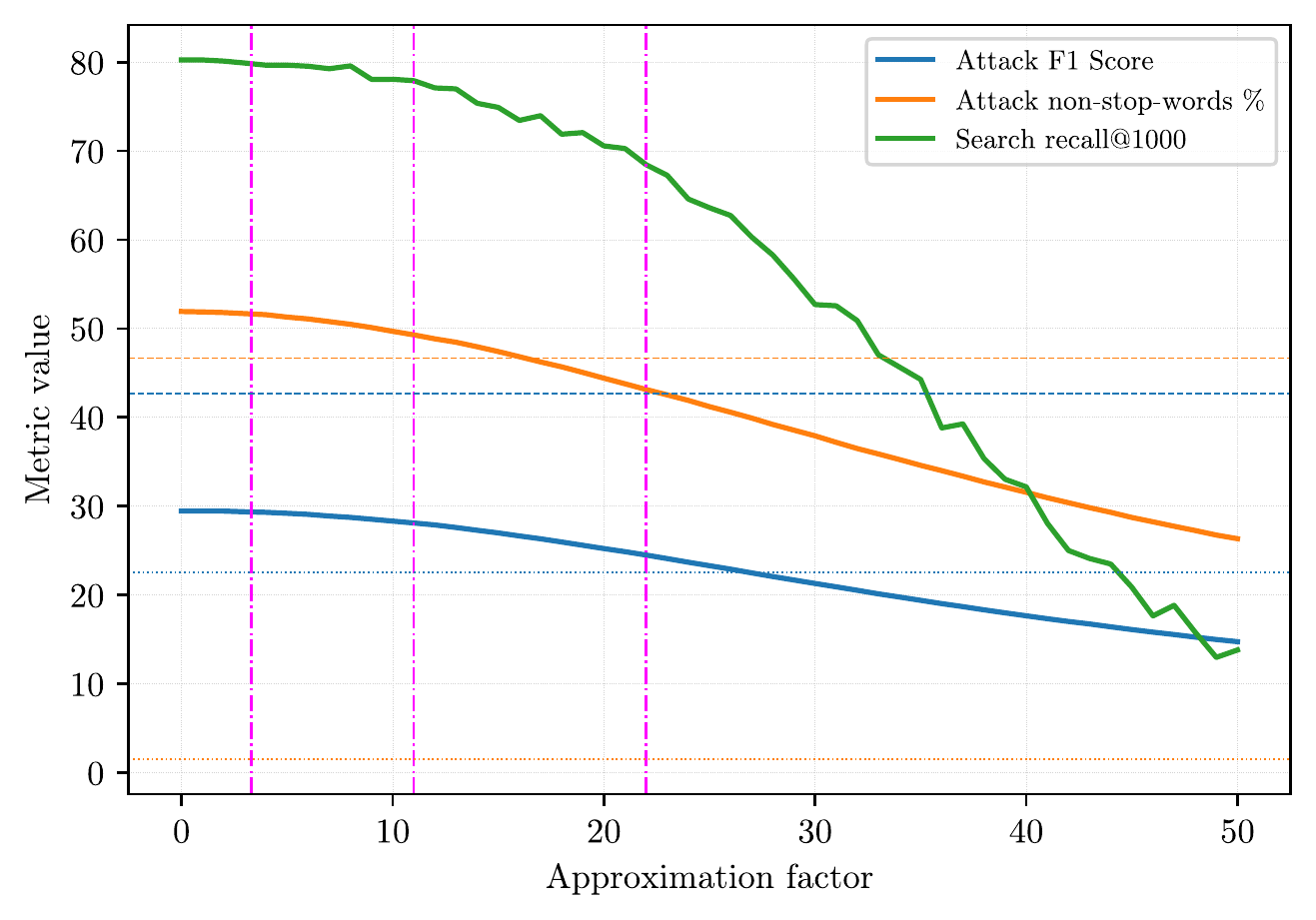} 
	\caption[The correlation of search accuracy and the attack efficiency with $\beta$]{
		The correlation of search accuracy (the {\color{MatplotlibThree}recall@1000}) and the attack efficiency (the {\color{MatplotlibOne}\FOne{} score} and the {\color{MatplotlibTwo}percent of non-stop-words}) with the approximation term $\beta$.
		The {\color{magenta}vertical bars} show special values of $\beta$: $\sqrt{\max N}$, $\max N$ and $2 \cdot \max N$ for $\max N \approx 11$ being the length of the longest vector in the dataset.
	}\label{figure:knn-tradeoff}
\end{figure}

		We note that the search accuracy degrades faster than attack efficiency, which implies that there is a tradeoff between functionality and security.
		Depending on the accuracy loss that the application can tolerate, the plot on \cref{figure:knn-tradeoff} can tell what level of protection against the inversion attack would be at that search accuracy.
		This plot also demonstrates the levels of accuracy and security for the $\beta$ as a derivative of the dataset, in particular, the length of the longest vector embedding $\max N$.

		We observe that at $\beta = \sqrt{\max N}$, the search accuracy corresponds to a plaintext dataset and the attack efficiency has dropped significantly compared to the plaintext version.
		We also see that at $\beta = \max N$ both search accuracy and attack efficiency drop insubstantially compared to $\beta = \sqrt{\max N}$.
		Finally, we note that at $\beta = 2 \cdot \max N$, although both measures drop significantly, after that point the accuracy goes down much faster, which implies that increasing $\beta$ beyond twice the longest vector size is pessimal.
		This value of $\beta$ is also close to a point where the attack \FOne{} score intercepts the \FOne{} score of a random embedding attack, which further confirms that this value of $\beta$ is optimal.

		\emph{Given the significant drop of attack efficiency for smaller $\beta$ while retaining almost optimal search accuracy, we conclude that \kanon{} (applying a \acrlong{dcpe} to an embedding to protect \acrshort{knn} queries) is efficient.}
		\kanon{} developed is also highly tuneable, with $\beta$ corresponding to the application-specific accuracy and security requirements.
		Finally, the construction is cheap in terms of performance and functionality --- the encryption of inputs is very fast and is done only once as a preprocessing step, and the existing algorithms work naturally over the encrypted data.

	\section{Conclusions}

		In this work, we developed and analyzed \kanon{}, a system that answers \acrshort{knn} queries securely in an outsourced setting.
		We adapted a \acrshort{dcpe} scheme, ran experiments on search accuracy and inversion attack efficiency over encrypted inputs.
		We analyzed the correlation between the accuracy and security, and concluded that the approach provides meaningful and tunable security and attack resiliency guarantees for a comparatively small penalty in search accuracy.

		\subsubsection{Future Work}

			An immediate deeper analysis of inversion attack over encrypted embeddings may include an evaluation of the words that the attack model returns beyond simple matching against the set of stop-words.
			It is reasonable to assume that with stronger encryption the \emph{quality} of the predicted words may degrade, where the quality may be defined as the frequency of a word in the vocabulary, for example.

			Another direction can be running other attacks against the embedding, beyond the inversion attack.
			These may be some of the \textcite{embedding-attacks} attacks or adaptations of lots of general attacks against \acrshort{ml} models, like membership attacks \cite{membership-attacks,enhanced-membership-attacks}.

			Finally, the \acrshort{dcpe} adapted from \cite{dcpe} preserves Euclidean distance.
			It would be interesting to explore which results the inner-product distance comparison preserving encryption would produce.

	\cleardoublepage%
	\chapter{Conclusions and Future Work}
\thispagestyle{myheadings}

	In this thesis, we covered the concept of an outsourced database system and two types of adversaries --- snapshot and persistent --- that have different capabilities on an untrusted server.
	We focused on three query types --- point, range and \acrlong{knn} --- and we have gone over the many works that propose systems that execute the relevant query types in a presence of an adversary.
	For the case of range queries in a snapshot adversary model, we provided an in-depth theoretical and practical analysis, an evaluation framework, a benchmark methodology, and its application to five \acrshort{ope} / \acrshort{ore} schemes and five secure range query protocols.
	For the case of point and range queries in a persistent adversary model, we offered an efficient and secure query mechanism, \epsolute{}, along with a novel definition of security, based on \acrlong{dp}.
	For the case of \acrlong{knn} queries in a snapshot adversary model, we offered a similarity search protocol, \kanon{}, and the analysis of its search accuracy and susceptibility to certain inversion attacks.

	The key take-away and future research directions that this thesis highlights are fourfold.

	\subsubsection{Practicality and reproducibility}

		First and foremost, future research in the area of secure outsourced database systems should focus more prominently on practicality and reproducibility.
		After analyzing a plethora of works in the literature (see \cite{ore-benchmark-17,epsolute}) we discovered that a large fraction of constructions either do not have experiments, or their code is unavailable or otherwise not suitable for inspection, or the experimental results are not reproducible.
		We firmly believe in the reproducibility mission (such as SIGMOD\footnote{\url{https://reproducibility.sigmod.org}} and pVLDB\footnote{\url{https://vldb.org/pvldb/reproducibility/}} efforts) and we encourage the works in the area to join the initiative.\footnote{
			We note that our work \cite{ore-benchmark-17} has received ``Most Reproducible Paper'' award.
		}

	\subsubsection{Practicality of property-preserving encryption}

		Second, our works \cite{ore-benchmark-17,k-anon} demonstrate the practical value of property-preserving encryption as a component of a secure database system.
		While an argument can be made that a property-preserving encryption is inherently less-than-ideally secure from a purely cryptographic perspective, we counter that its value is much greater in a practical outsourced database system, which may not necessarily require perfect secrecy.
		A construction using such encryption scheme can be practical as long as the scheme's performance is measured, its leakage is quantified and the effect of this leakage on the security of the entire system is properly analyzed.

	\subsubsection{Practicality of using ``heavy'' primitives and protocols}

		Third, as our work, \epsolute{}, demonstrates, the primitives and protocols that are (rightly) considered heavyweight, such as \acrshort{oram} and \acrshort{dp}-sanitizers, can still be used efficiently in an outsourced system.
		In \epsolute{}, we show that a clever parallelization and optimization on both macro and micro levels can result in a very fast system overall.\footnote{
			We have independently explored running \epsolute{} in a \acrlong{tee}, and we have observed even higher performance.
		}
		We encourage practitioners to revisit using ``heavy'' primitives and protocols, such as \acrshort{oram}, homomorphic encryption, garbled circuits, zero-knowledge proofs, in their systems.

	\subsubsection{More query types}

		Finally, while we have covered three query types for a secure outsourced database system, we need more types to build a full-featured database that can compete with existing mainstream \acrshort{rdbms} like PostgreSQL\@.
		The directions include \texttt{JOIN}, \texttt{GROUP BY}, \texttt{AGGREGATE} queries and custom predicates.

	\cleardoublepage%

	\begin{appendices}


\chapter{Abstract of \cite{bogatov-idemix-2020}}\label{section:appendix:idemix-abstract}
\thispagestyle{myheadings}

	In permissioned blockchain systems, participants are admitted to the network by receiving a credential from a certification authority.
	Each transaction processed by the network is required to be authorized by a valid participant who authenticates via her credential.
	Use case settings where privacy is a concern thus require proper privacy-preserving authentication and authorization mechanisms.

	Anonymous credential schemes allow a user to authenticate while showing only those attributes necessary in a given setting.
	This makes them a great tool for authorizing transactions in permissioned blockchain systems based on the user's attributes.
	In most setups, there is one distinct certification authority for each organization in the network.
	Consequently, the use of plain anonymous credential schemes still leaks the association of a user to the organization that issued her credentials.
	Camenisch, Drijvers and Dubovitskaya \cite{delegatable-creds} therefore suggest the use of a delegatable anonymous credential scheme to also hide that remaining piece of information.

	In this paper, we propose the revocation and auditability --- two functionalities that are necessary for real-world adoption --- and integrate them into the scheme.
	We present a complete protocol, its security definition and the proof, and provide its open-source implementation.
	Our distributed-setting performance measurements show that the integration of the scheme with Hyperledger Fabric \cite{fabric}, while incurring an overhead in comparison to the less privacy-preserving solutions, is practical for settings with stringent privacy requirements.

		\cleardoublepage%


\chapter{Abstract of \cite{dispot}}\label{section:appendix:dispot-abstract}
\thispagestyle{myheadings}

	\noindent \textbf{Motivation:}
		The complexity of protein-protein interactions (PPIs) is further compounded by the fact that an average protein consists of two or more domains, structurally and evolutionary independent subunits.
		Experimental studies have demonstrated that an interaction between a pair of proteins is not carried out by all domains constituting each protein, but rather by a select subset.
		However, finding which domains from each protein mediate the corresponding PPI is a challenging task.

	\noindent \textbf{Results:}
		Here, we present Domain Interaction Statistical POTential (DISPOT), a simple knowledge-based statistical potential that estimates the propensity of an interaction between a pair of protein domains, given their SCOP family annotations.
		The statistical potential is derived based on the analysis of more than \num{352000} structurally resolved protein-protein interactions obtained from DOMMINO, a comprehensive database on structurally resolved macromolecular interactions.

	\noindent \textbf{Availability and implementation:}
		DISPOT is implemented in Python 2.7 and packaged as an open-source tool.
		DISPOT is implemented in two modes, \emph{basic} and \emph{auto-extraction}.
		The source code for both modes is available on \href{https://github.com/korkinlab/dispot}{GitHub} and standalone docker images on \href{https://hub.docker.com/r/korkinlab/dispot}{DockerHub}.
		The web-server is freely available at \href{http://dispot.korkinlab.org/}{dispot.korkinlab.org}.

	\noindent \textbf{Contact:}
		\href{korkin@korkinlab.org}{korkin@korkinlab.org} or \href{onarykov@wpi.edu}{onarykov@wpi.edu}

	\noindent \textbf{Supplementary information:}
		\href{https://academic.oup.com/bioinformatics/article-lookup/doi/10.1093/bioinformatics/btz587#supplementary-data}{Supplementary data} are available at \textit{Bioinformatics} online.

		\cleardoublepage%

	\end{appendices}

	\newpage
	\singlespace%

	\phantomsection%
	\addcontentsline{toc}{chapter}{Bibliography}

	\printbibliography%

	%

\end{document}